\documentclass{aastex63}

\newcommand{\mm }{$\mu $m}

\newcommand{\SB}{erg~s$^{-1}$~cm$^{-2}$~arcsec$^{-2}$}
\newcommand{\EM}{cm$^{-6}$~pc}
\newcommand{\cmq}{cm{$^{-3}$}}

\newcommand{\kms}{km~s{$^{-1}$}}
\newcommand{\msol}{M{$_{\odot}$}}
\newcommand{\Msol}{M{$_{\odot}$}}

\newcommand{\Lsol}{L{$_{\odot}$}}

\newcommand{\Cplus}{{C$^+$}}

\newcommand{\Feii}{Fe~{\sc ii}}

\newcommand{\Hii}{H{\sc ii}}
\newcommand{\Nii}{N{\sc ii}}
\newcommand{\Sii}{S{\sc ii}}
\newcommand{\Oi}{O{\sc i}}

\newcommand{\Oiii}{O{\sc iii}}
\newcommand{\Ha}{H$\alpha$}
\newcommand{\Hb}{H$\beta$}

\newcommand{\Htwo}{H{$_2$}}

\newcommand{\dotMyr}{M$_{\odot}$~year$^{-1}$}
\newcommand{\rmaa}{RMxAA}

\received{\today}
\revised{\today}
\accepted{\today}

\submitjournal{ApJ}

\shorttitle{HH 80 \& 81}
\shortauthors{Bally et al.}

\begin{document}

\title{HH~80/81: Structure and Kinematics of the Fastest Protostellar Outflow.}

\correspondingauthor{John Bally}
\email{john.bally@colorado.edu}

\author[0000-0001-8135-6612]{John Bally}

\affiliation{Center for Astrophysics and Space Astronomy, 
     Department of Astrophysical and Planetary Sciences \\
     University of Colorado, Boulder, CO 80389, USA} 
     
\author{Bo Reipurth}
\affiliation{Institute for Astronomy, University of Hawaii at Manoa, 
      640 North Aohoku Place, Hilo, HI 96720, USA and \\
      Planetary Science Institute, 
      1700 East Fort Lowell, Suite 106, Tucson, AZ 85719}
     

\begin{abstract}
Hubble Space Telescope images obtained in 2018 are combined with archival HST data taken in 1995 to detect changes and measure proper motions in the HH~80/81 shock complex which is powered by the fastest known jet driven by a forming star, the massive object IRAS~18162-2048.   Some persistent features close to the radio jet axis have proper motions of $>$1,000 \kms\ away from IRAS~18162-2048. About 3 to 5 parsecs downstream from the IRAS source and beyond HH~80/81, \Ha\ emission traces the rim of a parsec-scale bubble blown by the jet.  Lower speed motions are seen in [\Sii ] away from the jet axis; these features have a large component of motion at right-angles to the jet. We identify new HH objects and \Htwo\ shocks in the counterflow opposite HH~80/81.  The northeastern counterflow to HH~80/81 exhibits an extended but faint complex of 2.12 \mm\ \Htwo\ shocks.  The inner portion of the outflow is traced by dim 1.64 \mm\ [\Feii ] emission.  The full extent of this outflow is at least 1,500\arcsec\ ($\sim$10 pc in projection at a distance of 1.4 kpc).  We speculate about the conditions responsible for the production of the ultra-fast jet and the absence of {\bf prominent} large-scale molecular outflow lobes.
\end{abstract}


\section{Introduction}

Most forming stars drive jets and bipolar outflows \citep{Bally2016}.  
Herbig-Haro (HH) objects are visual-wavelength, shock-excited nebulae associated with 
jets and outflows from forming stars \citep{ReipurthBally2001}.  HH~80/81, 
discovered by \citet{ReipurthGraham1988} and located 
in Sagittarius,  is an unusually high-excitation shock system dominated by 
strong 5007\AA\ [\Oiii ] and \Ha\ emission.  
This shock system is powered by a radio continuum jet emerging from  the luminous infrared source, 
IRAS 18162-2048, also known as GGD~27~MM1 \citep{RodriguezReipurth1989}.
This embedded source illuminates the near-infrared and visual-wavelength
reflection nebula GGD~27 \citep{Gyulbudaghian1978,Aspin1991,Aspin1994,Aspin1994b}.  
Previous estimates 
placed the host cloud, Lynds 291 (L291), at a distance of $\sim$1.7 kpc.  

The distance to the L291 cloud was constrained by \citet{AnezLopez2020}, who used both the increase of stellar polarization towards stars
near IRAS~18162-2048 and the abrupt increase of extinction as a function of decreasing
parallax angle.  These polarization data  
imply a distance of 1248$\pm$66 pc.  Their extinction measurements imply a distance 
1270$\pm$65 pc.  \citet{Zucker2020} used Gaia DR2 parallax and reddening measurements 
of stars in the L291 field  and found that the extinction increases dramatically at a 
distance of about  1,400$\pm$70 pc .  Because the \citet{Zucker2020} distance 
determination used a larger number of stars, we adopt this distance for the 
IRAS 18162-2048 / GGD~27 region and its outflow in the analysis presented here.

IRAS 18162-2048 is the most massive protostar in the L291 molecular cloud. 
The total flux scaled to a distance of 1.4 kpc yields a source luminosity 
L=$1.15 \times 10^4~$\Lsol ,  implying that the IRAS source contains an early-B or 
late-O type star in formation \citep{FernandezLopez2011}. 
ALMA observations reveal a massive, $\sim$5~\Msol\ circumstellar disk surrounding a 
$\sim$20 \Msol\ protostar \citep{AnezLopez2020,Girart2017,Girart2018,Fernandez_Lopez2023}.

The radio continuum jet driven by 
IRAS 18162-2048 is highly collimated \citep{Marti1993}.  
Several radio knots located along the jet axis imply a projected length of 
$\sim$10 parsecs \citep{Masque2012}.   The inner jet close to the source
has the fastest speed of any known protostellar 
outflow.  Radio proper motions indicate speeds up to $\sim$1200 \kms\ 
for the inner radio knots and $\sim$400 \kms\ for the outer knots when scaled 
to a distance of 1.4 kpc \citep{Marti1995,Marti1998,Masque2012,Masque2015}.
The jet exhibits both thermal and polarized non-thermal emission indicating
active acceleration of relativistic particles
\citep{CarrascoGonzalez2012,RodriguezKamenetsky2017}.
Low-frequency (325, 610, and 1,300 MHz) non-thermal emission from the HH~80/81 
radio knots was found by the Giant Meterwave Radio Telescope with a spectral index 
$-$0.7 \citep{Vig2018}.
X-rays were detected from HH~80/81 by \citet{Pravdo2004,Pravdo2009}.  This is
the first outflow from a forming star in which hard x-rays with energies up to
5 keV are detected \citep{LopezSantiago2013}.  
$\gamma$-rays have also been detected from a degree-scale region
containing HH~80/81 and IRAS 18162-2048 extending 
to an energy of 1 GeV by the LAT detector on the Fermi gamma-ray observatory 
\citep{Yan2022}.   
Spitzer 8 and 24 \mm\ images show a bi-conical infrared cavity 
(or a quasi-cylindrical cavity with a pinch)
centered on the IRAS source and surrounding the radio jet \citep{Qiu2008}.

In single-dish observations of the inner 1.5\arcmin\ region surrounding GGD~27, 
\citet{Qiu2019} report a two component, wide angle, low-velocity molecular 
outflow in CO with a spatial
extent of only about 1\arcmin\ - much smaller than the extent of the outflow
traced by HH objects, molecular hydrogen objects \citep{Mohan2023}, 
and radio continuum emission.    In these CO data, the line wings have Doppler 
shifts of less than 10 \kms\ 
with respect to the host cloud. This component of the outflow has a mass
of about 2 to 3 \Msol.  The redshifted CO lobe extends towards the
south-southwest.    The north-northeast CO lobe is very short with a low 
radial-velocity and wide opening angle.  

High resolution ALMA 1.14 mm observations reveal a cluster of at least 25 compact continuum sources within a 22\arcsec\ field-of-view centered on IRAS~18162-2048 \citep{Busquet2019}. 
Sub Millimeter Array interferometric spectral line maps of the central 1\arcmin\ diameter region show at least three monopolar outflow lobes emerging at angles that differ from the IRAS 18162-2048 radio jet \citep{Qiu2009,Fernandez_Lopez2013}.    These outflows originate from the embedded YSO binary MM2 located about 6\arcsec\ northeast of IRAS 18162-2048 and possibly from a molecular core located about 2\arcsec\ north of MM2.   The southeast facing jet-like flow from MM2 exhibits bullets with speeds up to 100 \kms\ and a total velocity extend of $\sim$190 \kms .   The northwest and northeast facing lobes are only seen at low radial velocities. These observations suggest that IRAS 18162-2048 star forming clump contains several cores in addition to the cluster of young stellar objects.   IRAS 18162-2048 (MM1) is the most luminous.  Given the sizes of their outflows, MM1 is likely older than  MM2 and the molecular core.

\citet{Heathcote1998} presented Hubble Space Telescope (HST) images of  
HH~80/81 and a combination of low- and high-dispersion
spectra.  These authors used older ground-based images to 
measure proper motions.    They found proper motions up to $\sim$800 \kms\ in 
some of the high-excitation knots in the HH 80/81 complex. Thus, this outflow 
exhibits the fastest known motions in an outflow from a forming star in both radio and visual-wavelength tracers.    The high-dispersion spectra show that the  HH~80/81 shock complex is redshifted with radial velocities ranging from -100  to over +600 \kms .   Using the highest radial velocities, proper motions, and a fit to a bow-shock model, \citet{Heathcote1998} derive an inclination angle of the outflow axis with respect to the plane of the sky of about 56$\pm$5\arcdeg\ at a distance of 1.7 kpc used in their analysis.  At a distance of 1.4 kpc this corresponds to  61$\pm$5\arcdeg.  Using the shape of the disk in a well-resolved 1.14 mm image, \citet{AnezLopez2020} derived an inclination angle of the disk axis of 49$\pm$5\arcdeg  . 

In this study, we combine the 1995 HST images with new HST observations 
taken in 2018 to investigate changes in the structure of the HH~80/81 shocks and
to measure new proper motions.    New ground-based images are used to identify
shocks in the counterflow which extends north-northeast of GGD~27.  These
images also reveal a parsec-scale \Ha\ bubble south-southwest of HH~80/81 beyond
the projected edge of the L291 molecular cloud along the outflow axis.   
The bubble is likely powered by the fast jet driven by IRAS~18162-2048.

\section{Observations}

\subsection{HST}
 
HST was used to image HH~80/81
in 1995 using WFPC2 (Program ID 6128; PI Reipurth) and again 
in 2018 using WFC3 (Program ID 15353; PI Reipurth).   In 1995,
multiple exposures were obtained in the WFPC2 narrow-band 
F502N ([\Oiii ]), F656N (\Ha ), and F673N ([\Sii ]) filters.  
In 2018,  the WFC3 narrow-band F547N (\Hb ), 
F502N ([\Oiii ]), F656N (\Ha ), and F673N ([\Sii ]) filters 
were used.  [\Oiii ] and the \Ha\ and \Hb\ hydrogen recombination
lines trace fast shocks with speeds in excess
of 100 to 150 \kms .   [\Sii ] traces slower shocks with speeds 
between 10 to 100 \kms .   Unfortunately, a guiding failure  
resulted in the loss of the \Ha\ image during the 2018 
observations.   The \Ha\ images from 1995 are combined with 
the new \Hb\ image from 2018 for the analysis of nebular evolution 
and  proper motions traced by hydrogen line emission.
Table 1 lists the observations used in this analysis. 

 \subsection{Apache Point Observatory}
 
\Ha\ images were obtained with the Apache Point 
Observatory (APO) 3.5 meter telescope between 14 May 2018
and 23 September 2022 using the 4096 by 4096 pixel  
ARCTIC CCD camera binned 3$\times$3 to give an effective pixel-scale of 
0.342\arcsec\ per pixel at the f/10.3 focus of the APO telescope.   
During the earlier observing runs, we used a narrow-band filter with a 30\AA\
bandpass centered at 6570\AA\ (\Ha) which excludes emission from the 6584\AA\
[\Nii ] emission line.  The 30\AA\ filter is only 4~inches in diameter and caused 
significant vignetting.  In 2021, 5~inch filters with pass-bands of 78 \AA\ 
centered on the 5007\AA\ [\Oiii ],  76\AA\ centered at 6726\AA\ to cover the 
6717+6731\AA\ [\Sii ] lines, and 78\AA\ centered at 6563\AA\ to cover the \Ha\ 
line were acquired.   This broader \Ha\ filter also transmits the
6548 and 6584\AA\ [\Nii ] lines. 
The HH~80/81 outflow was re-observed in 2021 and 2022  with these filters.
Exposure times, filters, and observation dates are listed in 
Table 2.  Three to twenty frames were acquired at each exposure time and median 
combined to remove cosmic rays.   Standard procedures were used for 
bias and dark current removal, and flat-fielding was done using twilight flats.

Narrow-band near-infrared images presented here were obtained using the APO 3.5 meter 
telescope with the NICFPS camera on the dates indicated
in Table~2. NICFPS uses a 1024 $\times$ 1024 
pixel Rockwell Hawaii1-RG HgCdTe detector.   The pixel  scale of this instrument is 
0.273\arcsec\ per pixel with a field of view  4.85\arcmin\ on each side. 
The narrow-band filters have band-passes of $\sim$0.4\% of the central wavelength.
Narrow-band filters centered off-line were used to obtain an off-line continuum frame
to remove stars and reflection nebulosity.  Images with 180 second exposures were 
obtained in the 2.122 $\mu$m S(1) line of H$_2$.   The central-wavelengths and 
band-passes  are listed in Table~2.  Separate off-source sky frames in each filter 
were interspersed with on-source images using the same exposure time at locations 
offset by at least 5\arcmin .   

During each observation, a set of 5 dithered images were obtained on both on-source 
and off-source positions. A median-combined set of unregistered, 
mode-subtracted sky frames were used to 
form a master sky-frame that was subtracted from each individual image.   The reduced 
images were corrected for optical distortions.   Field stars were used to align 
the frames, which were median-combined to produce the final images.  Atmospheric 
seeing produced  $\sim$0.9 to 1.5\arcsec\ FWHM stellar images.   The
observations are summarized in Table~2.
Continuum subtracted images showing only \Htwo\ emission were formed by subtracting
the 2.13 $\mu$m images from the 2.12 $\mu$m images.

\begin{table}
	\centering
	\caption{HST Observations Used For This Analysis}
	\label{tab:HST}
	\begin{tabular}{llcccr} 
		\hline
		Field	& Date                      & MJD           & Instrument    & Filter            & Exposure   \\
		\hline
		HH80/81 & 25 August 1995            & 49954         & WFPC2          & F502N [\Oiii ]    & 10,700 s  \\
		HH80/81 & 26 Jul.-19 Sept. 2018     & 58325-58380   & WFC3          & F502N [\Oiii ]    &  8,332 s  \\
		HH80/81 & 25 August 1995            & 49954         & WFPC2          & F656N \Ha         &  5,200 s  \\
		HH80/81 & 20 September 2018         & 58381         & WFC3          & F487N \Hb         &  5,590 s  \\
		HH80/81 & 25 August 1995            & 49954         & WFPC2          & F673N [\Sii ]     &  8,000 s  \\
		HH80/81 & 26 Jul.-23 Sept. 1018     & 58325-58384   & WFC3          & F673N [\Sii ]     &  8,332 s  \\
		\hline
	\end{tabular}
\end{table}

\begin{table}
	\centering
	\caption{Apache Point Observations Used For This Analysis}
	\label{tab:APO}
	\begin{tabular}{lccrl} 
		\hline
		 Date               & Instrument    & Filter            & Exposure  &  Comments\\
		\hline

14 May  2018    &  ~~~~"    & \Ha\ UNM 657  & 5 $\times$ 300 s  &  Five 5\arcmin\ fields.  $\Delta \lambda$ = 3 nm  \\		
18 June 2018    &  ~~~~"    & \Ha\ UNM 657  & 5 $\times$ 300 s  &  Five 5\arcmin\ fields.  $\Delta \lambda$ = 3 nm  \\
5  July 2019    &  ~~~~"    & SDSS, g, r, i & 5 $\times$ 120 s  &  Five 5\arcmin\ fields.    \\
26 July 2020    &  ~~~~"    & \Ha\ UNM 657  & 5 $\times$ 900 s  &  Five 5\arcmin\ fields.  $\Delta \lambda$ = 3 nm  \\	     
17, 19, 27 Sept. 2021  & APO/NICFPS    & \Htwo (2.12) ,off (2.13)   & 20 $\times$ 180 s & Four 5\arcmin\ fields.  $\Delta \lambda$ = 0.33\%.  \\
17 Sept. 2021 & APO/ARCTIC   & [\Feii ] (1.64), off (1.65)  & 10 $\times$ 180 s &  Two 5\arcmin\ fields.  $\Delta \lambda$ = 0.33\%.  \\

9 Oct.  2021    &  ~~~~"    &  [\Sii ]      & 3 $\times$ 300 s  & Four 8\arcmin\ fields.  $\Delta \lambda$ = 8 nm. \\
	~~~~~~~"    &  ~~~~ "   &  [\Oiii ], \Ha\ & 6 $\times$ 300 s  & Central field only.  $\Delta \lambda$ = 8 nm. \\
23 Sept. 2022   &  ~~~~"    & \Ha\       &  3 $\times$ 300 s     & Two 5\arcmin\ fields.  $\Delta \lambda$ = 8 nm; S-SW bow  \\
		\hline
	\end{tabular}
\end{table}

\begin{figure*}
\center{
\includegraphics[width=4.5in]{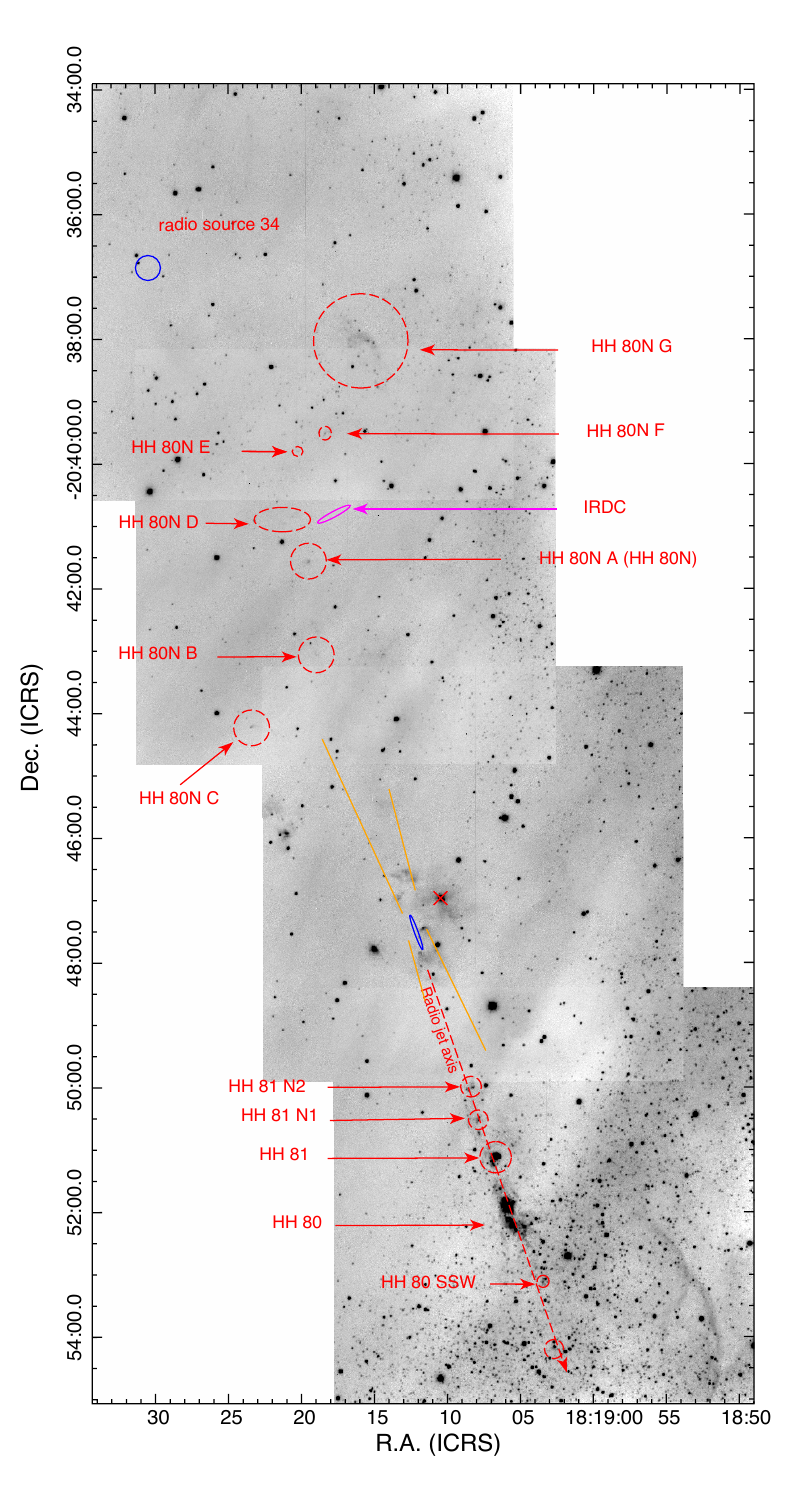  }
}
\caption{Overview of the HH~80/81 giant outflow in an \Ha\ mosaic from the APO 3.5m telescoe obtained with the 30\AA\ bandwidth filter.  This image has had a large-scale intensity gradient removed.  The blue oval shows the location of IRAS 18162-2018 (GGD~27) and the approximate orientation of the radio jet emerging from this source.  Yellow lines surrounding GGD~27 show the approximate location and spatial extent of the outflow cavity walls as traced in Spitzer 8~$\mu$m and 24~$\mu$m images.  Dashed red circles mark the locations of various shocks and radio features.  A blue circle near the top marks the location of radio source 34 at the suspected end of the radio jet.  The magenta oval marks the location of the IRDC discussed in the text. A {\bf red} `X' symbol near the source region marks the B2/B3 star located at the center of the circular near-IR \Htwo\ bubble discussed in the text.  The small, unlabeled, dashed circle near the bottom along the radio jet axis marks the location of a diffuse \Ha\ knot which is also shown in Figure~\ref{fig2_SSW_bow}.}
    \label{fig1_overview}   
\end{figure*}

\begin{figure*}
\center{
\includegraphics[width=6.5in]{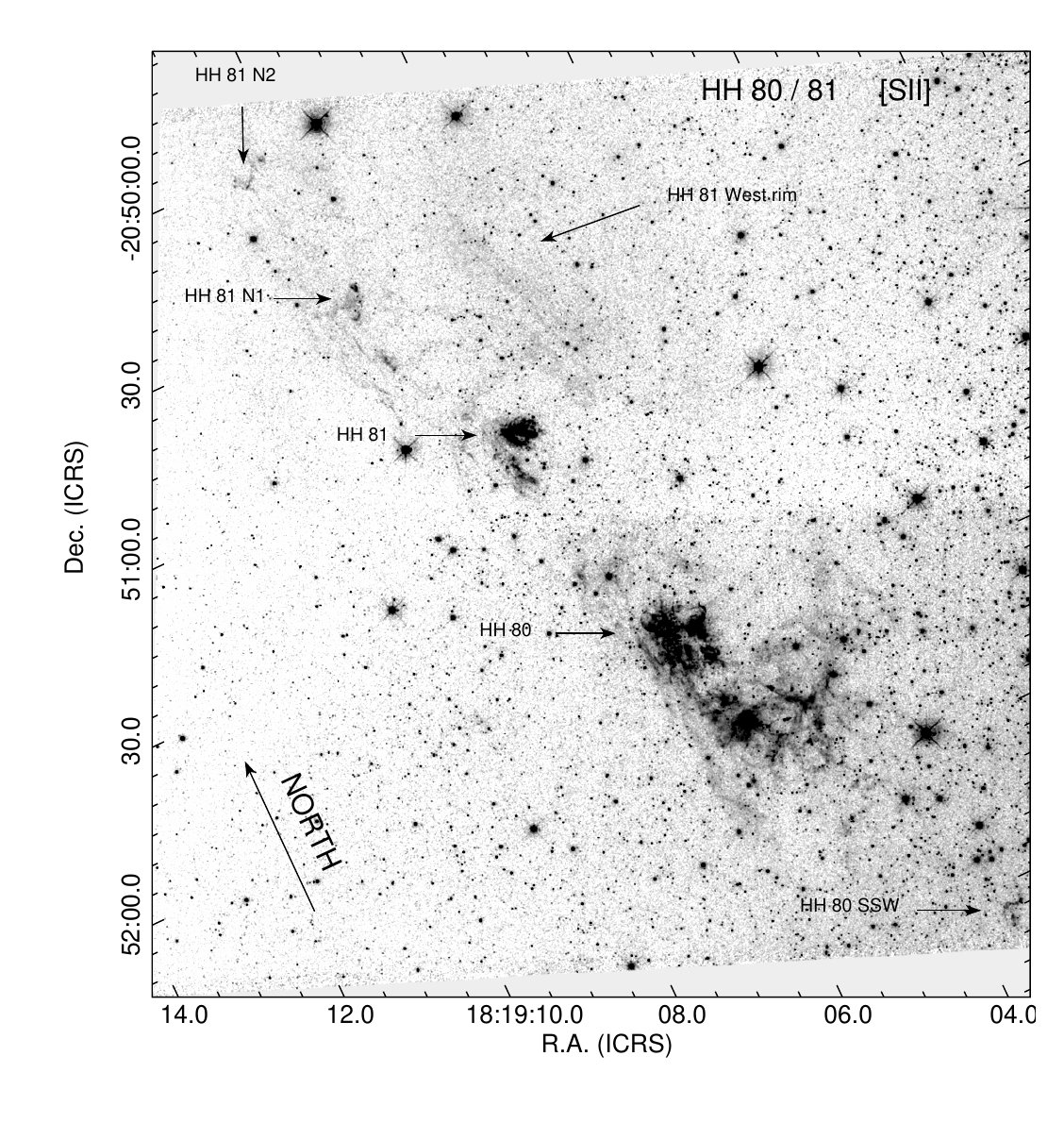  }
}
\caption{The full field observed in 2018 with HST showing HH~80/81 in [\Sii ]. }
    \label{fig_HH80_81_Sii_inverted}  
\end{figure*}

\begin{figure*}
\center{
\includegraphics[width=6.5in]{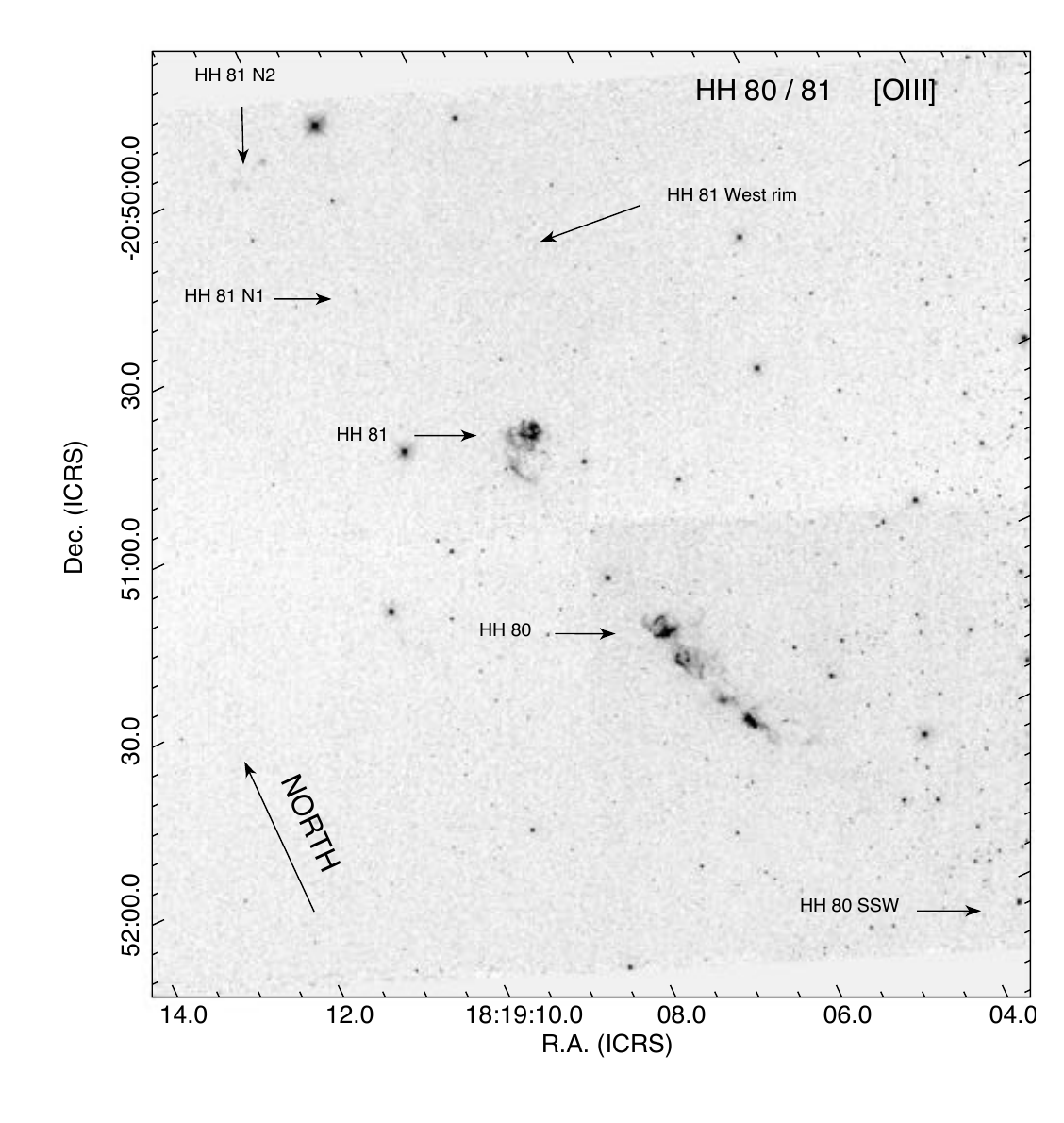}
}
\caption{The full field observed in 2018 with HST showing HH~80/81 [\Oiii ]. }
    \label{fig_HH80_81_Oiii_inverted}  
\end{figure*}

\begin{figure*}
\center{
\includegraphics[width=6.5in]{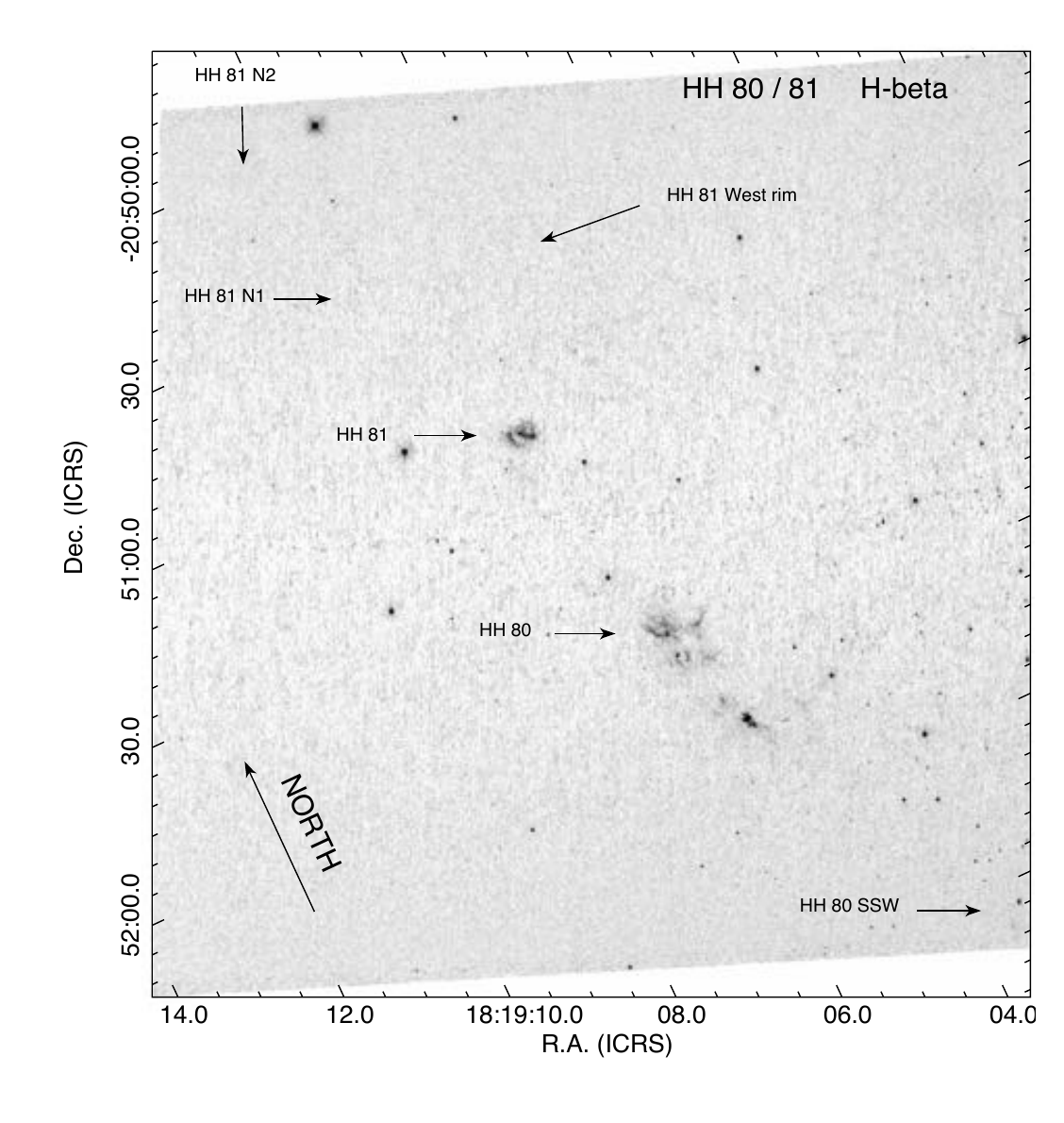  }
}
\caption{The full field observed in 2018 with HST showing HH~80/81 in \Hb . }
    \label{fig_HH80_81_Hbeta_inverted}  
\end{figure*}

\begin{figure*}
\center{
\includegraphics[width=5.0in]{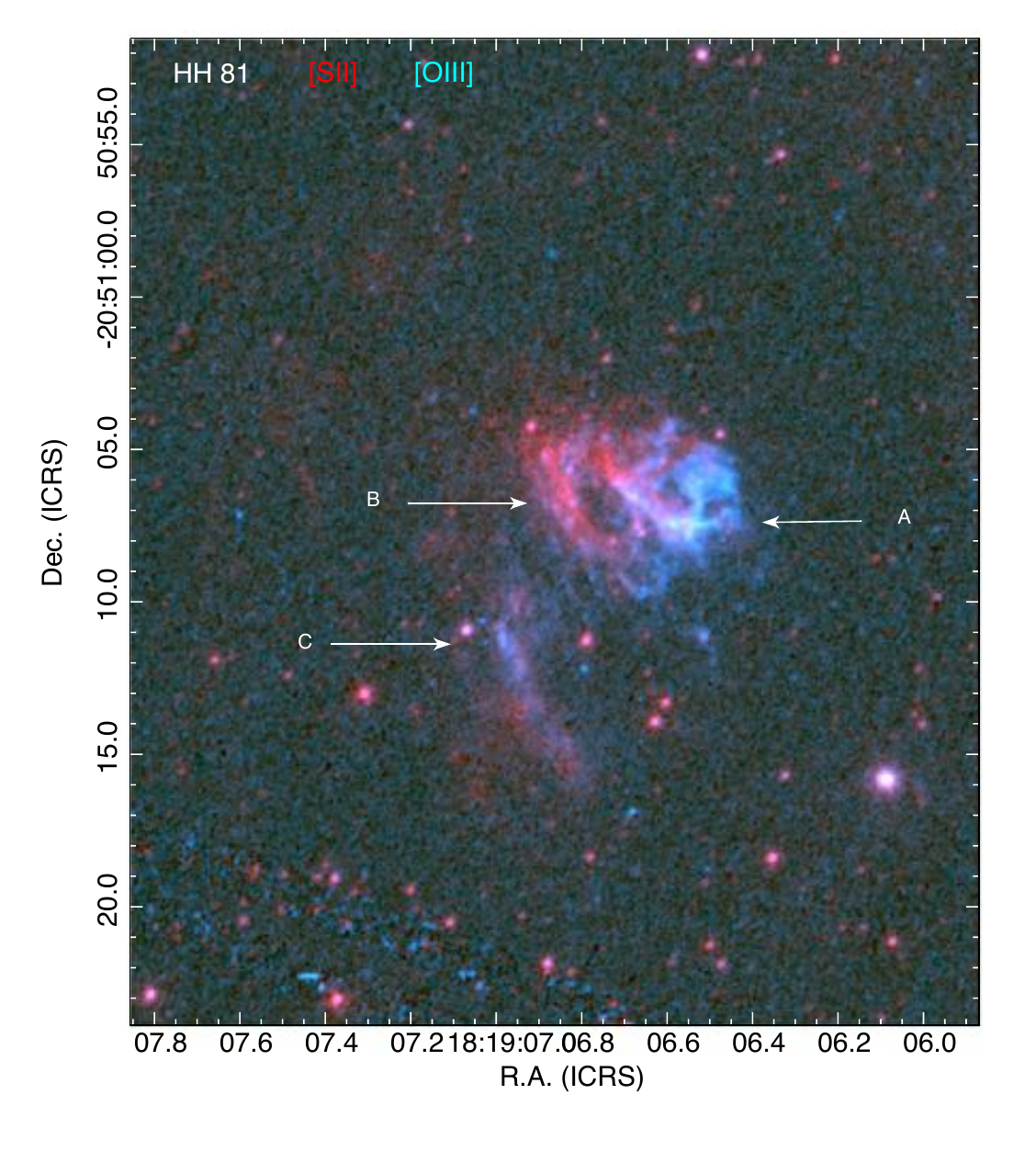  }
}
\caption{The HH~81 field observed in 2018 with HST showing [\Sii ] (red)
     and [\Oiii ] (blue and white).  The features identified by 
     \citet{Heathcote1998} are marked.} 
    \label{fig_HH81_color_Sii_Oiii_sub}  
\end{figure*}

\begin{figure*}
\center{
\includegraphics[width=7.0in]{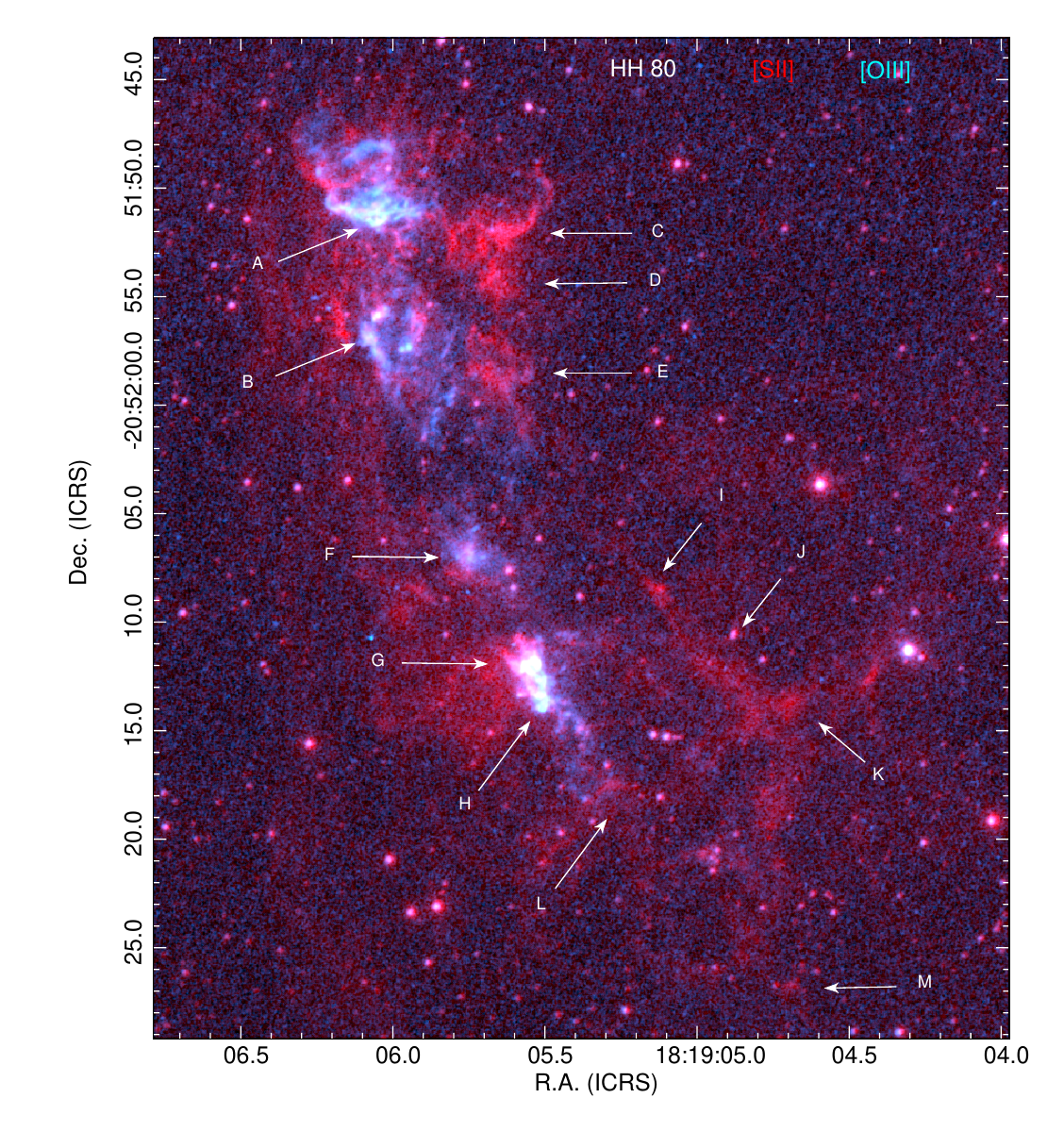  }
}
\caption{The HH~80  field observed in 2018 with HST showing [\Sii ] (red)
     and [\Oiii ] (blue and white).  The features identified by 
     \citet{Heathcote1998} are marked at their 2018 locations.} 
    \label{fig_HH80_color_Sii_Oiii_sub}   
\end{figure*}

\section{Results}

\subsection{Overview}

Figure \ref{fig1_overview} shows a mosaic of APO \Ha\ images acquired using a 
narrow-band filter with a pass-band of 30\AA .    In this figure, the
two brightest shocks, HH~80 and 81, are saturated.   The location and orientation
of the fast radio jet core is shown by a blue oval.  The location of the  walls of an 
infrared cavity surrounding the radio jet and visible in Spitzer 8 and 
24 \mm\ images is shown by orange lines \citep{Qiu2008}. 
Towards the south-southwest of the IRAS source, a dim chain of compact knots 
and filaments extends from near the IRAS source towards the bright, high-excitation
HH~80 and 81.   Beyond HH~81,  there is a faint but giant \Ha\ bow shock, the northern 
portion of which was first noted by \citet{Heathcote1998}.

Towards the north-northeast where the blueshifted counter-flow to HH~80/81 
is expected,  the APO observations reveal a chain of dim \Ha\ and [\Sii ] emission 
features extending up to 580\arcsec\ from IRAS 18162-2048. These objects lie close 
to the axis defined by the radio jet.     Because the flux ratio I[\Sii ]/I(\Ha ) 
is larger than 0.5, and {\bf since they} are located close to the axis of the outflow, we consider 
them to be Herbig-Haro objects.   In Figure \ref{fig1_overview}, dashed red circles 
show the locations of the new HH objects.   \Htwo\ emission is associated with some
of these objects.  The location 
and orientation of a prominent infrared dark cloud (IRDC) is marked by a magenta
ellipse.   This cloud, also known as HH~80N core,  was extensively studied by 
\citet{Girart2001} and \citet{Masque2011}.  The most distant radio feature thought to 
be associated with the IRAS 18162-2018 radio jet, the feature labeled 
`radio source 34' in Figure~\ref{fig1_overview} 
\citep{Masque2012}, has no visual or near-IR counterpart. 

We first discuss the changes in the structure of the HH~80 and 81 shocks in the 
fields observed during two epochs by HST. 
Then we present  and analyze proper motions in these fields. 
This is followed by a discussion of the parsec-scale H$\alpha$ bubble likely inflated by
the  fast ($>$800 \kms) flow as it breaks out of the L291 cloud.  After that,
we discuss the faint shocks located in the counter-flow north-northwest of the IRAS source traced by narrow-band \Ha , [\Sii ], 
[\Feii ], and \Htwo\ emission.  Finally, we discuss some ideas about the production of the ultra-fast jet and large proper motions in this extraordinary outflow.

\subsection{The 2018 HST Observations; Changes Since 1995}
 
The 2018 WFC3 images cover a larger field of view than the 1995 WFPC2 images with an angular resolution of $\sim$0.06\arcsec\ over the
entire field (the wide-field chips in WFPC2 have an image scale of
0.1\arcsec\ per pixel).   Of the three filters (\Hb , [\Oiii ], 
and [\Sii ]),  the [\Sii ] images show the most extensive nebulosity.
A nearly continuous chain of [\Sii ]-dominated shocks and filaments extends 
from the north-northeast corner of the field to the south-southwest 
corner along the axis of the radio jet.  

Figures \ref{fig_HH80_81_Sii_inverted}, \ref{fig_HH80_81_Oiii_inverted},
and \ref{fig_HH80_81_Hbeta_inverted} show the entire
field of view imaged in 2018 in the [\Sii ], [\Oiii ] and \Hb\ filters.   
Two faint HH objects are located between HH~80/81 and IRAS~18162-2048.   
These shocks are marked as
HH~81~N1 and HH~81~N2.  HH~81~N1 is mostly detected in [\Sii ] and is thus
a low-excitation shock.   However, the brightest component, which looks like an 
arc-second diameter bow-shock pointing directly away from IRAS 18162+2048, 
contains several unresolved [\Oiii ] knots at its tip.  Faint 
filaments of [\Sii ] emission connect the bright bow-shock in HH~81~N1
to the bright body of HH~81.

In the main HH objects, HH~80 and HH~81, the \Ha , \Hb , and [\Oiii ] 
images show bright, compact knots and filaments close to the axis of 
the radio jet.   The brightest features exhibit strong [\Sii ] emission.  Additionally, the [\Sii ] image shows an extended network of filaments and diffuse features not seen in \Hb\  or [\Oiii ].     A dim, diffuse region of [\Sii ] emission 
extends from HH~81 back towards HH~81~N2 in the upper-left corner of 
Figure \ref{fig_HH80_81_Sii_inverted}.  HH~81~N2 consists of a jumble 
of dim [\Sii ] knots and filaments over a 10\arcsec\ diameter region.

Parallel to the chain of bright HH objects extending from HH~80~N2 to HH~81, but about 
30\arcsec\ to the west,   there is a dim $\sim$10\arcsec\ wide filament of faint [\Sii ] emission.    This feature is marked as `HH~81~West rim' in Figure  \ref{fig_HH80_81_Sii_inverted}.  After a 45\arcsec\ gap in this structure between Declination -20:50:30 and -20:51:00,  it can be traced for about 130\arcsec\ to the south-southwest where it connects to the south-southwest edge of HH~80.   

Figures \ref{fig_HH81_color_Sii_Oiii_sub} and \ref{fig_HH80_color_Sii_Oiii_sub} 
show closeup views of the main HH~80/81 shock complex in 
[\Sii ] (red) and [\Oiii ] (cyan).    The high-excitation emission traced
by [\Oiii ] and, to some extent, \Hb, is confined to the core of the shock complex
close to the extrapolated radio jet axis.
In contrast, the lower-excitation [\Sii ] emission is much more widespread.  

Figure \ref{fig_HH81_color_Sii_Oiii_sub} shows a close-up view of 
the HH~81 shock complex in [\Sii ] (red) and [\Oiii ] (cyan).  Letter designations
correspond to features marked in \citet{Heathcote1998}.   This
color image shows a strong excitation gradient from west to east.
The brightest and highest-excitation region occurs in feature A
which resembled a letter `X' in 2018 on the [\Oiii ] image.  
A series of three arcs of emission suggests shocks moving sideways 
with respect to the jet axis
towards the southeast (features B and C).    Both features A and B
are [\Oiii ] dominated at their leading southern edges. The excitation 
gradient along filament C is reversed with the [\Oiii ] emission 
dominating its trailing northeastern edge. 

The high-excitation feature HH~80A near the top-left corner of 
Figure \ref{fig_HH80_color_Sii_Oiii_sub} exhibits a series of concentric
arcs whose radii shrink with increasing distance from the IRAS source.  
These features likely trace  `ripples' on the surface of a bow-shock whose apex is 
marked by  the $\sim$1\arcsec\ diameter, elliptical [\Oiii ]-bright ring 
at the south-southwest-facing tip of feature A.  At least three partial
ovoids of [\Oiii ] emission can be seen, followed by a low-excitation
partial ring of [\Sii ] emission on the trailing end of this structure.
If their true shapes are circular, their appearance favors an inclination
angle of the flow smaller than 60\arcdeg .

Feature HH~80B marks the brightest portion of a fainter, V-shaped [\Oiii ]
structure about 5\arcsec\ downstream from HH~80A.  A relatively diffuse
[\Sii ] envelope surrounds the [\Oiii ] emission.     Feature HH~80F
is a relatively dim [\Oiii ] region.   HH~80G and HH~80H mark the brightest
[\Oiii ] emission region in this complex and resembles a narrow cone-shaped
bow-shock.    The chain of four high-excitation [\Oiii ] objects  consisting of
features A, B, F, and G/H, trace the eastern wall of a cavity in
the [\Sii ] emission.    The western wall is marked by features C, D, 
E, I, J, and K (note that this is not `HH~81 West rim' marked in Figure
\ref{fig_HH80_81_Oiii_inverted}).

\subsection{Proper Motions}

Proper motions were measured by comparing the HST images taken in 1995 with 
the new set of images obtained in 2018.   Given a proper motion PM (in milli-arcseconds per year) of a feature, 
 the plane-of-the-sky speed is given by
$$ \rm
V_{PM} (km~s^{-1}) = 4.74 ~D_{kpc}~ PM (mas ~yr^{-1})
$$
where $\rm D_{kpc}$ is in units of 1 kpc.   Thus, at $\rm D_{kpc}$=1.4 kpc, 
$\rm V_{PM} (km~s^{-1}) = 6.6366 ~ PM (mas ~yr^{-1})$.

\begin{figure*}
\center{
\includegraphics[width=6.5in]{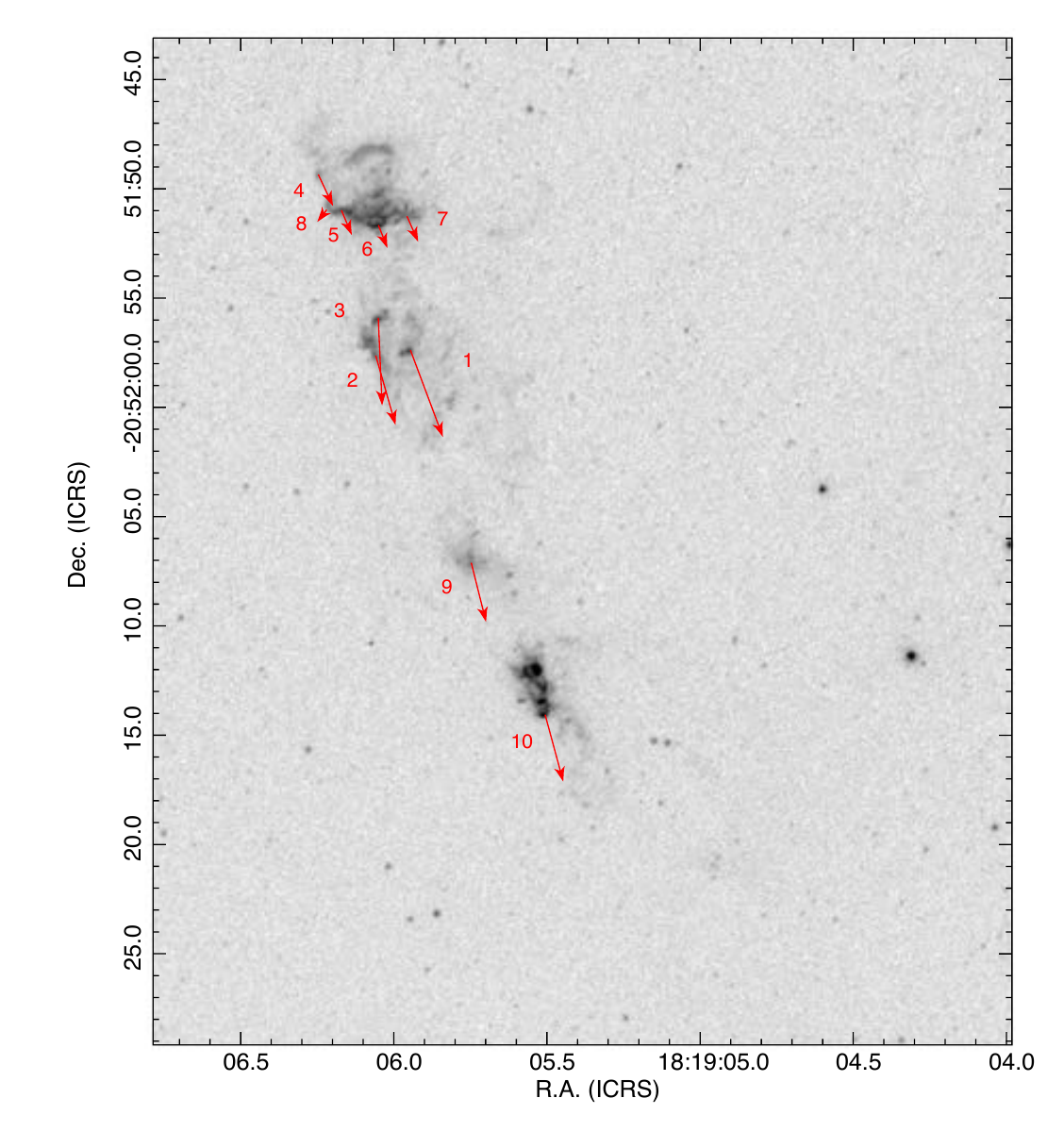 }
}
\caption{HH~80 in [\Oiii ] showing proper motions as vectors superimposed on the
         2018 image.  The vector lengths correspond to the motion in a 23 year 
         interval.  Feature B discussed in the text consists of the three knots
         whose proper motions are labeled as 1, 2, and 3 in this figure.  The
         A complex consists of features 4 through 8.
         An interactive version of this figure is available. Clicking on the image will switch between the 1995 and 2018 images to show the changes and motions between the 23 year span.
         }
    \label{fig_HH80_Oiii_PM_vectors}  
\end{figure*}

\begin{figure*}
\center{
\includegraphics[width=6.5in]{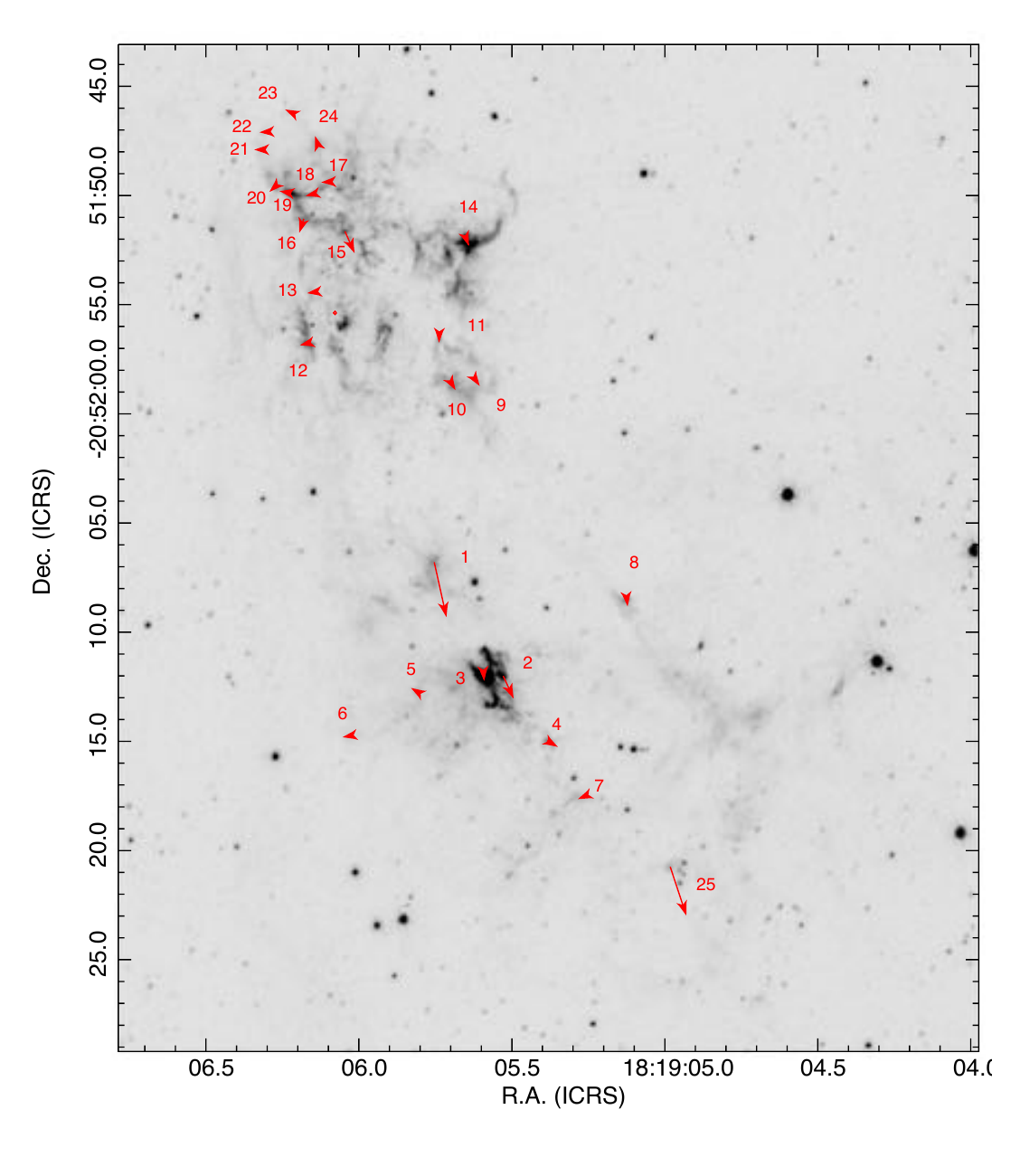 }
}
\caption{HH~80 in [\Sii ] showing proper motions as vectors superimposed on the
         2018 image.  The vector lengths correspond to the motion in a 23 year 
         interval. An interactive version of this figure is available. Clicking on the image will switch between the 1995 and 2018 images to show the changes and motions between the 23 year span.
         }
    \label{fig_HH80_Sii_PM_vectors}  
\end{figure*}


\begin{figure*}
\center{
\includegraphics[width=6.5in]{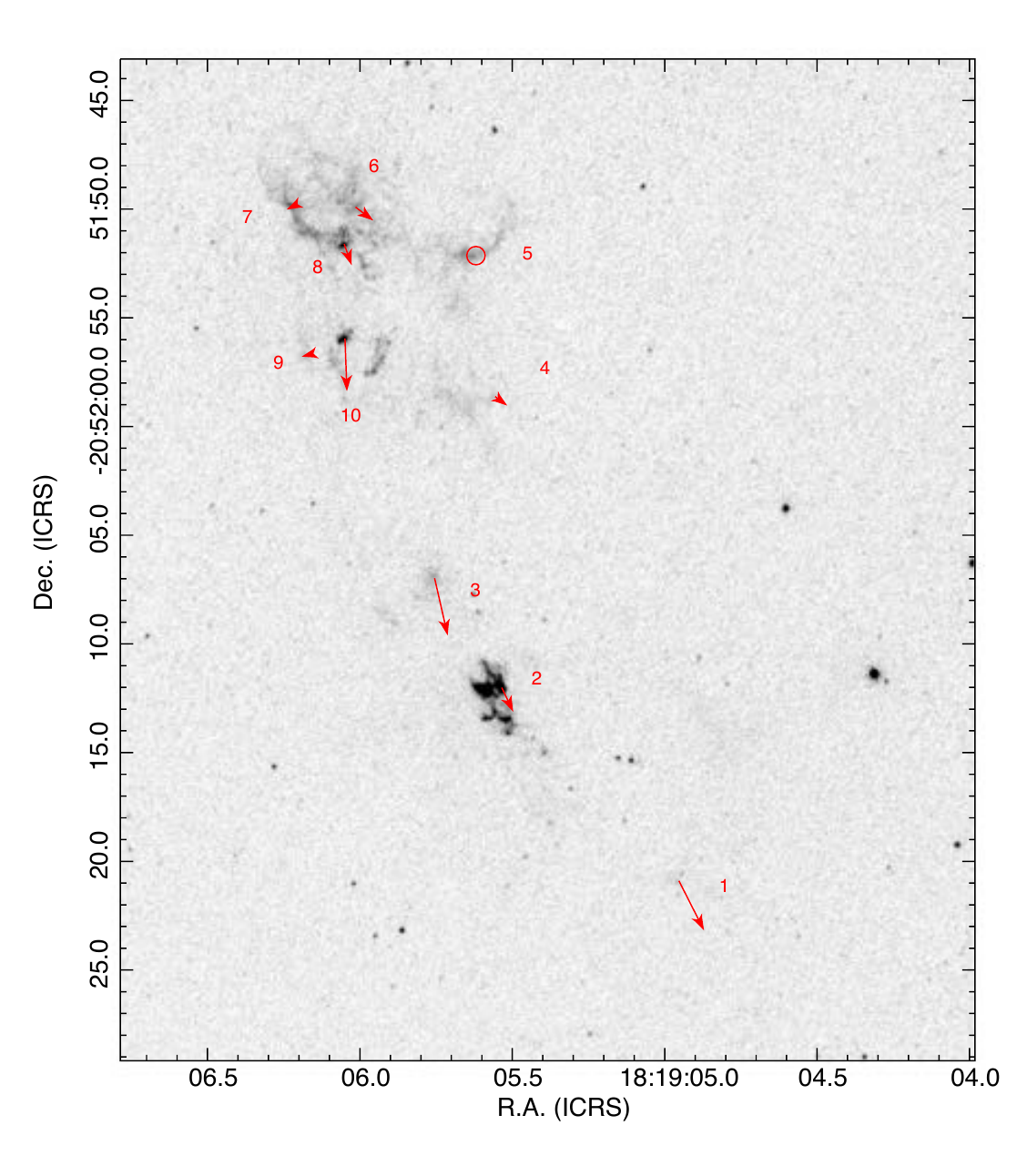 }
}
\caption{HH~80 in \Hb\ showing proper motions as vectors superimposed on the
         2018 image.  The vector lengths correspond to the motion in a 23 year 
         interval. An interactive version of this figure is available. Clicking on the image will switch between the 1995 and 2018 images to show the changes and motions between the 23 year span.
         }
    \label{fig_HH80_Hb_PM_vectors}  
\end{figure*}

\begin{figure*}
\center{
\includegraphics[width=6.5in]{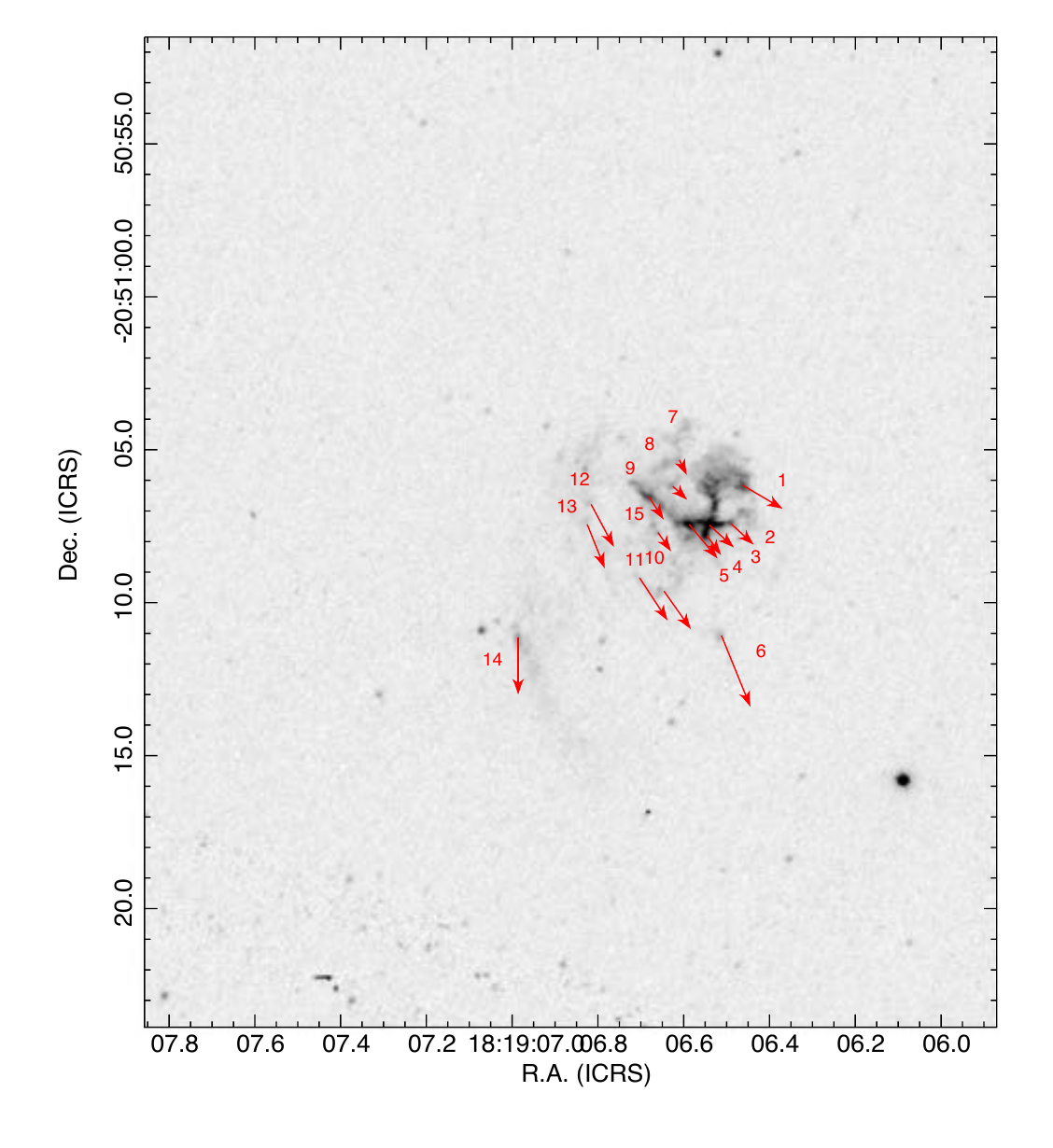 }
}
\caption{HH~81 in [\Oiii ] showing proper motions as vectors superimposed on the
         2018 image.  The vector lengths correspond to the motion in a 23 year 
         interval. An interactive version of this figure is available. Clicking on the image will switch between the 1995 and 2018 images to show the changes and motions between the 23 year span.
         }
    \label{fig_HH81_Oiii_PM_vectors}  
\end{figure*}

\begin{figure*}
\center{
\includegraphics[width=6.5in]{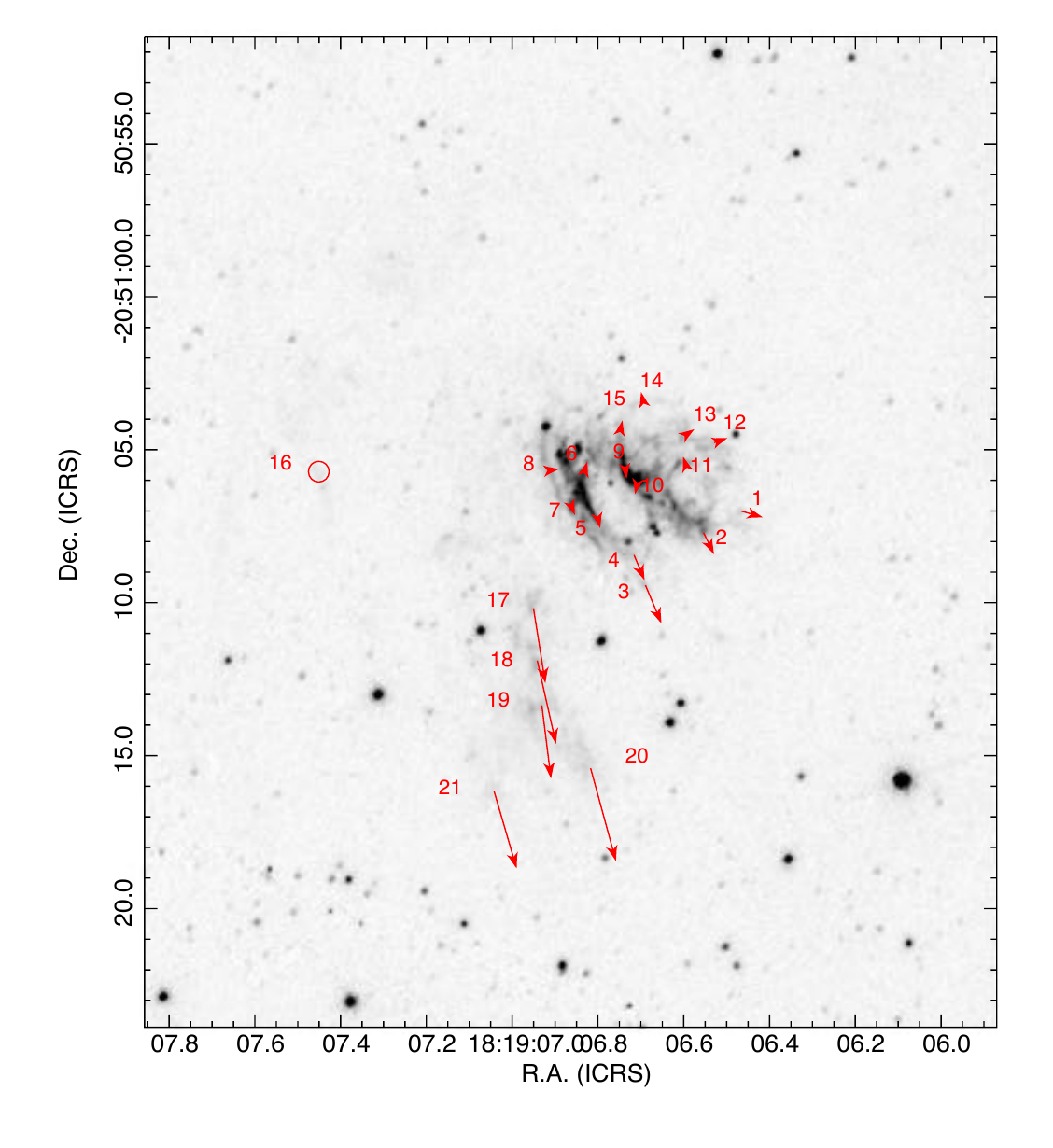 }
}
\caption{HH~81 in [\Sii ] showing proper motions as vectors superimposed on the
         2018 image.  The vector lengths correspond to the motion in a 23 year 
         interval. An interactive version of this figure is available. Clicking on the image will switch between the 1995 and 2018 images to show the changes and motions between the 23 year span.
         }
    \label{fig_HH81_Sii_PM_vectors}  
\end{figure*}

\begin{figure*}
\center{
\includegraphics[width=6.5in]{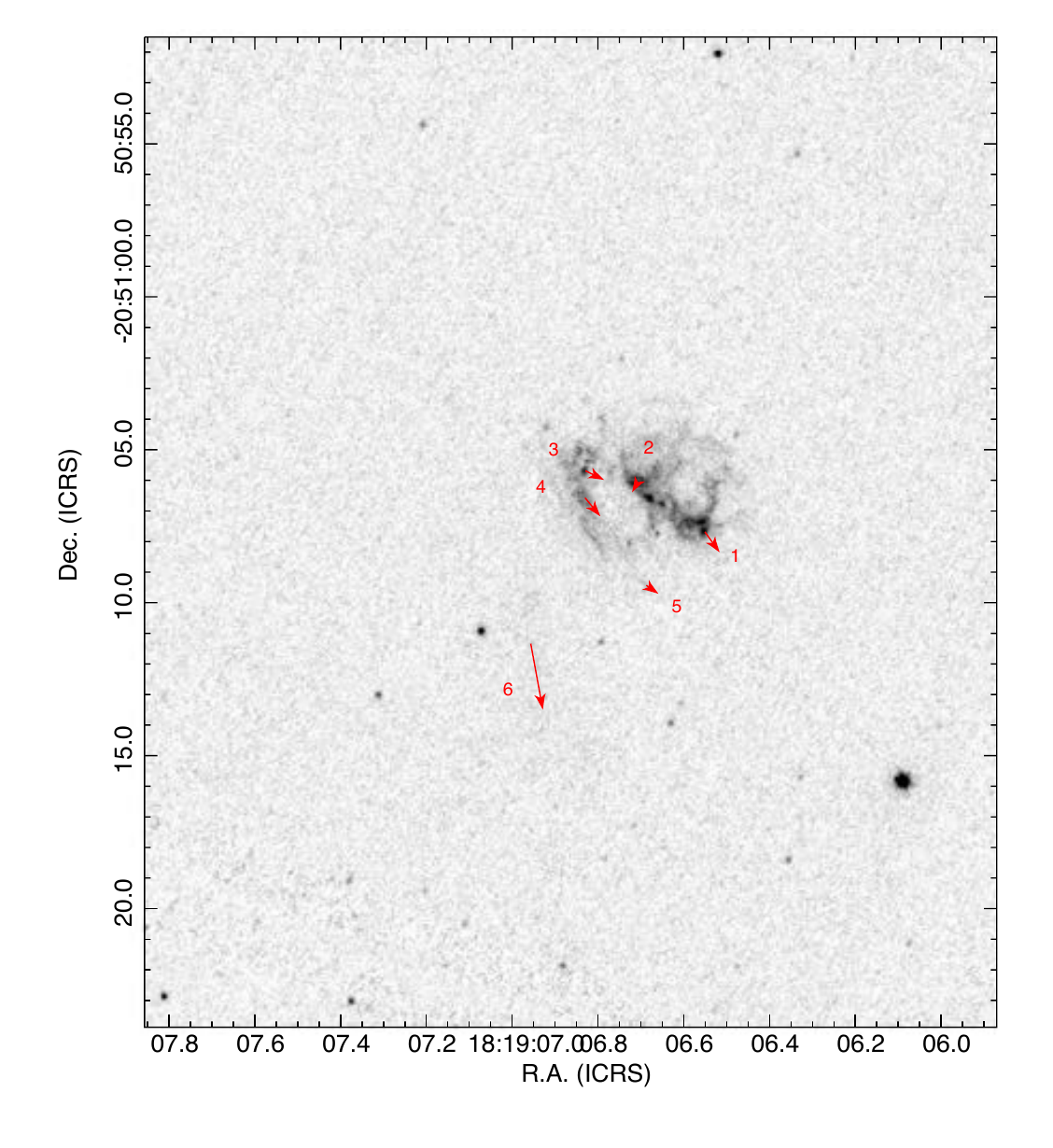 }
}
\caption{HH~81 in \Hb\ showing proper motions as vectors superimposed on the
         2018 image.  The vector lengths correspond to the motion in a 23 year 
         interval. An interactive version of this figure is available. Clicking on the image will switch between the 1995 and 2018 images to show the changes and motions between the 23 year span.
         }
    \label{fig_HH81_Hb_PM_vectors}  
\end{figure*}

Changes and proper motions are most obvious when the images taken 
in 1995 and 2018 are blinked.    Several methods of proper motion 
analysis and display were utilized:  visual inspection and measurement 
of displacements, use of a Python-based cross-correlation code 
\citep{Bally2022}, difference images, and images in which the first 
epoch is shown in red and the second in cyan.    Visual 
inspection suffers from personal bias.   The complex structure of HH~80/81, 
the rich star field, especially in the [\Sii ] and \Ha\ images, and the 
large magnitude of the motions over the 23 year interval between images
have made the use of the Python code difficult.  While intensity difference 
and ratio images show the motions well, they hide regions where there 
were no changes between the epochs.  We found that color displays
provide the best rendering of the complex motions in this shock system.
Such color images are presented in the Appendix.  

Figures 7 to 12
show the measured proper motions as vectors.   
Tables 3 to 5  lists the positions, proper motions, proper motion position
angles, and the resulting speeds on the plane of the sky assuming a distance 
of 1.4 kpc.    Table~3 presents [\Oiii ] proper motions, Table~4 presents 
[\Sii ] proper motions, and Table~5 presents motions measured on the \Ha\ 
and \Hb\ images. In each table the first part presents motions in HH~80, 
the second part presents motions in HH~81.  The numbering starts at 1 in each 
table section. The fastest motions are in excess of 1,200 \kms .

The cooling time of post shock plasma is
$\rm \tau_{cool} \approx  7000 ~n_H^{-1} V_{S,100}^{3.4} $
where $n_H$ is the hydrogen volume density (in \cmq ),  
$\rm V_{S,100}$ is the shock speed in units of 100 \kms , and
$\rm \tau_{cool}$ is the cooling time in years \citep{Draine2011}.   
Thus, at a density $\rm n_H$ = 1000 \cmq , typical for bright HH objects, 
the cooling time is about 7 years for a shock speed
of 100 \kms , less than the  23 year interval between the HST images.    
Thus, it is not surprising that the appearance of the shocks has changed 
considerably between 1995 and 2018.

A short cooling time 
of the post-shock plasma in the denser and brighter regions of the HH~80/81  
shock system makes the measurement of some proper motions ambiguous.  
Inspection of the images shows that some features have faded or disappeared 
altogether while new features have appeared.  Nevertheless, in all three 
species imaged with HST, there are many features whose motion away 
from GGD~27 is obvious when patterns and complex shapes are considered.   

Unlike in some HH objects where the motions of persistent knots can be easily 
measured by cross-correlation methods, the complex structure, rich field of 
background stars, and noise in the images led to the failure of our automated 
technique.   Comparison of the 1995 and 2018 images show the existence of 
some persistent patterns such as partial rings, arcs, and filaments among 
the jumble of structure.   Prior knowledge of the general direction of
motion helps in the identification of such features.  Although in the future,
artificial intelligence programs may be trainable to recognize such 
structure, here we use visual inspection to measure their apparent motion.

Figures \ref{fig_HH80_Oiii_PM_vectors},  \ref{fig_HH80_Sii_PM_vectors},
and \ref{fig_HH80_Hb_PM_vectors} show proper motion vectors in HH~80 in
[\Oiii ], [\Sii ], and \Hb\ superimposed on the 2018 epoch HST images.
A particularly striking example of large 
proper motions is seen in HH~80 feature B
in Figure \ref{fig_HH80_color_Sii_Oiii_sub} (knots and proper motion vectors
1, 2, and 3 in Figure \ref{fig_HH80_Oiii_PM_vectors}).   
This feature consists of an east-facing half-circle of [\Oiii ] emission 
with a pair of 
compact knots separated by about 1\arcsec\ close to the center of the ring
located just below the bright southern tip of the A complex.   Between 1995
and 2018, feature B shifted south by over 4\arcsec.    
In the Appendix, the 1995 image is shown in red while the 2018 image 
is shown in cyan. 
Feature B, which exhibits the highest proper motion in the entire 
HST field, contains multiple knots.   Entries 1, 2, and 3 in Table~3 show that 
these features have proper motions ranging from 950 to over 1,200 \kms\ 
in [\Oiii ], the fastest motions seen so far in any HH object.  

In contrast, the very bright feature A (Figure \ref{fig_HH80_color_Sii_Oiii_sub})  
exhibits considerably slower, and more complex, behavior.  As discussed above, 
feature A consists of a series of arcs and filaments which may trace ripples 
on the surface of a cone-shaped bow shock with a
high-excitation tip.  While the smaller, leading arcs are seen best
in [\Oiii ], the larger trailing arcs are best seen in the 
lower-excitation [\Sii ] image.  Proper motions indicate that this structure
is being compressed, with smaller motion at the leading edge than on the 
trailing side.  This suggests that ejecta is running into a slower moving
obstacle located ahead of the leading edge of A.    Support for this
hypothesis is provided by the presence of non-thermal radio emission 
and hard X-rays emerging from HH~80 \citep{Pravdo2004,Pravdo2009,
LopezSantiago2013,Vig2018}.  Analysis of Fermi/LAT data shows 
the presence of hard $\gamma$-rays with energies up to 1 GeV \citep{Yan2022}.
However, the source can only be localized to about 1 degree.  Thus, 
it {\bf is} unclear 
where the $\gamma$-rays originate in the IRAS~18162-2048 outflow complex.

Because feature B has
larger proper motions and is located ahead of feature A, we suggest  that
features A and B are located at slightly different distances along our 
line-of-sight.  Feature A may be interacting with slower moving, or even
stationary material along the walls of an outflow cavity drilled by a 
fast jet while feature B may be closer to the center of the flow channel.   

There is a relatively isolated arcsecond-scale blob, labeled F in Figure
\ref{fig_HH80_color_Sii_Oiii_sub}, about two-thirds of the way from feature A to
the bright features G and H at the southern tip of HH~80.  This is one of the 
few simple, blob-like features where the motion over 23 years is easy to
see.   Figures \ref{fig_HH80_Oiii_PM_vectors} and \ref{fig_HH81_Oiii_PM_vectors}
show this feature marked with vectors (number 9 in the [\Oiii ] image and
number 1 in the [\Sii ] image)
superimposed on the 2018 [\Oiii ] and [\Sii ] images of HH~80.

For most features, the [\Sii ] proper motions are much slower and  more chaotic 
than those in [\Oiii ] or \Ha /\Hb .   Blinking the HH~80 [\Sii ] images shows
overall expansion of the [\Sii ] emission orthogonal to the extrapolated radio
jet axis and the high-excitation core of the shock complex.   
The Appendix presents color images of the
[\Sii ] emission in 1995 (red) and 2018 (cyan).   
Figure \ref{fig_HH80_Sii_PM_vectors} shows the vector field.  Relatively high 
[\Sii ] proper motions are seen in the high-excitation core of HH~80 A
(features 15 and 16 in Table 4 and in Figure \ref{fig_HH80_Sii_PM_vectors}),  
HH~80 knot F (feature 1), and feature 25 farther downstream near the bottom of 
Figure \ref{fig_HH80_Sii_PM_vectors}.  Feature 25's motion is, however, 
uncertain;  it may represent a case where a knot seen in 1995 had faded 
or disappeared and a new feature appeared in 2018.  But since 
the apparent motion is along the jet axis at the expected extrapolated 
position of the radio jet, and close to the location of high proper motion
features seen in the high-excitation species, it is included in the 
figure and associated table.

The most striking aspect of the [\Sii ] proper motions is the expanding arc
of emission in the wake of HH~80-A (features 17 to 23 near the 
top-left in Figure \ref{fig_HH80_Sii_PM_vectors}).    The ring's motion is
not only orthogonal to the jet axis but, near the top, it exhibits motions
back towards the IRAS source.  This behavior can be understood when the 
large inclination of the jet axis is considered.  Assume the ring is
a ripple on the surface of a cone-shaped bow shock receding along the 
outflow axis.   If the ring is expanding with respect to the outflow axis, 
the portion closest to the IRAS source can exhibit 
backward proper motion  on the plane of the sky (apparent motion towards 
the source).  Conversely, portions of the ring farthest from the
source can exhibit proper motions higher than the tip of the bow. 

Figures \ref{fig_HH81_Oiii_PM_vectors},  \ref{fig_HH81_Sii_PM_vectors},
and \ref{fig_HH81_Hb_PM_vectors} show proper motion vectors within HH~81 in
[\Oiii ], [\Sii ], and \Hb\ superimposed on the 2018 epoch HST image.
The brightest [\Oiii ] emission is associated with relatively slow 
proper motions compared to fainter, faster features to the south and east.  
In Figure \ref{fig_HH81_Oiii_PM_vectors}, the
brightest emission consists of an `X-shaped' emission region delineated
by vectors 2 to 5.    The proper motions here are much lower than
in the relatively fainter diffuse features located south and east and
marked by vectors 6 and 10 through 14.   

This pattern is repeated in the
\Hb\ image in Figure \ref{fig_HH81_Hb_PM_vectors}. 
Hydrogen recombination line proper motions were measured by comparing the 
1995 \Ha\ image with the 2018 \Hb\ image.  Because of more than two 
magnitudes of visual extinction, features are much dimmer
in the \Hb\ image.  Thus, motions of only the brightest features were
measured;  these are listed in Table \ref{tab:PMs_H} and 
Figures \ref{fig_HH80_Sii_PM_vectors} and \ref{fig_HH81_Sii_PM_vectors}.
Overall, the hydrogen proper motions are similar to those seen in
[\Oiii ].

Figure \ref{fig_HH81_Sii_PM_vectors} shows the [\Sii ] proper motions
in HH~81.  The bright [\Oiii ] X-shaped feature is absent, and replaced
by dim [\Sii ] knots.      Bright [\Sii ] filaments  appear in the
wake of the bright [\Oiii ] X-shaped feature (towards the jet source).   
However,  the proper motions in this filament are relatively low.  

The [\Sii ] emission from HH~81 exhibits even more stunning expansion
away from the jet axis.  Table \ref{tab:PMs_Sii} lists motions measured
in [\Sii ] at a set of representative locations in HH~81 and indicated in
Figure \ref{fig_HH81_Sii_PM_vectors}.   A color representation of the changes
and motions between 1995 and 2018 similar to that shown for HH~80
is presented in the Appendix.

\begin{deluxetable*}{ccccccl}
\tablenum{3}
\tablecaption{Proper Motions in HH~80/81:  [\Oiii ]  \label{tab:PMs_Oiii}}
\tablewidth{0pt}
\tablehead{
\colhead{Feature} & \colhead{RA} & \colhead{Dec} & \colhead{PM} &
\colhead{V$_{PM}$ } & \colhead{position angle} & \colhead{Comments } \\
\colhead{} & \colhead{(ICRS)} & \nocolhead{(ICRS)} & \colhead{(mas/yr} &
\colhead{(\kms )} & \colhead{degrees} & \colhead{} }
\decimalcolnumbers
\startdata
    &               &     HH~80     &       &       &       &   \\
1   & 18:19:05.949  & -20:51:57.42  & 184   & 1218  & 201   & Knot inside east-facing arc  \\
2   & 18:19:06.056  & -20:51:57.62  & 143   & 950   & 196   & East-edge of east-facing arc \\
3   & 18:19:06.050  & -20:51:55.91  & 174   & 1153  & 183   & North-end of east-facing arc \\
4   & 18:19:06.249  & -20:51:49.26  &  69   & 457   & 205   & East-rim of A \\
5   & 18:19:06.169  & -20:51:50.99  &  53   & 351   & 203   & S-end of arc containing 4 \\
6   & 18:19:06.052  & -20:51:51.62  &  49   & 322   & 201   & Bright apex of HH~80 A \\
7   & 18:19:05.958  & -20:51:50.31  &  55   & 366   & 204   & West rim of HH~80 A \\
8   & 18:19:06.211  & -20:51:51.31  &  30   & 201   & 142   & Sideways splash from HH~80 A \\
9   & 18:19:05.748  & -20:52:07.11  & 121   & 803   & 194   & Knot F \\
10  & 18:19:05.510  & -20:52:14.08  & 138   & 914   & 195   & knot G; S end of HH~80 \\
    &               &     HH~81     &       &       &       &   \\
1   & 18:19:06.462  & -20:51:06.16  & 665   & 433   & 240   & Northwest side of HH~81 \\
2   & 18:19:06.487  & -20:51:07.43  &  44   & 292   & 227   & West end of `X' \\
3   & 18:19:06.539  & -20:51:07.48  &  47   & 311   & 228   & Center of `X' \\
4   & 18:19:06.544  & -20:51:07.85  &  34   & 226   & 216   & South end of `X' \\
5   & 18:19:06.588  & -20:51:07.45  &  63   & 415   & 220   & Knot in `X' \\
6   & 18:19:06.510  & -20:51:11.11  & 109   & 723   & 202   & Knot below HH~81 \\
7   & 18:19:06.610  & -20:51:05.35  &  24   & 156   & 206   & North side of HH~81 \\
8   & 18:19:06.624  & -20:51:06.20  &  28   & 184   & 226   & Center of HH~81 \\
9   & 18:19:06.681  & -20:51:06.52  &  39   & 260   & 213   & NE edge of `X' \\
10  & 18:19:06.652  & -20:51:09.58  &  66   & 437   & 215   & Arc at South edge of HH~81 \\
11  & 18:19:06.702  & -20:51:09.20  &  73   & 480   & 213   & Northeast edge of Arc \\
12  & 18:19:06.815  & -20:51:06.80  &  69   & 455   & 208   & East side of HH~81 \\
13  & 18:19:06.826  & -20:51:07.46  &  67   & 444   & 202   & East side of HH~81 \\
14  & 18:19:06.986  & -20:51:11.16  &  80   & 533   & 180   & Tip of filament SE of HH~81 \\
15  & 18:19:06.659  & -20:51:07.73  &  33   & 216   & 215   & Dim knot SE of `X' 
\enddata
\tablecomments{Velocities assume a distance of 1.4 kpc} 
\end{deluxetable*}

\begin{deluxetable*}{ccccccl}
\tablenum{3}
\tablecaption{Proper Motions in HH~80/81: [\Sii ]  \label{tab:PMs_Sii}}
\tablewidth{0pt}
\tablehead{
\colhead{Feature} & \colhead{RA} & \colhead{Dec} & \colhead{PM} &
\colhead{V$_{PM}$ } & \colhead{position angle} & \colhead{Comments } \\
\colhead{} & \colhead{(ICRS)} & \nocolhead{(ICRS)} & \colhead{(mas/yr} &
\colhead{(\kms )} & \colhead{degrees} & \colhead{} }
\decimalcolnumbers
\startdata
    &               &   HH~80       &       &       &       &      \\
1   & 18:19:05.753  & -20:52:06.79  & 113   & 739   & 193   & HH~80 knot F \\
2   & 18:19:05.537  & -20:52:11.94  &  49   & 326   & 205   & HH~80 knot G \\
3   & 18:19:05.592  & -20:52:11.92  &  14   & 94    & 185   & knot G SE \\
4   & 18:19:05.390  & -20:52:14.97  &  24   & 163   & 240   & HH~80 knot H \\
5   & 18:19:05.809  & -20:52:12.72  &  12   &  82   &  59   & East of knot G \\
6   & 18:19:06.033  & -20:52:14.75  &  12   &  82   & 100   &    "     "   \\
7   & 18:19:05.276  & -20:52:17.58  &   5   &  32   &  32   &    113?  \\
8   & 18:19:05.123  & -20:52:08.52  &   7   &  45   & 186   & West of G \\ 
9   & 18:19:05.615  & -20:51:58.50  &  11   &  76   & 214   & knot E \\
10  & 18:19:05.691  & -20:51:58.74  &   9   &  61   & 211   &  " \\
11  & 18:19:05.737  & -20:51:56.55  &   8   &  54   & 179   &  " \\
12  & 18:19:06.171  & -20:51:56.78  &  14   &  91   & 104   & filament B \\
13  & 18:19:06.152  & -20:51:54.44  &  10   &  68   &  98   & N end of filament B \\
14  & 18:19:05.642  & -20:51:52.22  & $<$   & $<$13 &  -    & Arc C \\
15  & 18:19:06.045  & -20:51:51.65  &  47   & 312   & 204   & Tip of A \\
16  & 18:19:06.181  & -20:51:51.08  &  29   & 192   & 162   & East side of A \\
17  & 18:19:06.099  & -20:51:49.38  &  14   &  93   &  93   & In ring North of A \\
18  & 18:19:06.149  & -20:51:49.95  &  14   &  92   & 103   &  "  \\
19  & 18:19:06.219  & -20:51:49.91  &  24   & 158   &  80   &  "  \\
20  & 18:19:06.272  & -20:51:49.56  &  17   & 112   & 133   &  "  \\
21  & 18:19:06.320  & -20:51:47.90  &  13   &  84   &  90   & East rim of ring A \\
22  & 18:19:06.307  & -20:51:47.10  &  10   &  68   &  93   &  "  \\
23  & 18:19:06.225  & -20:51:46.17  &  10   &  65   &  65   & North rim of ring A \\
24  & 18:19:06.136  & -20:51:47.59  &  14   &  91   &  18   & Center of ring A \\
25  & 18:19:04.981  & -20:52:20.78  & 102   & 676   &  198  & NE of M \\
    &               &   HH~81       &       &       &       &      \\
1   & 18:19:06.464  & -20:51:07.01  &  32   & 210   &  254  & W side of HH~81  \\
2	& 18:19:06.554	& -20:51:07.73	& 34	& 224	& 206	&  Bow on SW side of HH~81 \\
3	& 18:19:06.688	& -20:51:09.43	& 59	& 389	& 203	&  South end of filament \\
4	& 18:19:06.718	& -20:51:08.45	& 39	& 256	& 203	&  knot in filament \\ 
5	& 18:19:06.798	& -20:51:07.28	& 12	&  79	& 193	&  knot in filament \\	
6	& 18:19:06.829	& -20:51:05.64	& 12	&  80	& -16	&  N end of filament \\
7	& 18:19:06.861	& -20:51:06.89	& 11	&  76	& 202	&  East spur in filament \\
8	& 18:19:06.913	& -20:51:05.67	& 13	&  85	& -83	&  Eastern filament \\
9	& 18:19:06.737	& -20:51:05.60	& 15	& 100	& 187	&  East side of core \\
10	& 18:19:06.709  & -20:51:06.10	& 15 	& 103	& 169	&  Bow in core \\
11	& 18:19:06.599	& -20:51:05.34	&  5	&  35	&  17	&  Ring on W side \\
12	& 18:19:06.530	& -20:51:04.77	& 20	& 131	& -70   &  NE corner \\
13	& 18:19:06.596	& -20:51:04.52	& 15	& 101	& -56	&   "   \\
14	& 18:19:06.694	& -20:51:03.43	& 13	&  86	&  13	&  North edge of ring \\
15	& 18:19:06.744	& -20:51:04.14	&  4	&  29	& -12	&  NE edge of ring \\
16	& 18:19:07.450	& -20:51:05.72	&  - 	&  -   	&  -    &  East filament \\
17	& 18:19:06.954	& -20:51:10.18	& 108	& 713	& 189	&  Fast SE filament \\
18	& 18:19:06.941	& -20:51:11.89	& 121	& 804	& 193	&   "       "  \\
19	& 18:19:06.934	& -20:51:13.29	& 105	& 693	& 187	&   "       "  \\
20	& 18:19:06.816	& -20:51:15.44	& 136	& 904	& 195	&  South end of SE filament \\
21	& 18:19:07.044	& -20:51:16.16	& 115	& 761	& 196	&  Faint SE knot  
\enddata
\tablecomments{Velocities assume a distance of 1.4 kpc} 
\end{deluxetable*}

\begin{deluxetable*}{ccccccl}
\tablenum{3}
\tablecaption{Proper Motions in HH~80/81: \Ha\ and \Hb   \label{tab:PMs_H}}
\tablewidth{0pt}
\tablehead{
\colhead{Feature} & \colhead{RA} & \colhead{Dec} & \colhead{PM} &
\colhead{V$_{PM}$ } & \colhead{position angle} & \colhead{Comments } \\
\colhead{} & \colhead{(ICRS)} & \nocolhead{(ICRS)} & \colhead{(mas/yr} &
\colhead{(\kms )} & \colhead{degrees} & \colhead{} }
\decimalcolnumbers
\startdata
    &               &   HH~80       & H          &   &       &      \\
1   & 18:19:04.954  & -20:52:20.88  & 110   & 732   & 207 & Ahead of HH~80 \\
2   & 18:19:05.535  & -20:52:12.02  &  53   & 353   & 205 & HH~80 core \\
3   & 18:19:05.756  & -20:52:06.99  & 117   & 775   & 193 & Knot F \\ 
4   & 18:19:05.555  & -20:51:58.63  &  30   & 198   & 233 & Knot E ? \\
5   & 18:19:05.619  & -20:51:52.14  &   -   &  -    &  -  & Arc C; no detected motion \\
6   & 18:19:06.016  & -20:51:49.94  &  44   & 293   & 233 & HH~80 West side  \\
7   & 18:19:06.222  & -20:51:49.86  &  12   & 83    & 117 & East side of HH~80 \\
8   & 18:19:06.053  & -20:51:51.64  &  45   & 297   & 199 & Tip of HH~80 A \\    
9   & 18:19:06.269  & -20:51:57.18  &  10   & 67    & 104 & Sideways motion of filament \\
10  & 18:19:06.050  & -20:51:55.98  & 105   & 698   & 182 & Bright knot; center of fast arc \\
    &               &   HH~81       & H          &   &       &      \\
1   & 18:19:06.551  & -20:51:07.69  & 36    & 236   & 216 & West side of HH~81 \\
2   & 18:19:06.711  & -20:51:06.15  & 12    & 78    & 151   & Center of HH~81 \\
3   & 18:19:06.827  & -20:51:05.72  & 29    & 195   & 244   & NE corner of filament \\
4   & 18:19:06.830  & -20:51:06.59  & 34    & 225   & 220   & Center of filament \\
5   & 18:19:06.686  & -20:51:09.43  & 21    & 138   & 235   & S of Center \\
5   & 18:19:06.957  & -20:51:11.35  & 95    & 630   & 190   & SE filament 
\enddata
\tablecomments{Velocities assume a distance of 1.4 kpc} 
\end{deluxetable*}

\subsection{A Parsec-scale Bubble Blown by a Fast Protostellar Jet}

\begin{figure*}
\center{
\includegraphics[width=6.0in]{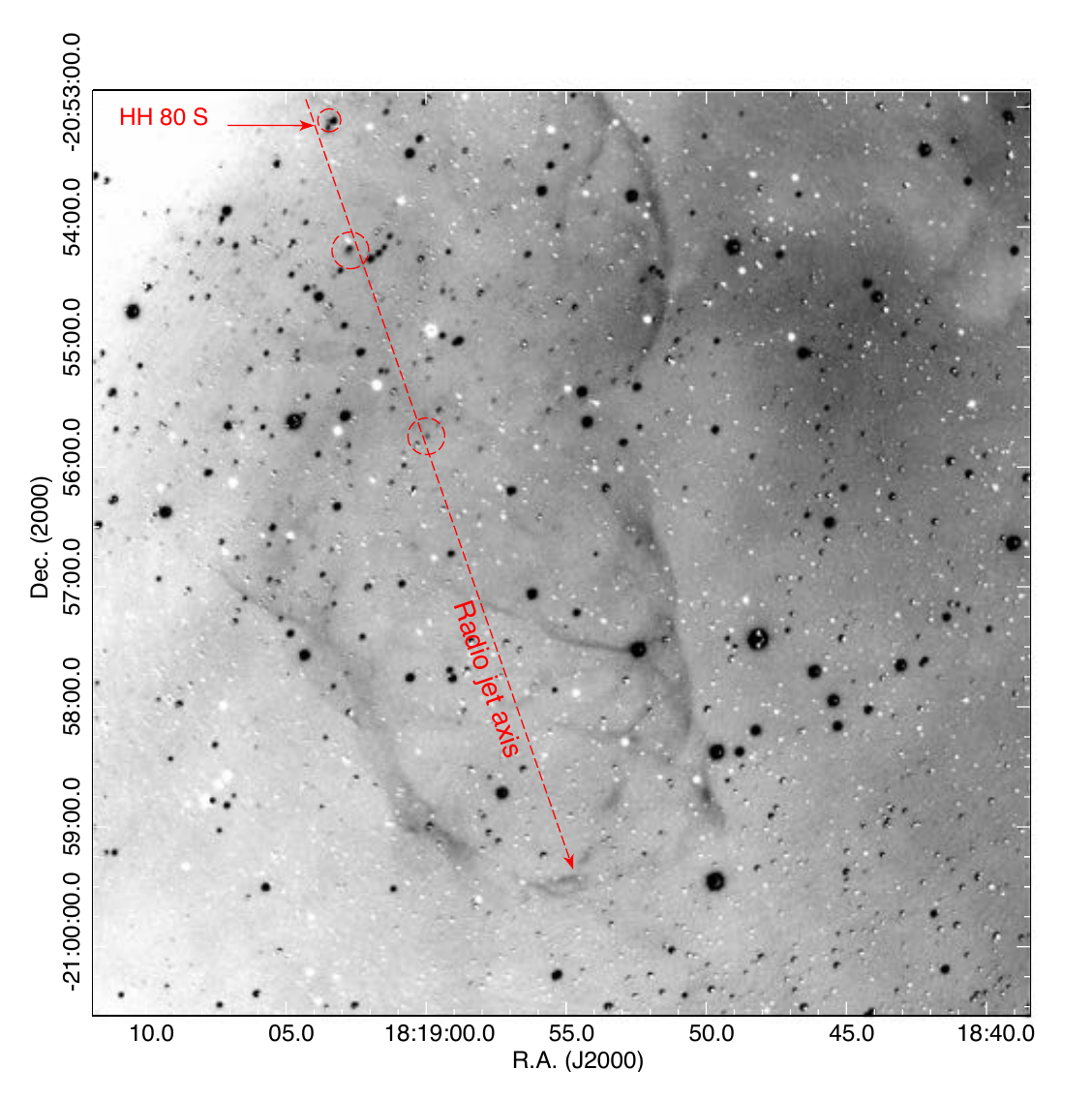  }
}
\caption{The giant \Ha\ bow shock inflated by the radio jet south-southwest
of GGD~27 in a continuum-subtracted image formed by subtracting a scaled SDSS~i
band image from the narrow-band \Ha\ image.  The \Ha\ image was obtained using
the 78\AA\ band-pass filter.  Dashed arrow marks the orientation 
and extrapolated location of the radio jet from IRAS 18162-2018. 
Dashed circles mark the locations of compact \Ha\ knots close to the radio jet axis.    
See text for details.
         }
    \label{fig2_SSW_bow}  
\end{figure*}

Beyond HH~80, towards the expected south-southwest terminus of this giant outflow   
there is a faint but giant \Ha\ bow shock, the northern 
portion of which was first noted by \citet{Heathcote1998}. 
APO images obtained with an 80\AA\ filter in September
2022 show that the feature seen by \citet{Heathcote1998} is the northern side of a parsec-scale bubble located where the radio jet from IRAS~18162-2048 is expected 
to break out into the low density ISM.   
Figure \ref{fig2_SSW_bow} shows an image formed by subtracting a
broad-band image taken with an SDSS~i filter from the \Ha\ image.   
The SDSS~i image was scaled to provide the best average subtraction 
of stellar images in the \Ha\ image.   Stars which are relatively 
brighter in the \Ha\ filter are black; stars which are relatively brighter 
in the longer wavelength SDSS~i filter are white. 
  
The \Ha\ bubble is at least 7.7\arcmin\ (3.1 pc)
long with a width ranging from $\sim$5\arcmin\ (2 pc) at the top of 
Figure \ref{fig2_SSW_bow} to less than 2\arcmin\ (0.7 pc) at the bottom.   
Its axis of symmetry lies close to the extrapolated 
location of the fast radio jet and closely aligned with the direction of the 
proper motion vectors in HH 80/81.  The full extent of the outflow, measured
from radio source 34 in the north-northeast to the tip of the \Ha\ bubble
is about 1,500\arcsec\ which corresponds to a physical, projected end-to-end 
length of about 10.2 pc on the plane of the sky for a distance of 1.4 kpc.  
Since the redshifted, south-southwest lobe is inclined away from the Sun, 
this is a lower bound on the true length.   A $\sim$60\arcdeg\ inclination
with respect to the plane of the sky implies an astounding physical length of 20 pc.

The outer edge of the \Ha\ bubble is sharp while the inner edge fades into the
noise on an angular scale of a few to about 10 arcseconds.   Thus, the \Ha\ 
emission comes from a limb-brightened, {\bf filamentary} structure.  
We estimate that the typical line-of-sight path length through the bubble, 
within an arcsecond or so of the outer edge is of order 0.07 pc.

Field stars with known SDSS r-band magnitudes were used to determine a 
photometric zero-point for the stacked \Ha\ image of the bubble.  This
zero-point was used to measure the observed \Ha\ surface brightness.   
The peak surface brightness is about  
$\rm SB(H \alpha) \approx 2 \times 10^{-16}$ \SB .
The noise level is about $\rm 1.5$ to $\rm 2 \times 10^{-17}$ \SB , dominated
by diffuse \Ha\ emission and airglow.   Assuming a foreground
extinction of about $\rm A_V = 1.4$ magnitudes (note that this is lower than
the extinction towards HH~80 and HH~81 because the bubble is farther
away from the L291 cloud.  We here assume a mean extinction of 
A$\rm _V$=1 mag/kpc).   If the extinction at
the wavelength of \Ha\ is 0.7808 $\rm A_V$, the extinction corrected peak 
surface brightness is 
$\rm SB_{cor}(H \alpha) \approx 5.5 \times 10^{-16}$ \SB .  The dimmest parts
of the bubble have $\rm SB(H \alpha) \approx 6 \times 10^{-17}$ \SB , which 
when corrected for extinction corresponds to 
$\rm SB_{cor}(H \alpha) \approx 1.6 \times 10^{-16}$ \SB .

The emission measure is related to the \Ha\ surface brightness by 
$\rm EM ($\EM$) = 4.86 \times 10^{17}~ I_{H \alpha }$ \SB  
{~\bf (see Haffner et al. 1998)}.
The emission measure, EM,  of the limb-brightened bubble ranges 
from below 80 \EM\ to about 300 \EM .  
The emission measure can be related to the mean
electron density in the emission region by $\rm EM = \int n_e^{2} dL_{pc}$
where $\rm n_e$ is the mean electron density and $\rm L_{pc}$ is the 
line-of-sight path-length through the emission region in units of a
parsec.   Thus,
$\rm n_e \approx (EM / L_{pc})^{1/2}$.  For an assumed path length
$\rm L_{pc}$ = 10\arcsec ($\sim$0.07 pc), the electron densities
range from about 30 to 200 \cmq .   A very crude, order-of-magnitude
shell mass can be estimated by assuming that the mean shell thickness 
is about 1\arcsec , that its mean electron density is 100 \cmq , and
that it is a cylinder with a projected surface area of about 4.5 square
parsecs (diameter $\sim$1.5 pc; length 3 pc). 
This gives about 0.1 \Msol \ as a likely lower bound.   If the
interior surface brightness is at the noise level, and the LOS depth is 1.5 pc,
the emission measure would be 35 \EM\ and the mean electron density about
5 \cmq . The total mass would then be about 1 Mo. This is likely an
upper bound on the mass of bubble wall and interior.   Thus the mass is likely 
to be between 0.1 and 1.0 \Msol .   As discussed below, the interior of 
the \Ha\ bubble is likely filled with hot plasma whose cooling time may be 
longer than the dynamical age of the outflow.

\subsection{The Counterflow to HH~80/81}

\begin{figure*}
\center{
\includegraphics[width=6.0in]{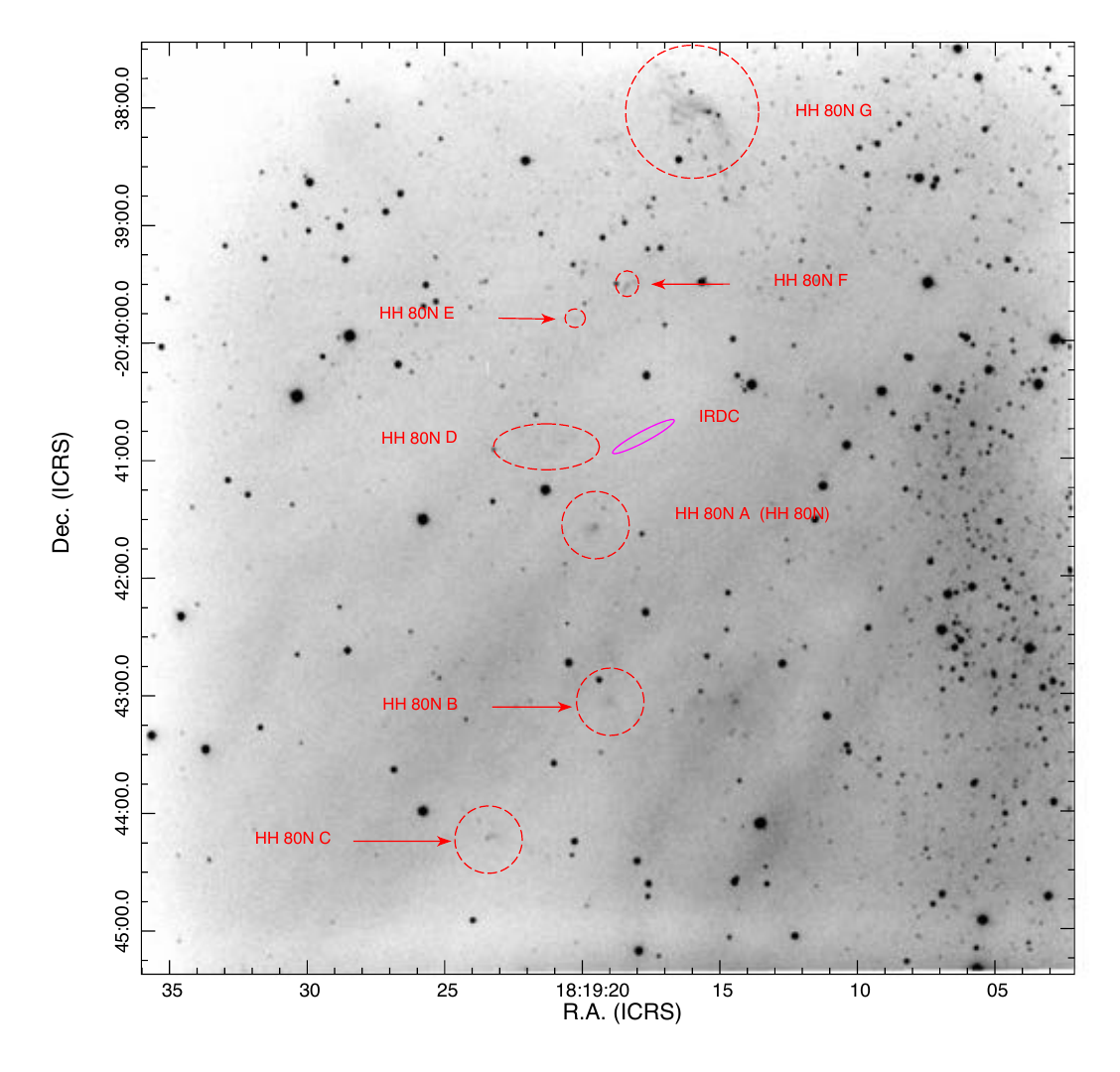}
}
\caption{An \Ha\ image taken with the 30\AA\ filter showing
the HH objects in the counterflow direction north of IRAS~18162-2048
{\bf The magenta oval marks the location of the IRDC north of
HH~80N A (previously HH 80N) which is discussed in the text.}
         }
    \label{fig_14_north_Ha_HHs}  
\end{figure*}

\begin{figure*}
\center{
\includegraphics[width=6.0in]{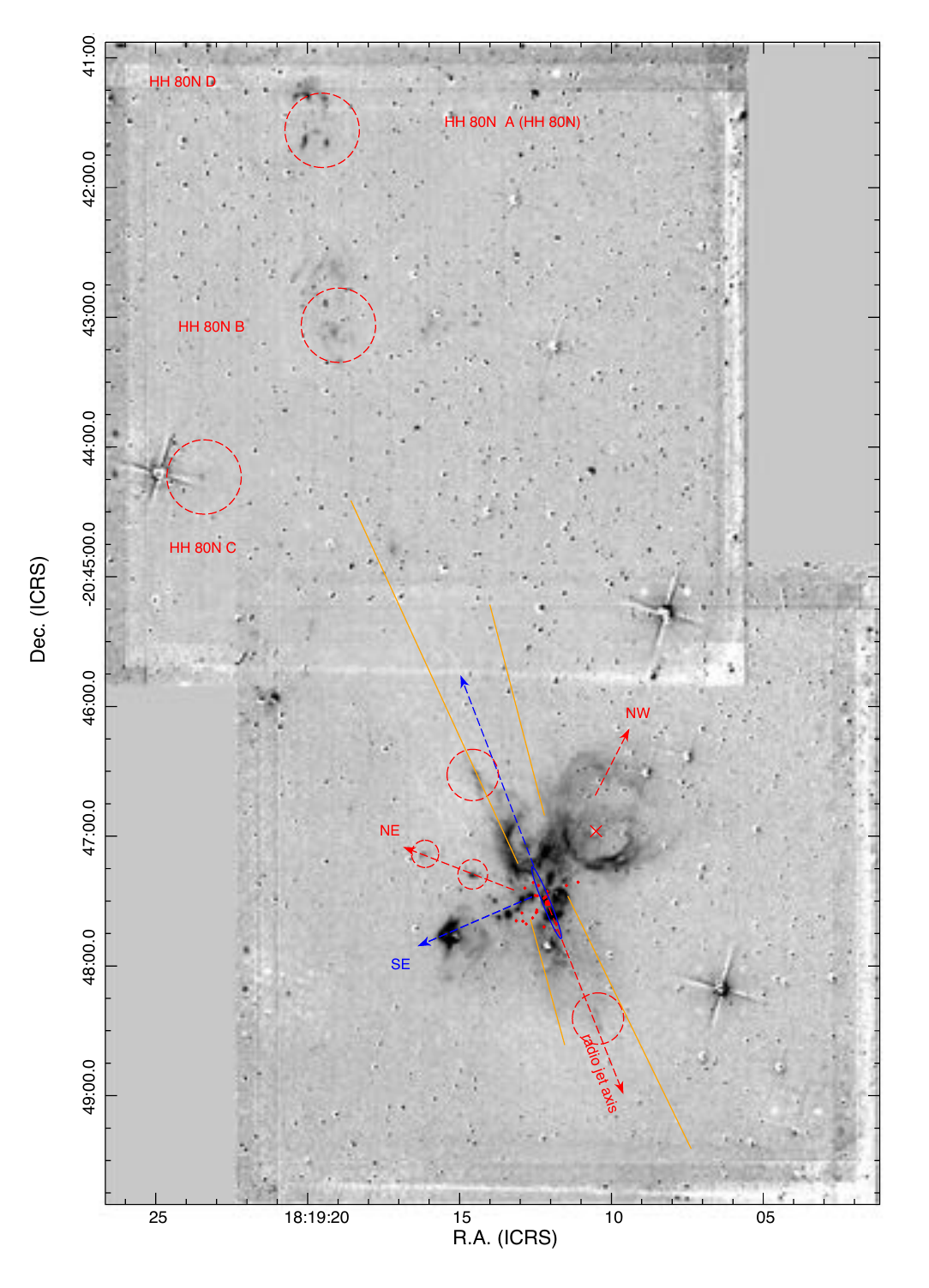}
}
\caption{A continuum subtracted 2.12 \mm\ \Htwo\ image showing the IRAS 18162-2048 field
and the field located to the northeast containing HH~80N~A (HH80N).  Small, red circles show the locations of the cluster of sources detected by \citet{Busquet2019}.  Dashed blue and red arrows show the orientations of multiple molecular outflows emerging from the core containing IRAS 18162-2048 identified by \citet{Fernandez_Lopez2013}.  Small, dashed, red circles in the southern field show some of the MHOs in the southern field.  The yellow lines mark the locations of the mid-IR cavity walls as in Figure 1. The larger, dashed, red, circles in the northern field are the HH objects marked in Figure 1. }
    \label{fig_H2_mosaic}  
\end{figure*}

\begin{figure*}
\center{
\includegraphics[width=7.2in]{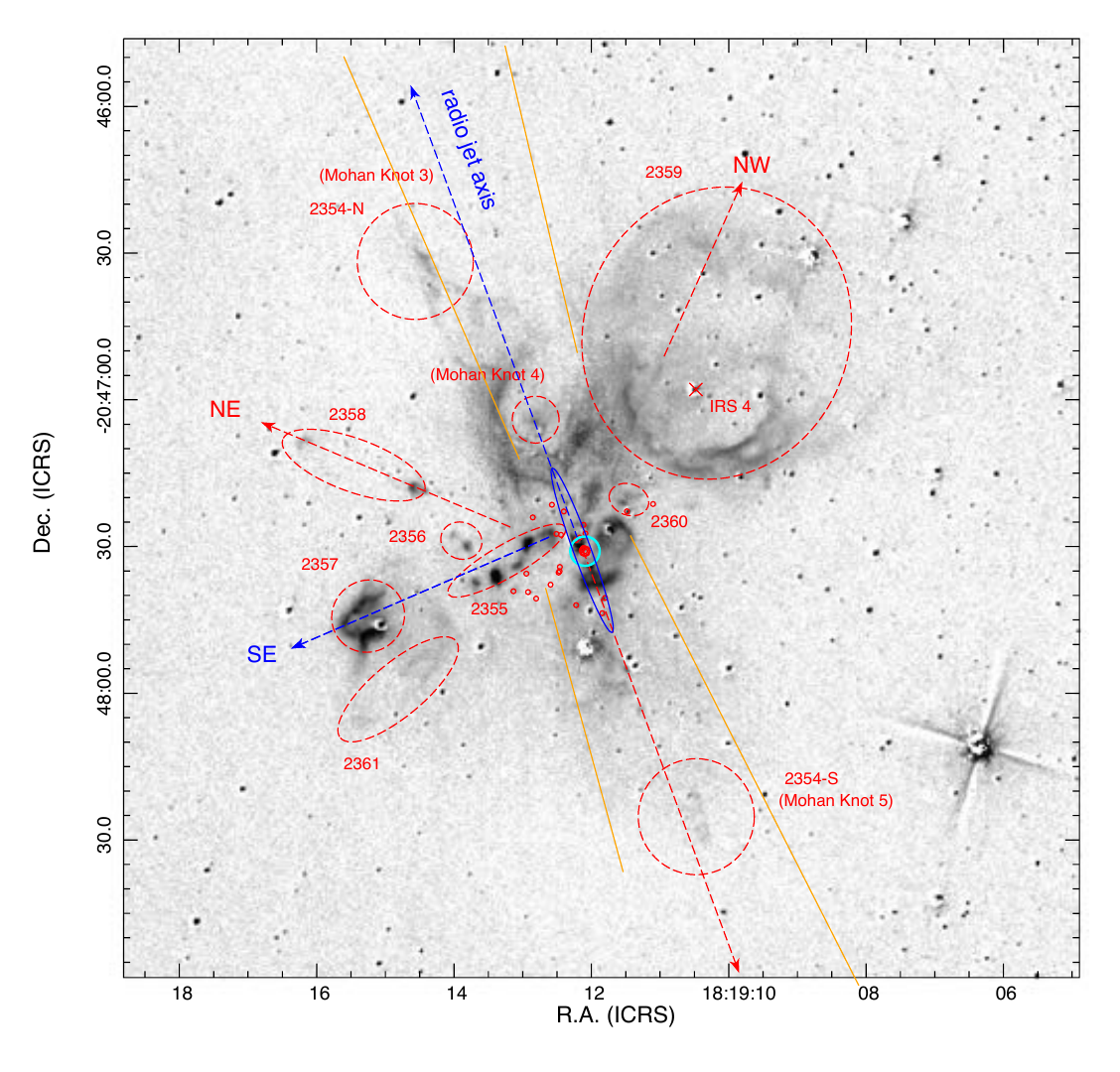  }
}
\caption{A continuum subtracted 2.12 \mm\ \Htwo\ image showing the immediate vicinity of  IRAS 18162-2048.  The dashed, red ovals mark \Htwo\ emitting MHOs with the designations given in 
\citet{Mohan2023}.   Small, red circles show the locations of the cluster of sources 
detected by \citet{Busquet2019}.  Dashed blue and red arrows show the orientations of multiple outflows emerging from the core containing IRAS 18162-2048 {\bf (Fernandez-Lopez et al. 2013)}.   }
    \label{fig_H2_core}  
\end{figure*}

\begin{figure*}
\center{
  \includegraphics[width=7.2in]{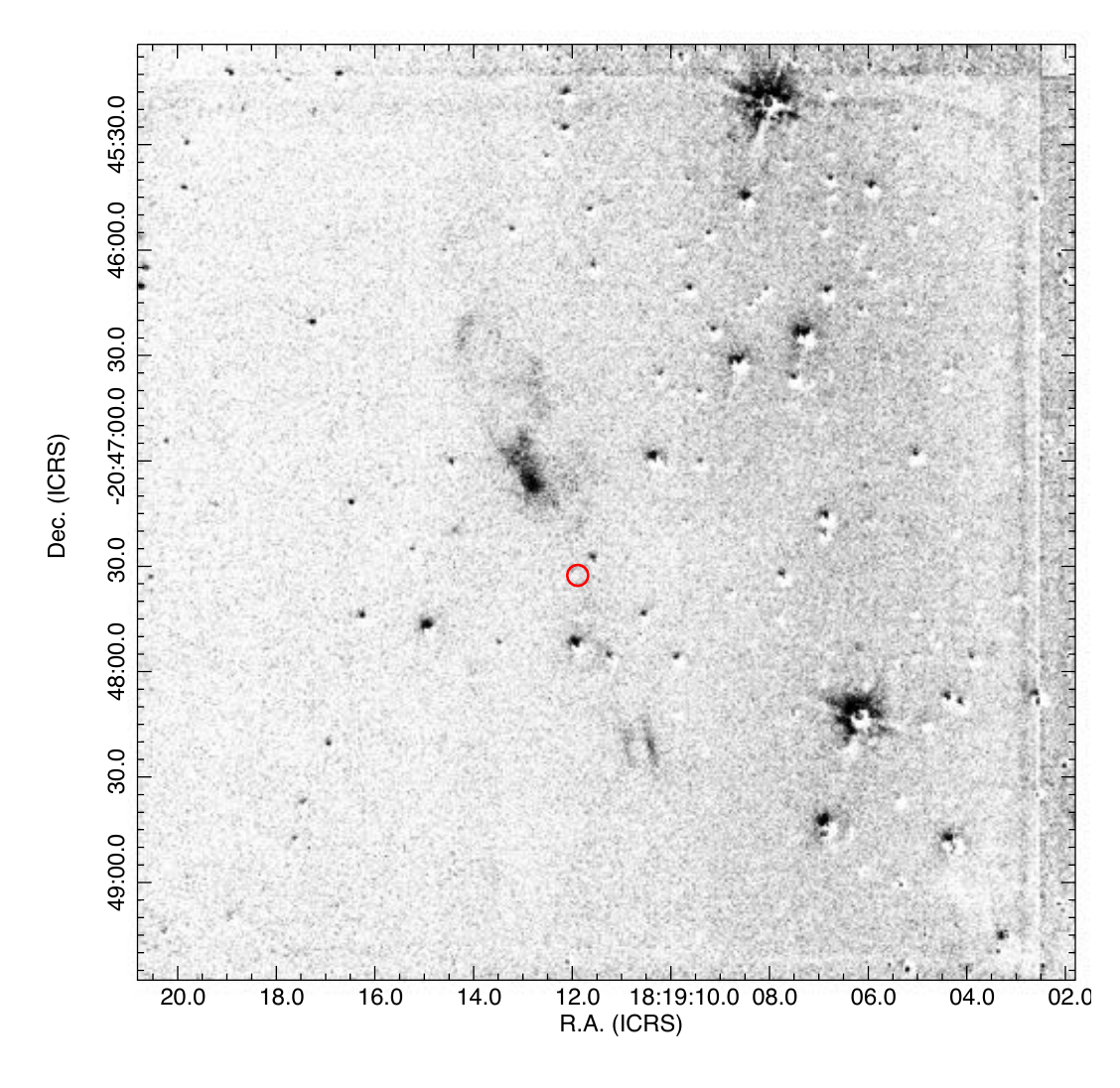}
}
\caption{The immediate vicinity of  IRAS 18162-2048 in a 
continuum-subtracted [\Feii ] image.  
     {\bf  The red circle marks the location of IRAS~18162-2048.}}.
    \label{fig_Feii_core}  
\end{figure*}

\begin{figure*}
\center{
  \includegraphics[width=7.2in]{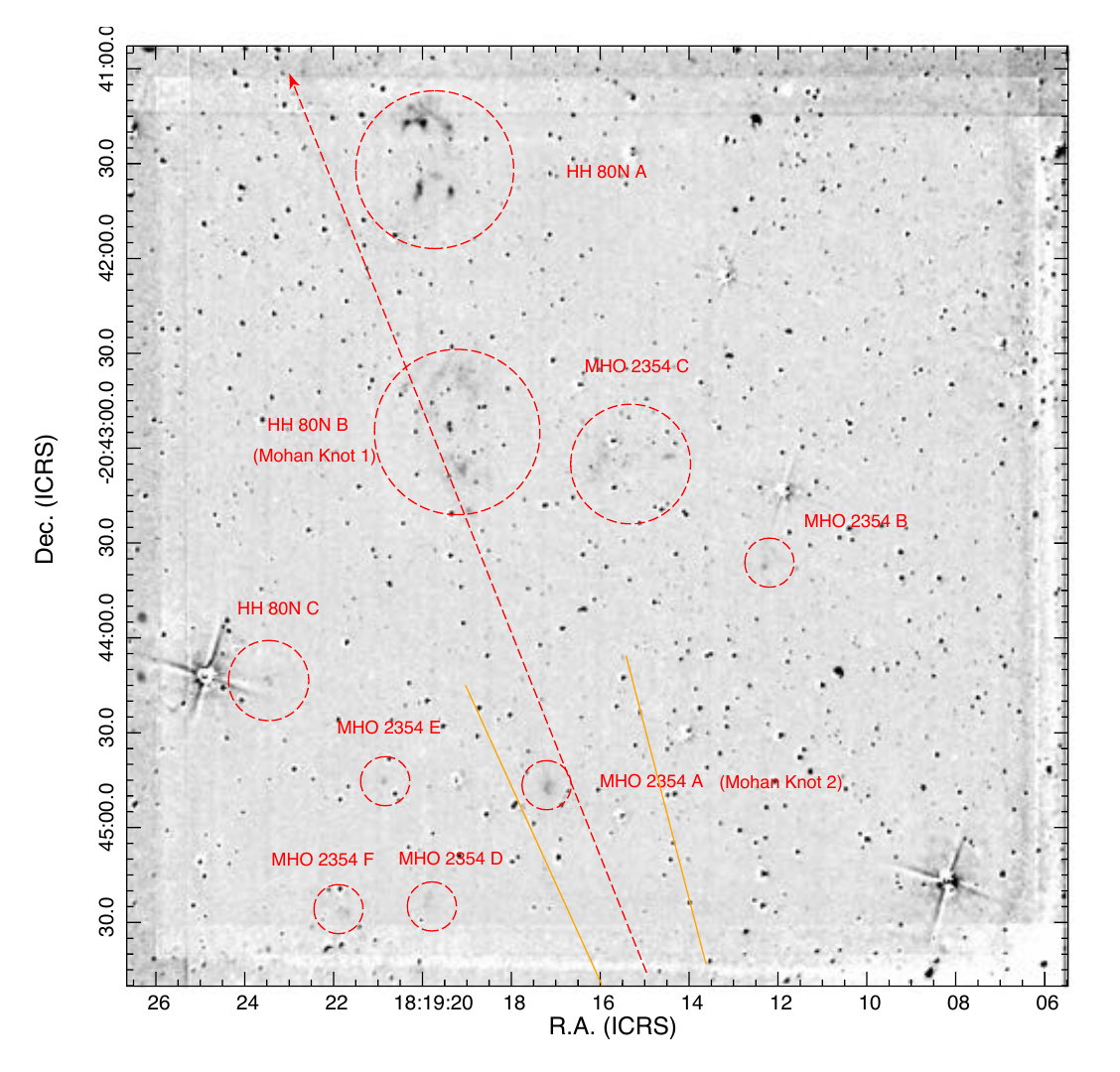}
}
\caption{A continuum subtracted 2.12 \mm\ \Htwo\ image of the region north 
of IRAS 18162-2048 containing HH~80N. Dashed cyan lines show the outline 
of the cavity walls seen in the Spitzer 24 \mm\ images.
{\bf The yellow lines mark the approximate location of the outflow cavity
walls centered on the IRAS source as seen in the Spitzer 8 and 24 \mm\ 
images.  These are also as shown in Figure 1.}}
    \label{fig_HH80N_H2} 
\end{figure*}

An unusual aspect of the outflow from IRAS~18162-2048 is that the brightest
HH objects, HH~80 and HH~81, are highly redshifted.   These shocks are likely
visible because they are seen toward the low-obscuration region beyond the
western edge of the L291 cloud.   Since HH~80/81 are behind the driving source,
the counterflow is expected to be located in the foreground where one might naively 
expect lower extinction.  Yet, no {\bf comparably} bright Herbig-Haro (HH) objects or 
molecular hydrogen objects (MHOs) exist north-northeast of the IRAS source.   

Our images reveal a faint chain of HH objects and MHOs
in the expected counterflow direction. The locations of these features are marked in
Figure \ref{fig1_overview}.   Their positions are listed in Table \ref{tab:HHs}.
Figure \ref{fig_14_north_Ha_HHs} shows an \Ha\ image
of the sub-field located north-northwest of IRAS~18162-2048.
Many of these objects are also visible in the 2.122 \mm\
\Htwo\ S(1) transition.  

The first object suspected to be a shock in the counterflow was a relatively 
bright  radio frequency emission knot, designated HH~80N by \citet{Marti1993}.    
HH~80N is located about one arcminute south of the 
IRDC marked in Figure \ref{fig1_overview} and was thought to mark the site where
the radio jet slammed into the outer parts of this cloud.   
\citet{Molinari2001} presented far-IR spectra of 63 \mm\ and 145 \mm\ [\Oi ]
and 157 \mm\ [\Cplus ] emission detected by the ISO satellite 
to show that this visually obscured region is shock-excited
and thus likely to be a Herbig-Haro object.  
\citet{Molinari2001}
detected extended [\Cplus ] emission along the entire length of the radio jet. 
They argue that the radio jet is surrounded by an extended photon-dominated-region
(PDR).   UV radiation emitted by shocks in the fast jet may be sufficiently hard to
dissociate molecules and to produce a PDR.

In the discussion below, we designate all candidate HH objects 
in the counterflow with the designation HH~80N followed by a capital letter. 
A compact knot of  \Ha\ and [\Sii ] emission coincides with HH~80N 
which we here designate as  HH~80N~A.  This HH object also emits  in
the \Htwo\ lines and is thus an MHO.   The \Htwo\ emission consists of an irregular
bow-shaped structure containing three bright knots.  About 30\arcsec\ south, this 
object is trailed by a pair of knots which may mark the limb-brightened rim
of this bow shock.

Where the \Htwo\ emission coincides with or is near a candidate HH object, we use the HH object designation (HH~80N followed by a capital letter).  
Where only \Htwo\ emission is seen, with no \Ha\ or [\Sii ] counterpart,  we use the designation MHO~2354,  the name given to MHOs associated with the 
outflow from IRAS~18162-2048 in \citet{Mohan2023}, followed by a capital letter. 
Although HH~80N~A is also an MHO, it is outside the field imaged by \citep{Mohan2023}.

HH~80N B is the southernmost visual wavelength feature in a chain of \Ha\ and [\Sii ] emission knots located on the radio jet axis in the counterflow direction.  It is the base of an extended network of \Htwo\ filaments that can be dimly traced to HH~80N~A.    HH~80N~B was designated MHO~2354~Knot~1 by \citet{Mohan2023} because it was the most northern MHO in their observed field.

HH~80N C is a compact, $\sim$4\arcsec\ long, cone-shaped feature located 
about 2\arcmin\ east of the radio jet axis with a point like-apex facing 
southeast with similar fluxes in \Ha\ and [\Sii ].  A faint trail of \Ha\ emission
extends west toward the jet axis.   The conical nebula is also visible in 
\Htwo\ emission.  Its not clear if this is a side-shock associated with the 
IRAS~18162-2048 outflow, or powered by another YSO.

HH~80N~D consists of a pair of dim \Ha\ blobs about 35\arcsec\ and 24\arcsec\ 
due east of the IRDC (`HH~80B-core' - see below).  These features are directly on
the radio jet axis and separated by about 13\arcsec\ along a SE-NW line.

HH~80N~E is a compact \Ha\ knot close to, but slightly west of the radio jet axis. 

HH~80N~F is a brighter and compact  \Ha\ and and [\Sii ] knot between HH~80N~E
and HH~80N~G.

HH~80N~G is the brightest and most extended HH object in the counterflow direction. 
It is located several arc-minutes west of the radio jet axis defined by a 
line from IRAS~18162-2048 and radio source \#34.   HH~80N~G 
resembles a partial bow shock $\sim$20\arcsec\ by 40\arcsec\
in extent which faces northwest.  The mean [\Sii ] surface brightness 
is about 1.1 times larger than
the \Ha\ surface brightness, indicating excitation in a shock.
The $\sim$2\arcmin\ offset from the jet axis
raises the possibility that it is powered by a source other than
IRAS~18162-2048.   However, given the giant bubble located south-southwest of 
the IRAS source, it is possible that HH~80N~G also traces a mostly obscured 
giant bubble wall located in the counterflow.  It may be seen through a 
particularly translucent part of the L291 cloud. 

Spitzer images at 8 and 24 \mm\ show the presence 
of a compact, 6\arcsec\ by 36\arcsec , Infra Red Dark Cloud (IRDC) 
at ICRS=18:19:17.77, -20:40:48 
about 160\arcsec\ south of HH~80N~G.  This cloud was studied extensively by
\citet{Masque2011} who  called it the `HH~80N~core'.  
The minor axis of this cloud is towards the northeast at PA$\sim$30\arcdeg. 
\citet{Masque2011} found a total mass of $\sim$20 \Msol\ for the clump.  
They found three young stellar objects embedded within which are  
potential sources for exciting HH objects in this region.  \citet{Girart2001}
found a bipolar molecular flow emerging from the HH~80N~core.  This flow is
oriented nearly east-west at position angle $\sim$80\arcdeg .  It is
possible that HH~80N~D is associated with this flow.  However, the orientation
of this outflow is inconsistent with being the driver of HH~80N~G.   

We searched the vicinity of radio source \#34 for both HH objects and MHOs, but
none were detected at our sensitivity limit.  Although this portion of the 
IRAS~18162-2048 flow is expected to be blueshifted and approaching us, it is 
nevertheless still highly obscured judging from the relatively low-density
of stars in our image.  This implies that the L291 cloud, if it is a sheet,
is even more inclined to our line-of-sight than the giant outflow from 
IRAS~18162-2048.

Near-IR imaging shows the presence of faint and extended 2.12 \mm\ molecular
hydrogen emission between IRAS~18162-2048 and HH~80N.   Figure \ref{fig_H2_mosaic}
shows a continuum subtracted \Htwo\ image of the field from 2\arcmin\ south of
IRAS~18162-2048 to HH~80N~A (previously HH~80N).  Figure \ref{fig_H2_core} 
shows a closeup view of the immediate vicinity of IRAS~18162-2048 
{\bf in a continuum subtracted \Htwo\ image.  
Figure \ref{fig_Feii_core} shows
the core region in a continuum subtracted [\Feii ] image.}
Figure \ref{fig_HH80N_H2}  shows the field north of IRAS~18162-2048
containing HH~80N~A (HH80N) {\bf in a continuum subtracted \Htwo\ image}
with the various MHOs marked.  The
continuum subtraction removes most of 
the bright reflection nebulosity.  Thus, these images primarily show pure
2.12~\mm\ \Htwo\ emission in the v=0-0 S(1) line.  

In Figure \ref{fig_H2_core}, the location of 
IRAS~18162-2048 is indicated by a cyan circle.   
The bubble on the upper right traces a cavity created
by a moderate mass star at  ICRS=18:19:10.457, -20:46:58 also
known as GGD~27~IRS4, which is thought to be a Herbig AeBe star with
spectral type B2.  There is a faint, jet-like feature extending from 
near the IRAS source through the cavity.  The orientation and location is indicated
in Figure \ref{fig_H2_core} by a dashed red arrow.  This feature lines-up with
the `NW' redshifted CO outflow found by \citet{Fernandez_Lopez2013}. 

The U-shaped cavity in the counterflow opening towards the northeast 
traces the inner walls of the GGD~27 reflection nebula where the radio jet and
surrounding outflow has created a cavity in the clump hosting IRAS~18162-2048.
A knot of \Htwo\ emission, designated knot 3 in \citet{Mohan2023},  
in the middle of the U-shaped cavity lies along the 
radio jet axis and may mark a shock in the outflow from the IRAS source. 

The \Htwo\ images (Figures \ref{fig_H2_mosaic} and \ref{fig_H2_core}) 
show a 45\arcsec\ long chain of 
knots extending to the  east-southeast of the IRAS source and terminating in a bow 
shock.  This flow  may be the \Htwo\ counterpart of the blueshifted 'SE' 
CO outflow emerging at position angle PA = 126\arcdeg\ detected by 
\citet{Fernandez_Lopez2013}.  \citet{Mohan2023} give these features the designations
MHO~2355 and MHO~2357.   MHO~2355 consists of a wiggly chain of five knots.
MHO~2357 is the bright \Htwo\ bow shock.  MHO~2360 may trace an \Htwo\ knot
in the counterflow to MHO~2355 / MHO~2357. 
Two \Htwo\ knots, marked by small, dashed circles in Figure \ref{fig_H2_mosaic}
mark the locations of \Htwo\ knots along the redshifted `NE' CO outflow lobe
found by \citet{Fernandez_Lopez2013}.   \citet{Mohan2023} give these features 
the designation MHO~2358.   A pair of knots designated MHO~2356 are located 
between these two flows indicating yet another outflow.  

Figure \ref{fig_H2_core} shows two wedge-shaped \Htwo\ features close to the 
radio jet axis, marked by the larger, red dashed circles.   These two objects are 
symmetrically placed about IRAS~18162-2048 defining a line close to the radio jet axis. 
But the northern \Htwo\ wedge is slightly east of the radio jet axis while 
the southern wedge is offset slightly west of the radio jet axis.    These
features are designated MHO~2354-N and MHO~2354-S.  MHO~2354-N,
designated knot 3 in \citet{Mohan2023}, coincides with
the northeastern rim of the mid-IR cavity seen in Spitzer 8 and 24~\mm\ images.

Faint 1.644 \mm\ [\Feii ] emission is seen along the jet axis around the 
location of the \Htwo\ wedges (Figure \ref{fig_Feii_core}).  
Figure \ref{fig_H2_core} shows a continuum-subtracted zoom-in
view of the region around the IRAS source along with the several dozen YSOs
detected by ALMA \citep{Busquet2019}.   Along the northern direction, the
[\Feii ] emission resembles an elongated bubble.  Only dim [\Feii ] emission
is seen near the southern \Htwo\ wedge in the form of a pair of parallel
streaks.  The relative dimness of the southern [\Feii ] feature is consistent
with more extinction the redshift of the southern outflow lobe and blueshift 
of the northern lobe,

\begin{deluxetable*}{cccl}
\tablenum{4}
\tablecaption{Counterflow HH Objects and MHOs \label{tab:HHs}}
\tablewidth{0pt}
\tablehead{
\colhead{Feature} & \colhead{RA (ICRS)} & \colhead{Dec (ICRS)} & \colhead{Comments}  \\ 
\colhead{mas/yr} & \colhead{position angle} & \colhead{description} \\
\colhead{} & \colhead{(ICRS)} & \nocolhead{ICRS} & \colhead{} }

\decimalcolnumbers
\startdata
HH~1180      &   18:19:00     & -20:55:30 &  Giant \Ha\ bow shock. \\
HH~80N A    &  18:19:23.4    & -20:44:14 &  \Ha\ knot 75\arcsec\ E of jet axis. \\
HH~80N B    &  18:19:19.0    & -20:43:04 &  \Ha\ knot on radio jet axis (Mohan Knot 1). \\
HH~80N C    &  18:19:19.5    & -20:41:34 &  HH80N; tip of \Htwo bow. \\
HH~80N D    &  18:19:21.3    & -20:40:54 &  Pair of \Ha\ knots separated by $\sim$10\arcsec . \\
HH~80N E    &  18:19:20.3    & -20:39:48 &  Compact knot. \\
HH~80N F    &  18:19:18.4    & -20:39:30 &  Compact knot. \\
HH~80N G    &  18:19:15.9    & -20:38:02 &  Large bow.  \\
MHO~2354 A  &  18:19:17.0    & -20:44:48 &  Knot on radio jet axis (Mohan Knot 2) \\
MHO~2354 B  &  18:19:12.0    & -20:43:38 &  Dim knot W of jet axis \\
MHO~2354 C  &  18:19:15.1    & -20:43:07 &  Pair of diffuse knots NE of MHO~2354 B \\
MHO~2354 D  &  18:19:19.6    & -20:45:27 &  Dim knot E of jet axis \\
MHO~2354 E  &  18:19:20.7    & -20:44:47 &  Dim knot E of jet axis, N of MHO~2354 D \& F \\
MHO~2354 F  &  18:19:21.7    & -20:45:27 &  Dim knot E of MHO~2354 D \\
\enddata
\end{deluxetable*}

\section{Discussion}

With speeds in excess of 1,000 \kms , the HH~80/81 radio jet exhibits the 
fastest known proper motions of any outflow from a young stellar object.   
The high-excitation Herbig-Haro objects are mechanically illuminated by this 
jet.  Our proper motion 
measurements imply that such fast motions persist at a projected distance 
$\sim$300\arcsec\ (2 pc) from the source{\bf , the massive young 
stellar object (MYSO) IRAS~18162-2048.}    

Rapidly accreting MYSOs with masses between $\sim10$~\Msol\ to $\sim$20~\Msol\ 
develop bloated and cool photospheres which 
prevent them from emitting hydrogen-ionizing, extreme ultraviolet (EUV) 
radiation \citep{Hosokawa2009}.   However, for accretion rates 
below $10^{-3}$~\dotMyr , as they approach $\sim$20 \Msol, their photospheric 
radii and temperatures approach zero-ago main sequence (ZAMS) values.   They 
start to ionize their surroundings.  The disk rotation curve suggests that 
IRAS~18162-2048 may have a mass of $\sim$20~\Msol\ and may be approaching the 
stage where it emits EUV at a rate expected for a ZAMS star.  

If IRAS~18162-2048 is on the ZAMS, its massive disk may trap the Lyman 
continuum emitted by the star, preventing the growth of an \Hii\ region 
\citep{Hollenbach1994}.  For a stellar mass,  $\rm M_{star}$=20~\msol , 
and a sound speed in photo-ionized plasma, $\rm c_s$=10~\kms , the 
gravitational radius is $\rm r_G = G M_{star} / c_s^2$=178~AU .  
Photo-ionized plasma will be bound by the gravity of the star as long 
as the ionization front is closer to the star than $\rm r_G$.  

The observed peak radio continuum emission at 1.4, 5, and 15 GHz from 
IRAS~18162-2048 is about $\rm S_{\nu} = 3.8 ~ \nu ^{+0.2 \pm 0.1}$ mJy in 
beam-matched observations where $\nu$ is in units of 1 GHz
\citep{Marti1993}.   Most of this emission likely originates from 
optically thin free-free emission from the jet and thus places an upper-bound on 
the flux from a hyper-compact \Hii\ region.

ALMA  1.14 mm (263 GHz) 
observations at the longest baselines indicate the presence of 
a compact source less than $\sim$40~mas (56 AU) in diameter with a flux of
19 mJy,  interpreted by \citet{AnezLopez2020} to trace ionized gas
from a hyper-compact \Hii\ region.   
Extrapolating the above formula for the flux density from the centimeter
regime to 1.14 mm implies a peak flux to
be in the range 6.6 to 20.2 mJy with a most likely value of 11.6 mJy.
Thus, in addition to the emission from the thermal plasma in the jet, 
there may be a hyper-compact \Hii\ region surrounding this MYSO which may
be optically thick at the centimeter wavelengths.  If $\sim$12 to 19 mJy flux
originates from a hyper-compact \Hii\ region which is optically thick at 263 GHz, 
its radius would be about 8 to 10 AU.  Such a compact \Hii\ region would be
bound by the gravity of the star and disk.

A 20~\Msol\  ZAMS star has a Lyman continuum luminosity of about 
$\rm 3 \times 10^{48}$  ionizing photons per second.  
A hyper-compact \Hii\ region with a spherical radius of 8 to 10  AU, 
uniformly filled with plasma in photo-ionization equilibrium with this 
ionizing luminosity will have an electron density of about $\rm n_e \sim 10^9$~\cmq . 
The ALMA 1.14 mm image constrains the hyper-compact \Hii\ region to be
smaller than 56 AU diameter.  
For a radius of $\sim$56~AU,  $\rm n_e \sim 7 \times 10^7$~\cmq .

If IRAS~18162-2048 grew via highly variable, episodic accretion, 
the growing protostar may have experienced periods during which it was 
accreting at low rates or not accreting at all
{\bf \citep{Galvan_Madrid2008} }.   
Such periods of quiescence may have allowed the photosphere to collapse and
heat-up.   The MYSO may have settled onto the ZAMS even at a mass 
well-below 20~\Msol.  

\subsection{The Jet and Disk Axis Inclination Angle Revisited}

The inclination angle of the jet or the disk axis is given
by $\rm {\it i} = arctan(V_r / V_{PM})$ where $V_r$ is the radial velocity and $V_{PM}$ is the
proper motion of a feature.  Here, $i=90$ corresponds to the disk being face-on and
the jet axis pointing directly at or away from the Sun.
\citet{Heathcote1998} used proper motion and radial velocity measurements to determine
the inclination angle of HH~80 and 81.   Their
spectra found a maximum radial velocity of $V_r$=+600~\kms\ and a mean proper motion of 
HH~80 and 81 of $\sim$350 \kms .  Their estimate of the inclination angle resulted in a
value of $i$=60\arcdeg , corrected to a distance of 1.4 kpc.

The high-angular resolution measurements presented here show that the proper motion 
vector field is complex.  Thus, it is unclear if a mean value of the proper motions 
is an appropriate estimator of the inclination angle of the HH object motions.  Using
a maximum proper motion, $V_{PM}$=1,200 \kms\ and the maximum radial velocity of 
600~\kms\ from \citet{Heathcote1998} gives $i\approx$27\arcdeg .   However, these
maximum values were determined for different features in the shock complex. 

The high angular resolution ALMA 1.14 mm images of the IRAS~18162-2048 disk presented
by \citet{Girart2017,Girart2018} and \citet{AnezLopez2020} allow a 
measurement of the orientation
of the disk axis.  The analysis of \citet{AnezLopez2020} gives an inclination for the
disk axis of $i_{disk}$=49$\pm$5\arcdeg .   For the remainder of the analysis, we assume
that the inclination angle is likely to be between 44\arcdeg\ and 65\arcdeg\ with
the south-southwestern lobe of the outflow containing the bright HH objects HH~80 and 81
receding away from the Sun.  The fastest proper motions in the jet and 
HH objects, $\sim$1,200~\kms , imply speeds ranging from 1,670 to 2,400 \kms\ 
depending on the actual value of the inclination angle.

\subsection{A Collimated, Fast, Line-driven Wind?}

Given the $\sim$44\arcdeg\ to~65\arcdeg\ inclination angle of the outflow axis from 
IRAS~18162-2048, the proper motions imply speeds similar to the line-driven 
stellar winds powered by main sequence O stars, about 2,000 \kms\ \citep{Vink2022}.  
If  IRAS~18162-2048 is a ZAMS O star, it must be so highly embedded that no 
visual or UV  light escapes towards our line-of-sight.  However, a fast, 
line-driven wind could form near the stellar photosphere.   

An isotropic,  fast wind from the central star may be collimated into a jet by 
a combination of density gradients and strong magnetic fields.   Such a field, 
anchored to a massive, differentially rotating disk is expected to develop 
magnetic hoop stress.  A lower bound on the strength of a collimating magnetic  
field can be obtained  by setting the magnetic pressure equal to the ram pressure
of the wind.   For a wind speed of 2,000~\kms\ and  mass loss-rate of 
$\rm \dot M = 10^{-5}$~\Msol ~$\rm yr^{-1}$, collimation at distances of 10
to 100 AU from the star requires a magnetic field of order 4 to 0.4 gauss, 
respectively.

A 20~\Msol\ ZAMS star has a luminosity of about $\rm 10^5$~\Lsol , 
nearly an order of magnitude larger than the measured luminosity
of IRAS~18162-2048 assuming an isotropic radiation field.  In this scenario, a
disk constrained \Hii\ region and jet cavity may allow most of the luminosity 
to be beamed along the jet axis.   This `flashlight effect' may explain the 
low measured luminosity of IRAS~18162-2048 from our vantage point
\citep{Kuiper2015}.

One problem is that the mass-loss rates from late O-type main sequence
stars are typically $\rm 10^{-8}$ to $\rm 10^{-7}$~\Msol\ ~year$^{-1}$, 
two to three orders of magnitude lower than the mass loss-rate estimated 
from the \Htwo\ emission by \citet{Mohan2023} or from the empirical 
correlation between stellar luminosity and  mass-loss rate \citep{Maud2015}.  
As discussed below, a strong magnetic field anchored to the massive disk may 
launch a magneto-centrifugal-wind on its own, adding to the 
mass-loss rate of the system if there is a line-driven component.

\subsection{A Hybrid, Ionization Initiated, Magneto-Centrifugal Wind?}

The massive, $\sim$5 \Msol\ disk distinguishes IRAS~18162-2048 from most other MYSOs. 
Continued accretion through a disk requires that the
rate at which it flows through the disk toward the MYSO be larger than the rate at which 
the disk surface layers are photo-ablated by Lyman continuum from the star.
\citet{Hollenbach1994} investigated mass-loss from a disk due to photo-ablation
by a massive central star.   The EUV field can drive a slow wind from the disk surface
at radii larger than the gravitational radius, $\rm r_G$. 
For a $\sim$20~\Msol\ star, photo-ablation-driven mass loss
rates tend be around $\sim 10^{-5}$~\dotMyr\, close to what is estimated for the
IRAS~18162-2048 outflow.    
{\bf However, the absence of an extended \Hii\ region on scales
larger than $\rm r_G $ rules out this type of model.}

\citet{KuiperHosokawa2018} presented models of massive star formation with radiative
feedback, including photo-ionization but they ignored the role of magnetic fields.
If the circumstellar disk is threaded by a magnetic field, differential rotation-induced 
shear can lead to field amplification.  Within the disk, or partially ionized disk
surface layers, the field is expected to have a toroidal geometry wrapped-up by shear. 
The expected geometry of open field-lines above and below the disk is a helical,  
hourglass-shaped field pinched at its waist by the disk.   
\citet{OlivaKuiper2023a,OlivaKuiper2023b}  present models with magnetic fields.  
But, this work ignores fields that might link the circumstellar disk to the forming 
massive star.  \citet{Tanaka2016,Tanaka2017} presented models of MYSOs with strong outflows,
showing that as they approach the ZAMS, their outflows are the first components to
become ionized.

Continued accretion from a magnetized disk onto the MYSO is expected to drag field 
lines from the disk onto the stellar photosphere, creating closed magnetic loops.  
These loops will funnel accreting matter onto high latitude regions of the star, in
`funnel flows', commonly thought to occur in low-mass star formation.  Accretion
mediated by funnel flows are thought to regulate the spin-rates of the forming stars. 
The kilo-gauss magnetic fields present  on some young ZAMS O-type stars such as 
$\theta ^1 \rm Ori$~C \citep{Donati2002} may be fossil remnants of such magnetic 
structures.    

Magnetically confined funnel 
flows will become ionized by EUV radiation if the MYSO is close to the ZAMS.  
But, sufficiently dense accretion columns are likely to be optically thick to 
EUV and may contain neutral cores.  In this case, only the surface layers will be ionized.   
As long as accretion columns don't fully cover the projected surface area of the 
MYSO, some, or most EUV will escape the inner magnetospheric 
closed loops to irradiate the disk surface penetrated by open field lines. 

Photo-ionization of the disk surface may provide an efficient mechanisms for mass-loading
the open magnetic field-lines.
Charged particles injected onto open field lines can be accelerated, 
forming a magneto-centrifugal wind \citep{Pudritz2019}.  Such  winds are thought 
to play important roles in driving jets and outflows from forming stars.  
We consider a scenario in which a strongly magnetized disk can support 
continued accretion onto the MYSO and at the same time drive a magneto-centrifugal 
wind launched by photo-ionization of the disk surface which loads plasma onto the 
open field lines. 

The electron density and radius of the hyper-compact \Hii\ 
region implied by the 1.14 mm ALMA  data is 
$\rm n_e \approx 10^8$ to $10^9$~\cmq\ at  $R_{ii} \approx$ 60 to 10 AU.    
If the photo-ablating 
plasma is lofted off the disk surface as a quasi-spherical 
wind with a velocity $\rm V_w$ 
the implied mass-loss rate is 
$\rm \dot M_{ii} = 2 \pi R^2_{ii} \mu m_H n_e V_w$
= $\mu m_H V_w R_{ii}^{1/2} (3 \pi  Q / \alpha_B)^{1/2}$.
In a frame co-rotating with a portion of the disk,  $\rm V_w$ will likely have a value
about 1.5 to 3 times the sound speed in the photo-ionized plasma, 
$\rm c_s \approx$10~\kms\ because of the acceleration by the large 
vertical density and pressure gradients. 
For $\rm V_w$=20 \kms , 
$\rm \dot M_{ii} \sim$
$\rm \approx 9 \times 10^{-6}$ to $1.6 \times 10^{-5}$~\dotMyr\ for 
$\rm R_{ii}$=10 to 30 AU, close to the estimated mass-loss rate in the 
radio jet.

In the rest frame of the star, the plasma loaded onto open field lines
will inherit the  roughly Keplerian orbit speed at the footprint of the 
field.    For a $\sim$20~\Msol\ star the orbit speed is
$\sim$420~\kms\ at 0.1 AU, 134 \kms at 1 AU, and 25~\kms\ at 
30~AU.     Plasma loaded onto the  field-line at 0.1 AU will reach a 
a speed of 2,000 \kms\ at a radial distance of 0.476 AU from the axis of the disk.
Plasma loaded onto a rigid, open field-line at 1 AU, will reach 2,000 \kms\
at a radial distance of $\sim$15 AU from the disk axis.  But, plasma launched
at 25 AU would require a magnetic lever arm {\bf of} nearly 80 to reach this speed at
R$\sim$2,000 AU.  

The plasma loaded onto open field-lines is likely to flow 
away from the disk on a nearly radial trajectory as seen from the star.  The
hoop-stress of the azimuthal component of the field may collimate this 
wide-angle flow into a jet at vertical distances of 10s to hundreds (for the 
launch radii of 0.1 to 1 AU ) to  thousands of AU (for the outer launch radii
near 30 AU).  In this  magneto-centrifugal launch scenario, 
it is likely that the jet is layered with a fast core surrounded by a sheath 
of slower ejecta reflecting a larger radius of the jet launch point and a 
slower orbit speed.   Only a small fraction of the jet mass-loss-rate is required
to have speeds in excess of 1,000 \kms\ to produce the observed ultra-fast 
motions in the radio jet and HH objects.   In a layered jet model, only the
jet core is required to have such fast motions.  The flow speed may decrease
with increasing distance from the jet axis.

A rough estimate of the field strength required to produce the outflow 
can be obtained by equating the kinetic energy density of the plasma with 
the magnetic energy density where the magnetic lever-arm causes the
photo-ablation flow from the disk surface to reach the observed
jet velocity.   Given a mass-loss (or mass-loading) rate $\dot M_w$, the
jet velocity, $V_j$, and radius where the magneto-centrifugal wind reaches
the jet velocity, $R_j$, the mean density of the flow is 
$\rho (R_j) \approx \dot M / (\pi R^2_j V_j)$.   
The mean kinetic energy density is then $0.5 \rho V^2_j$.   Setting this equal to
$B^2(R_j) / 8 \pi$ gives 
$B(R_j) \approx [4 \pi \rho (R_j)]^{1/2} V_j \sim 1.6~ \dot M^{1/2} _{-5}~ V^{1/2}_{j,2000}~R^{-1}_{j,30} $
gauss where 
$\dot M _{-5}$ is in units of $10^{-5}$ \Msol ~ yr$^{-1}$,
$R_{j,30}$  is in units of 30 AU, and
$V_{j,2000}$ is in units of 2000 \kms .   Thus, a field of a few gauss at 30 AU
and less than a kilo-gauss at 0.1 AU is sufficient to accelerate a photo-ablation
driven, low-velocity wind from the disk surface to the $\sim$2000 \kms\ flow
observed in the IRAS~18162-2048 jet by means of the magneto-centrifugal mechanism.

\subsection{Impact of a Fast, $>10^3$ \kms\  Jet on the Surrounding Medium}

The visibility of these fast motions in tracers
such as [\Sii ] and even [\Oiii ] is surprising in light of these extraordinary speeds.  
Shocks with speeds in excess of 1,000 \kms\ would lead to the production of considerably 
higher ionization states.
The presence of relatively low ionization states in the fastest moving knots suggests that
the observed species are excited by shocks formed where fast ejecta overtake slightly 
slower, but nonetheless fast moving debris.  Alternatively, if the shocks are formed by the
collision of fast ejecta with stationary or slow-moving material, the low-ionization states
trace forward shocks propagating into a much denser medium than the fast flow.  
Magnetic fields may also cushion shocks, enabling lower-ionization states to
survive and be excited into emission.

The HH~80/81 shocks are located beyond the western edge of the L291 cloud and are thus 
likely in a relatively low-density environment.  The rich, background star-field at 
visual wavelengths implies our line-of-sight can penetrate several kpc without more 
than a few magnitudes of extinction.   If the inter-cloud medium were uniform, 
this would imply a density $\rm n(H) < 1$~\cmq .  But the region just outside obvious 
obscuration may be somewhat denser,  but likely to have a density 
less than $\sim$10 to 100 \cmq .     
The electron density of the post-shock plasma in the HH~80/81 shocks has been measured 
to be  between $\rm n_e = 10^3$ to $10^4$~\cmq\ using the [\Sii ] 
doublet ratio \citep{Heathcote1998}.  Thus, the second model above (dense clumps moving 
supersonically into a lower density medium) is likely correct. 

As discussed in Section 3.3, the cooling time of post shock plasma is
$\rm \tau_{cool} \approx  7000 ~n_H^{-1} V_{S,100}^{3.4} $
where $n_H$ is the hydrogen volume density and $\rm V_{S,100}$ is the shock
speed in units of 100 \kms\ \citep{Draine2011}.
For $\rm n_H$ = 100 \cmq , likely an upper bound on the density of 
gas into which fast ejecta from IRAS~18162-2048 is running, 
$\rm \tau_{cool} \approx 1.8 \times 10^5$ years, sufficiently long to enable 
the formation of the \Ha\ bubble by fast shocks such as indicated by the 
large proper motions.  The post-shock, low-density plasma will 
likely expand adiabatically into the surrounding medium.  

The temperature immediately behind a 1,000 \kms\ shock is 
$\rm T_{ps} = 3 \mu m_H V_S^2 / 16 k $  
$\rm \approx 1.38 \times 10^5~ V^2 _{S,100}$ Kelvin for $\rm \mu  = 0.609$
appropriate for fully ionized plasma
\citep{Draine2011}.   Thus, behind a 1,000 \kms\ shock, the plasma 
reaches a temperature  of about $10^7$ Kelvin, sufficiently hot
to explain the observed X-ray emission from parts of HH~80 and 81.  If driven by the
impact of a jet or dense clumps of ejecta interacting with a lower density
medium, the hot plasma will expand adiabatically to power an energy-conserving
bubble, similar to the early phases of supernova remnant expansion.  
The expanding bubble sweeps up a shell from the surrounding ISM.  
The shell can be ionized by fast shocks, by EUV escaping along the jet axis
from IRAS~18162-2048, EUV radiation produced by the recombining and cooling
X-ray plasma, or the ambient Lyman continuum radiation field pervading the
exterior of the L291 cloud.   All four mechanisms may contribute to the visibility
of the giant \Ha\ bubble.

It is remarkable that no extensive molecular outflow exists around this fast 
radio jet and associated HH objects.
One possible interpretation is that molecules that were swept-up by 
the younger outflow produced by the IRAS source as {\bf it} was accreting was 
subsequently completely dissociated.  As discussed above, inside the parent cloud 
dissociation could have been caused by strong shocks, UV radiation emitted by the 
MYSO,  or by radiation emitted by shocks powered by the ultra-fast jet.   
The nearly 
1,000 \kms\ velocity variation shown by our proper motions shows that post-shock
plasmas can reach temperatures in excess of $10^7$ Kelvin, a result confirmed by
X-ray detection of the brightest HH objects in this outflow \citep{Pravdo2009}.  
The detection of
a parsec-scale bubble provides further evidence that the outflow cavity produced
by this outflow is likely filled with low-density EUV-ionized and soft-X-ray plasma,
with an emission measure too low to be detected by current methods.  The 
radiation produced as plasma recombines into neutral 
hydrogen may be capable of dissociating a pre-existing molecular outflow
produced during earlier evolutionary phases of IRAS~18162-2048.

The IRAS~18162-2048 outflow may be the best example of the extreme
feedback impacts of forming massive stars on their birth environment. 
As MYSOs grow from under 1 \Msol\  Solar mass to over 10 \Msol , they
likely create bipolar molecular flows as fast disk winds and jets 
sweep-up ambient material.  Their mechanical power increases with
source luminosity and mass \citep{Maud2015}.  The size of the entrained 
molecular outflow lobes will be limited to the size of their parent molecular 
clump or host cloud.  When outflows punch out of their host molecular 
clouds, the entrained gas may be predominantly atomic   
or even ionized as is the case with the parsec-scale 
HH~80/81 flow.   

If  the shock-ionized plasma produces
only one ionizing photon as it recombines, the minimum ionizing photon
luminosity of the shock will be given by the rate at which the medium is 
swept up.   The cross-sectional area of the shock, $\rm \pi R^2_s$, 
times the density $n$ of the medium into which it is running, times the
shock speed $\rm V_s$ gives a minimum on ionizing luminosity.
For a shock with a radius
$\rm R_s = 10^{17}$ (5\arcsec\ at a distance of 1.4 kpc), and speed
$\rm V_s = 10^3$ \kms , running into a medium with a hydrogen density 
$\rm n = 10^2$ \cmq , we get 
$\rm Q_{min} = \pi  R^2 _s n V \sim 3 \times 10^{44}$~photons~s$^{-1}$.

With a post-shock temperature of order 1 to 10 MK, much of this radiation
will emerge as X-rays and EUV radiation.  Reprocessing of X-ray
photons into softer {\bf EUV} photons by the surrounding medium can increase 
the Lyman continuum luminosity of the shock by more than an order of 
magnitude over the above rough estimate. Additionally, multiple shocks 
will increase  the UV luminosity.  Finally, fast shocks propagating 
into a previously swept-up molecular shell may directly contribute 
to the dissociation of bipolar molecular outflow lobes.  

The 2018 [\Sii ] and 1995 \Ha\ images show that a cylindrical region
with a  radius of order $\rm 10^{18}$ cm and a length of
more than a parsec contains shocks emitting in these species.  Unfortunately,
because of the failure of the 2018 \Ha\ images, the proper motions of
fainter features away from the jet axis could only be measured in [\Sii ]. 
The proper motions in [\Sii ]  show expansion away from the jet axis 
with speeds of between 100 to 300 \kms .  As discussed
above, it is likely that the \Ha\ and [\Sii ] emission arises in reverse shocks to
keep much of the sulfur in its first ionization stage.  The forward shocks
must then have higher excitation and likely dissociate any molecules
they encounter, including those associated with any previously swept-up
bipolar outflow.  Using the above formula, the Lyman continuum luminosity 
of the [\Sii ] / \Ha\ emission region is likely to be larger than
$\rm 10^{46}$ s$^{-1}$ due to its larger area.

Although we do not have any direct measurements of the total outflow mass, 
a crude estimate of its energetics is possible based on the empirical 
relations found for other MYSO outflows \citep{Maud2015}.  It is likely that
the fast speeds found here are what mostly distinguishes this flow
from other MYSO outflows.  If we assume that over its formation
IRAS~18162-2048 ejected a mass of 0.1 \Msol\ in a fast, $\rm 10^3$ \kms\
jet, the kinetic energy of this component would be $\rm E \sim 10^{48}$
ergs.  This is comparable to the energy required to dissociate about 50
to 100 \Msol\ of \Htwo\ by fast shocks and their UV radiation.   Thus,
this fast outflow could destroy its own fossil molecular outflow. If
this scenario is correct, then most of the outflow mass should be 
atomic or ionized.  Future sensitive 21 cm HI , 158 \mm\ \Cplus\ ,
or 63 \mm\ \Oi\ observations may be used to measure its mass.

There has been some  discussion of outflow-triggered star-formation in the 
literature with the case of HH~80N being an important potential example of 
this process \citep{Molinari2001,Girart2001,Masque2011}.  However, it remains 
unclear if the clump (IRDC) ahead of HH~80N~A was already forming stars before the 
IRAS~18162-2048 outflow impacted its environment.  Although shocks may 
alter the chemistry at the cloud surface, it is unclear if they exerted 
any significant dynamical influence.

\subsection{Comparison with Other Nearby MYSOs}

Other nearby (less than 1.4 kpc) massive-star forming complexes producing 15 to
20~MYSOs provide  interesting comparisons to  the IRAS~18162-2048 radio jet, 
MHO, and HH outflow complex.   We briefly comment on the Orion OMC1, Cepheus A, 
and Sh2-106 regions. 

The BN/KL outflow complex from Orion OMC1 located $\sim$0.1 pc behind the Orion nebula
contains a $\sim$15~\Msol\  MYSO, radio source I \citep{Ginsburg2018,Wright2023}, 
the $\sim$10~\Msol\  Becklin-Neugeabauer (BN) object, and the $\sim$3~\Msol\ source x.
These three protostars were ejected by a dynamic interaction about $\sim$550
years ago with speeds of $\sim$10, 30, and 55 \kms , respectively.  This event was 
associated with a $\sim 10^{48}$ erg explosion which launched about 10~\Msol\ of 
molecular gas. The explosion created  hundreds of molecular
streamers seen in CO, many of which are associated with shock-excited fingers of
molecular hydrogen.  The fastest proper motions in the fingertips are in excess of
400 \kms .  For a recent discussion of this outflow and associated  runaway stars, 
see \citet{Bally2020}.  Source I (Src I) drives a very young outflow
(dynamic age $<$300 years) which may be powered by an ionized jet 
launched along the axis of the circumstellar disk which survived the dynamic 
interaction \citep{Wright2023}.  The speed of this jet is not yet measured.

The $\sim$15~\Msol\ MYSO, HW2 in Cepheus A,  drives a radio jet
exhibiting proper motions of $\sim$500~\kms\  \citep{CarrascoGonzelez2021}. 
Cepheus A contains a spectacular shock-excited molecular hydrogen and CO
outflow complex.   The multiple chains of MHOs and HH objects originate from
the vicinity of HW2 and have been interpreted in terms of a pulsed, precessing
jet launched by this massive protostar \citep{Cunningham2009}. 

Sh2-106 contains about 3~\Msol\ of plasma, indicating that it is a well 
developed \Hii\ region. Sh2-106 is ionized by an embedded O9 star, S106IR,  
which is obscured by over A$_V \approx$20 magnitudes, and may be surrounded by a circumstellar disk \citep{Bally2022}.   Instead
of a fast wind, S106~IR drives a relatively slow, $\sim$100 to 400 \kms\ wind and 
there is no evidence of a jet.   The estimated masses of S106~IR and 
IRAS~18162-2048 are $\sim$15 and $\sim$20 \Msol , similar to within a factor of two.  
A key difference between these two stars may be the masses of their circumstellar 
disks and their evolutionary stages. 
The S106~IR disk has a mass less than 0.8 \Msol\ while the disk surrounding 
IRAS~18162-2048 may have a mass of $\sim$5~\Msol .  
Additionally, the accretion histories of these MYSOs 
may have been different.

\section{Conclusions}

A comparison of narrow-band images obtained with the Hubble Space Telescope
taken in 1995 and 2018 in [\Oiii ], \Ha\ and \Hb , and [\Sii ] reveals
significant and complex changes in the shock morphologies of HH~80 and HH~81.   Some features
have disappeared and others have appeared.  Where persistent patterns can be
recognized in the two epochs, proper motions were measured.  Proper motions
ranging up to 1,200 \kms\ are found.  However, there is a huge dispersion
in the measured values.  

As a rule, the fastest motions, up to 1,200 \kms\ are seen close to the 
extrapolated radio jet  axis in relatively dim features.   The brightest 
shocks tend to exhibit slower motions in the range of 200 to 400 \kms .
We interpret this as evidence that the bright shocks are produced where
fast flows are impacting slower or stationary obstacles.  The images reveal 
expanding arcs or partial rings around these regions.  In these rings, 
some proper motions are backwards toward the source.   We use 
published spectra to infer the radial velocities of the brightest shocks.
The ratio of radial velocity to proper motions of these shocks are used to
re-derive the inclination angle of the flow with respect to the plane of
the sky.  This angle is found to be about {\bf 44 to 65\arcdeg }, consistent with 
previous measurements.   The backwards and sideways motions are thus 
consistent with the splash of post-shock gas as it moves around a slower 
obstacle.

Away from the jet axis, [\Sii ] emission exhibits extensive but slow
proper motions orthogonal to the jet, indicating that the outflow is
creating an expanding bubble.   As this ultra-fast  flow bursts out 
from behind the L291 cloud into the surrounding, lower density ISM, 
it has inflated a parsec-scale bubble seen in new images as a network
of \Ha\ filaments.  

We identify a chain of faint HH objects and MHOs in the expected
counterflow located on the opposite side of IRAS~18162-2048 to
HH~80/81.    Dim \Ha\ and [\Sii ] trace these HH objects.  More
extensive but dim 2.12 \mm\ \Htwo\ emission is seen between the GGD~27
reflection nebula and star-forming clump located ahead of HH~80N~A.   
Given the 10 parsec projected length of this outflow as traced by radio 
continuum, HH objects, and MHOs, it is remarkable that only a tiny molecular 
outflow is seen in the immediate vicinity of the IRAS source.
The outflow may have burst out of the host cloud core and is primarily
interacting with mostly atomic or ionized inter-cloud or inter-clump
gas.  Slower moving debris may represent left-over fragments of
a bipolar outflow produced when IRAS~18162-2048 was much less massive and
drove a less powerful molecular outflow.  Shocks could have directly
destroyed molecules in such flows.  Additionally UV radiation 
fields produced by the ultra-fast shocks in this outflow may have 
contributed to the dissociation of molecules.  

We briefly discuss possible models for the ultra-fast radio jet and proper
motions observed in HH~80 and 81.  It is possible that a magneto-centrifugal 
wind is launched by a strong magnetic field anchored to the $\sim$5~\Msol\
disk.  Photo-ionization of the inner disk surface layers may load 
plasma onto open field lines whose footprints co-rotate with the
disk.  If these field lines rotate rigidly as they expand to larger
radii above and below the disk plane, they can accelerate the plasma 
to the observed $>$1,000 \kms\ speeds.  Hoop stress in the azimuthal component
of the field can collimate the accelerated, radial flow into a jet.

Sensitive future soft X-ray, 21 cm atomic hydrogen, deep radio continuum,
and mid-IR tracers such as the 158 \mm\ \Cplus\ and 63 \mm\ [\Oi ]
may reveal components of this outflow that may better illuminate its
powerful feedback impacts by completing the inventory of mass, 
momentum, and energy in this giant outflow.


\section*{Acknowledgments}

J.B.  acknowledges support by National Science Foundation through grant No. 
AST-1910393 and AST-2206513.   
BR acknowledges support by NASA through grant HST-GO-15353.

Based on observations made with the NASA/ESA Hubble Space Telescope and obtained 
from the Hubble Legacy Archive, which is a 
collaboration between the Space Telescope Science Institute (STScI/NASA), 
the European Space Agency (ST-ECF/ESAC/ESA), and the Canadian Astronomy 
Data Centre (CADC/NRC/CSA).  Some/all of the data presented in this paper 
were obtained from the Mikulski Archive for Space Telescopes (MAST). STScI 
is operated by the Association of Universities for Research in Astronomy, Inc., 
under NASA contract NAS5-26555.

Some of the work presented here is based on observations obtained with the 
Apache Point Observatory 3.5-meter telescope, which is owned and operated 
by the Astrophysical Research Consortium.    We thank the Apache Point Observatory 
Observing Specialists for their assistance during the observations. 
We thank Mr. Zen Chia for assisting with some of the 
observations and data reduction.  

We thank Nayab Gohar who did the initial 
proper motion measurements on the HH~80/81 system using our multi-epoch HST data.

\subsection{Appendix:  Proper Motions Shown as Color and Animated gif Images}

In this Appendix, we show proper motions as color images with the 1995
epoch images shown in red and the 2018 epoch image shown in cyan.

\begin{figure*}
\center{
\includegraphics[width=6.5in]{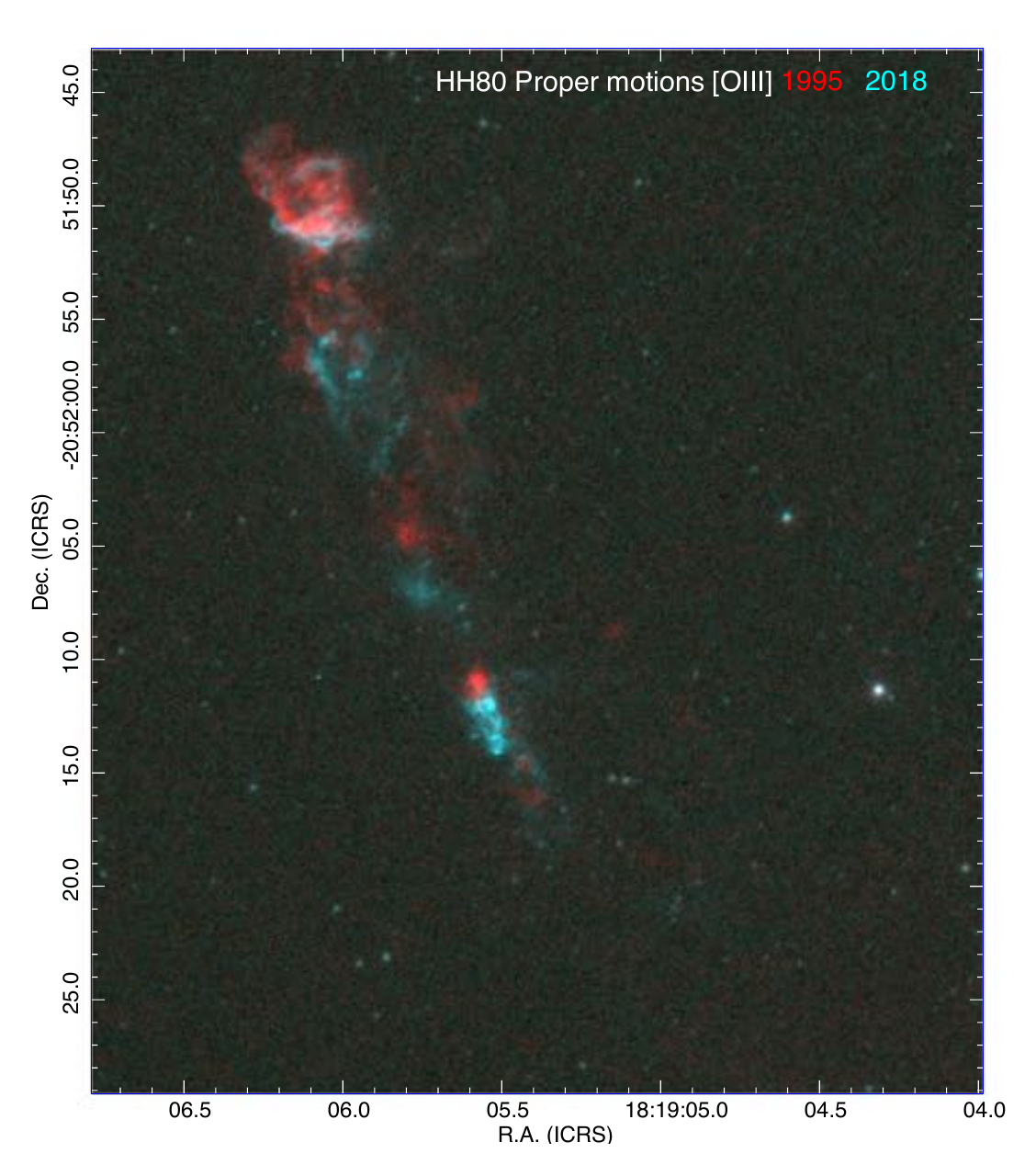  }
}
\caption{{\bf Proper motions of features in  HH~80 in [\Oiii ]} .  Red shows 1995 epoch \Ha\ observed with WFC2's WF channel.  Cyan shows 2018 epoch [\Oiii ].}
    \label{fig_PMs_HH80_color_Oiii_sub}
\end{figure*}

\begin{figure*}
\center{
\includegraphics[width=6.5in]{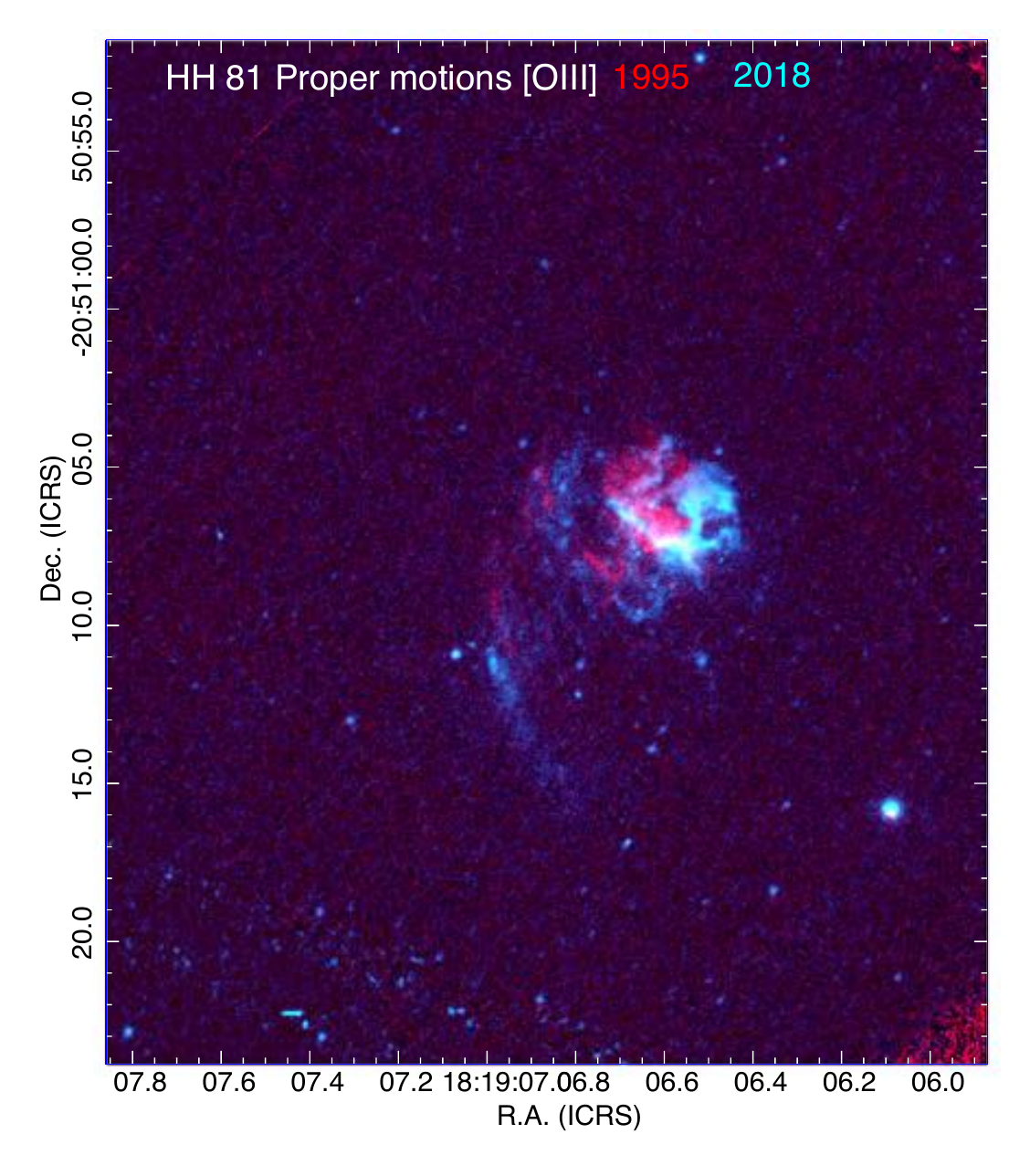 }
}
\caption{{\bf Proper motions of features in  HH~81 in [\Oiii ]}.  Red shows 1995 epoch \Ha\ observed with WFC2's PC channel.  Cyan shows 2018 epoch [\Oiii ].}
    \label{fig_PMs_HH81_color_Oiii_PC}
\end{figure*}

\begin{figure*}
\center{
\includegraphics[width=6.5in]{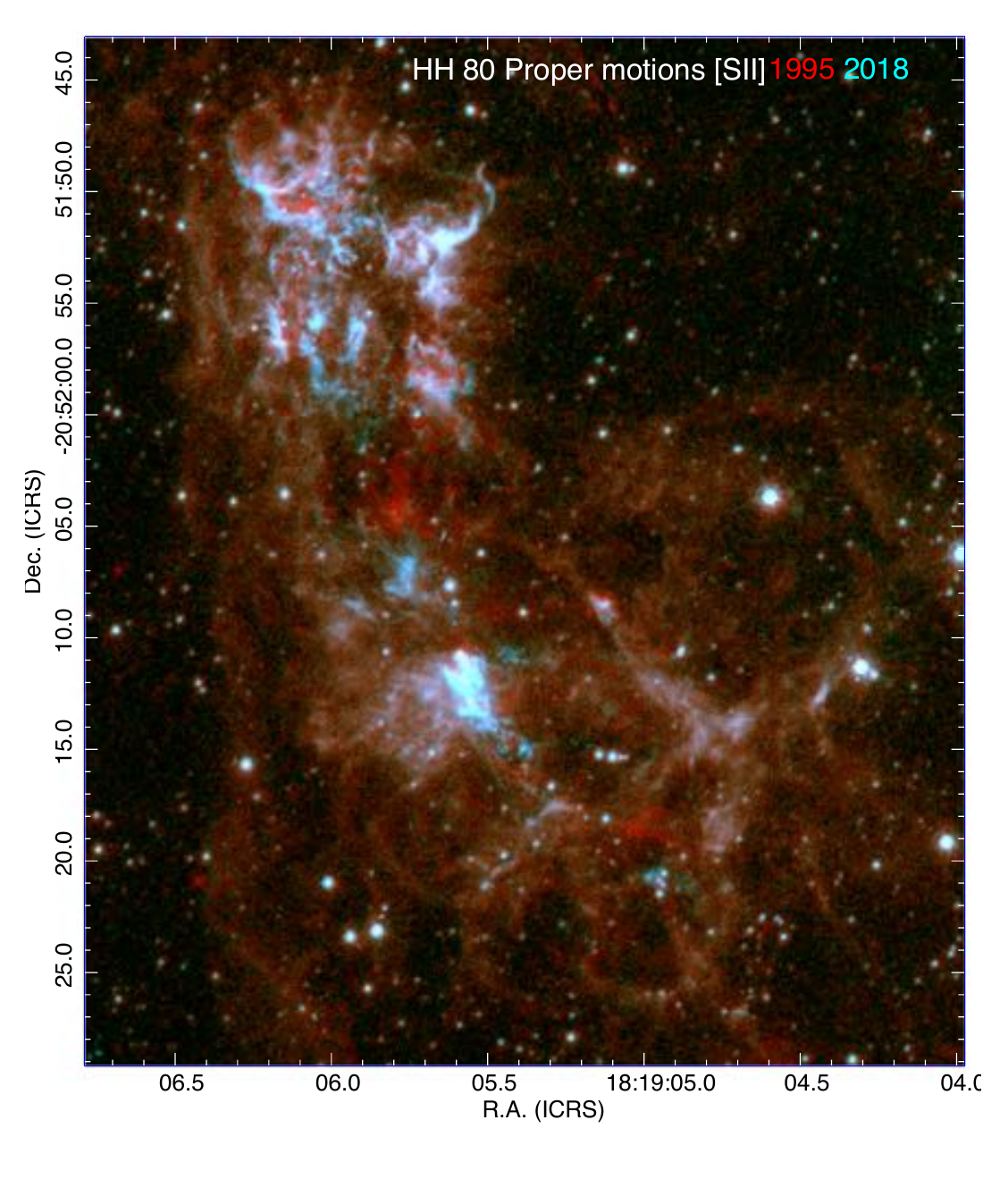  }
}
\caption{{\bf Proper motions of features in HH~80 in [\Sii ]}.  Red shows 1995 epoch \Ha\ observed with WFC2's WF channel.  Cyan shows 2018 epoch [\Sii ].}
    \label{fig_PMs_HH80_color_Sii_sub}
\end{figure*}

\begin{figure*}
\center{
\includegraphics[width=6.5in]{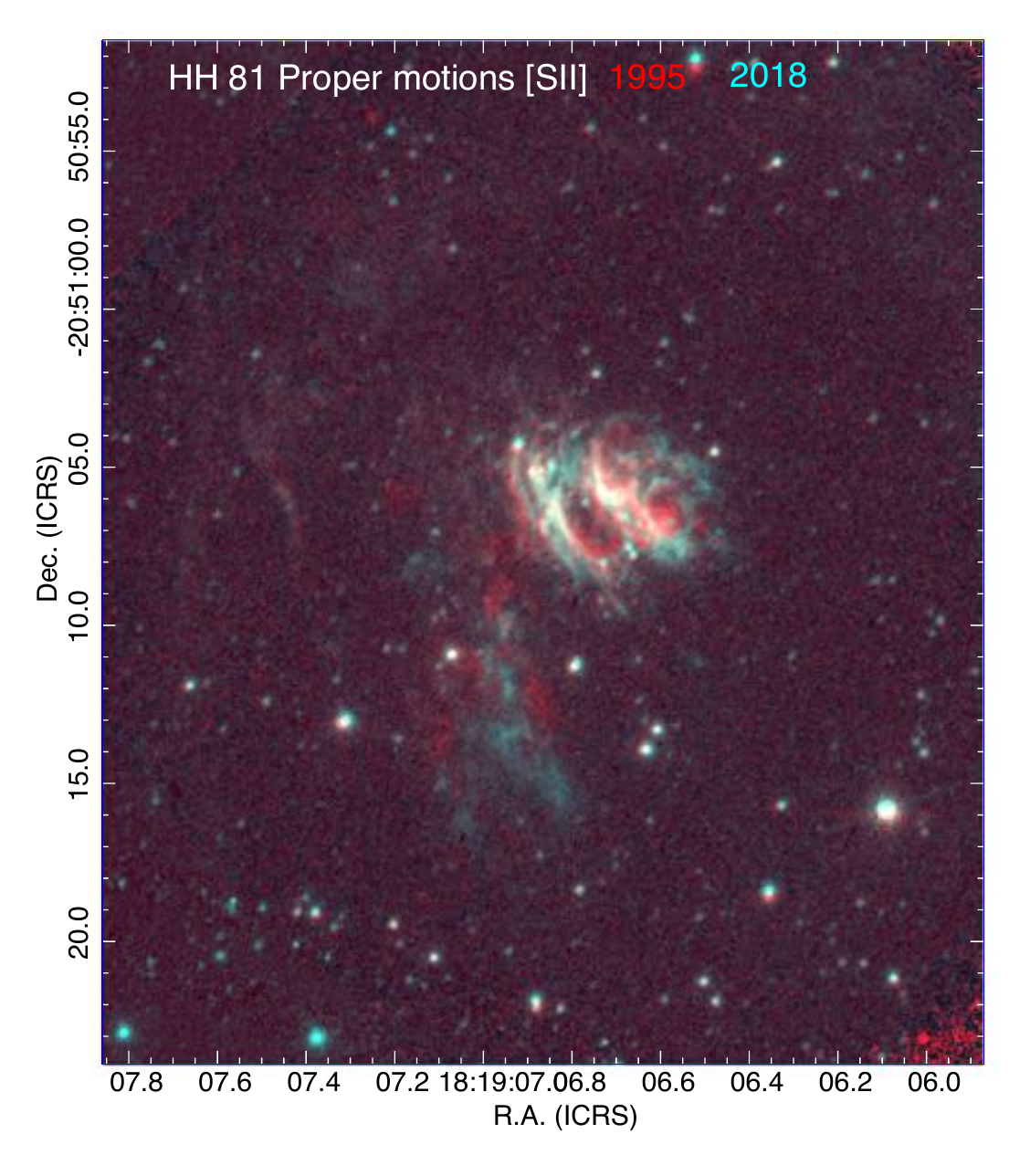  }
}
\caption{{\bf Proper motions of features in HH~81 in [\Sii ]}.  Red shows 1995 epoch \Ha\ observed with WFC2's PC channel.  Red shows 1995 epoch Cyan shows 2018 epoch.}
    \label{fig_PMs_HH81_color_Sii_PC}
\end{figure*}

\begin{figure*}
\center{
\includegraphics[width=6.5in]{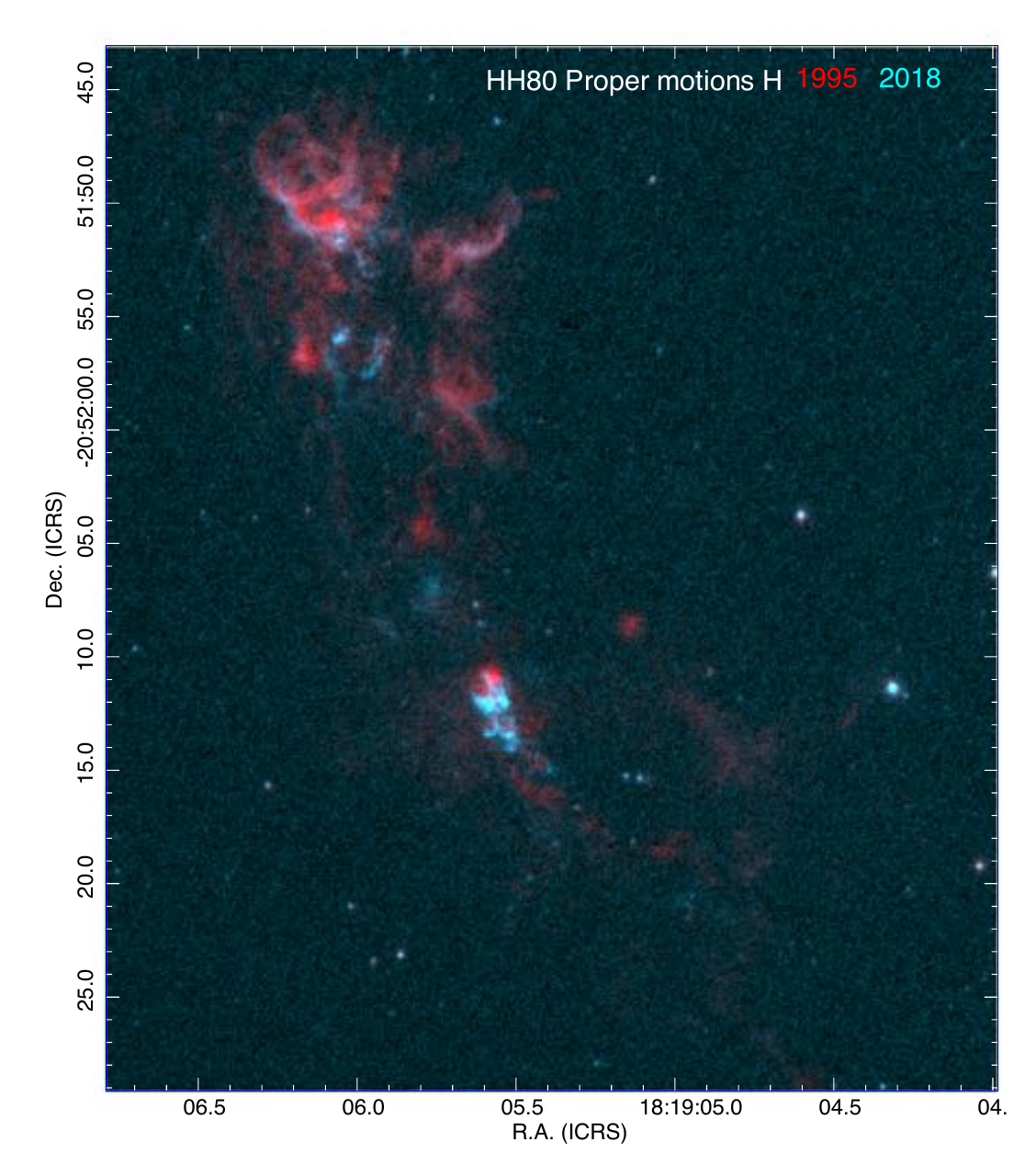  }
}
\caption{{\bf Proper motions of features in HH~80}  in hydrogen.  Red shows 1995 epoch \Ha\ observed with WFC2's WF channel.  Cyan shows 2018 epoch \Hb .}
    \label{fig_PMs_HH80_color_H_sub}
\end{figure*}

\begin{figure*}
\center{
\includegraphics[width=6.5in]{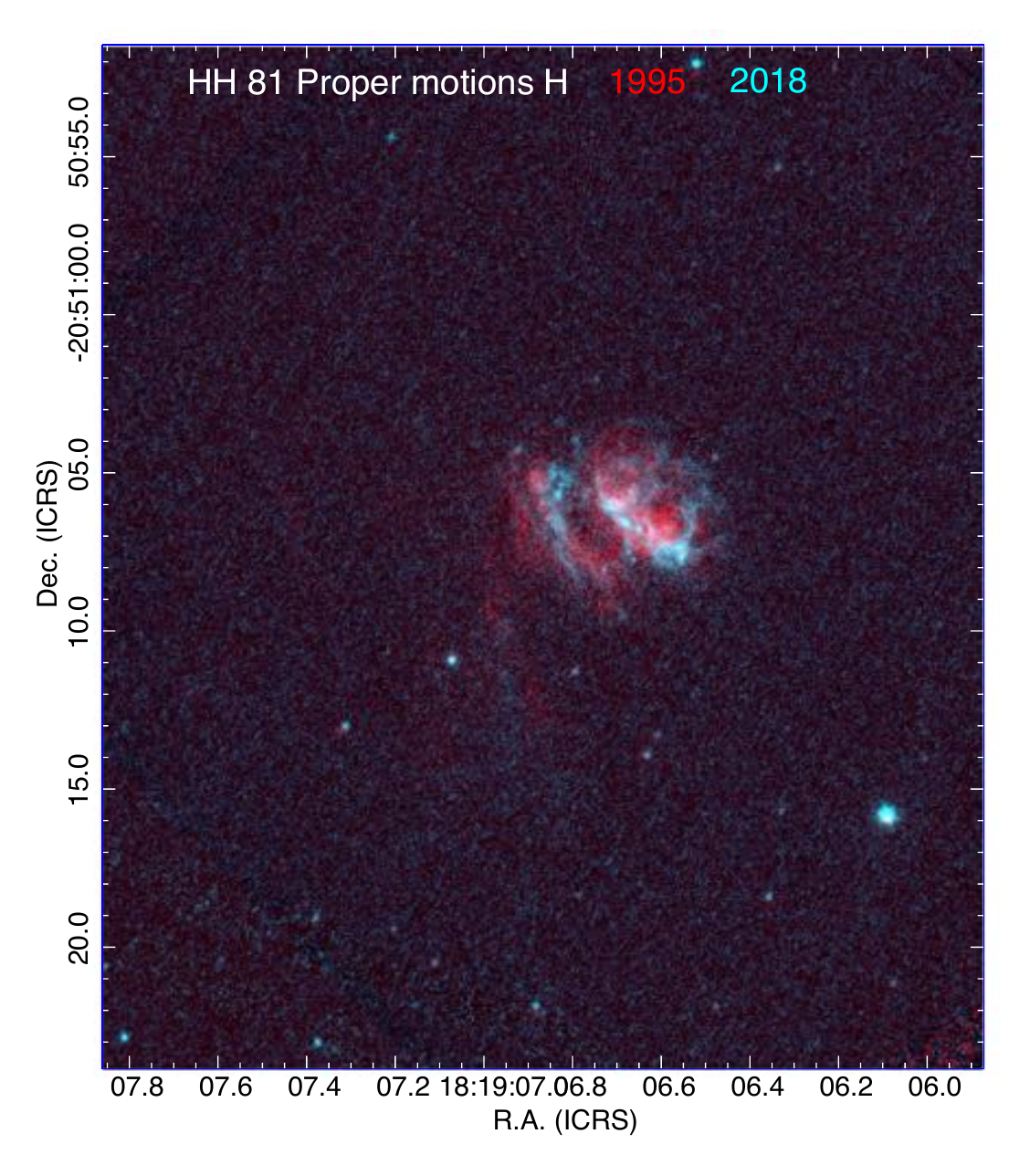  }
}
\caption{{\bf Proper motions of features in HH~81} in hydrogen.  Red shows 1995 epoch \Ha\ observed with WFC2's PC channel.  Cyan shows 2018 epoch \Hb .}
    \label{fig_PMs_HH81_color_H_PC}
\end{figure*}

Figure \ref{fig_PMs_HH80_color_Oiii_sub} shows HH~80 in [\Oiii ] in 1995 (red) and 2018 (cyan).
Figure \ref{fig_PMs_HH81_color_Oiii_PC} shows HH~81 in  [\Oiii ] in 1995 (red) and 2018 (cyan).
Figure \ref{fig_PMs_HH80_color_Sii_sub} shows HH~80 in [\Sii ] in 1995 (red) and 2018 (cyan).
Figure \ref{fig_PMs_HH81_color_Sii_PC} shows HH~81 in [\Sii ] in 1995 (red) and 2018 (cyan).
Figure \ref{fig_PMs_HH80_color_H_sub} shows HH~80 in \Ha\ in 1995 (red) and 2018 (cyan).  
Figure \ref{fig_PMs_HH81_color_H_PC} shows HH~81 in \Hb\ in 1995 (red) and 2018 (cyan).


\begin{figure*}
\center{
\plottwo{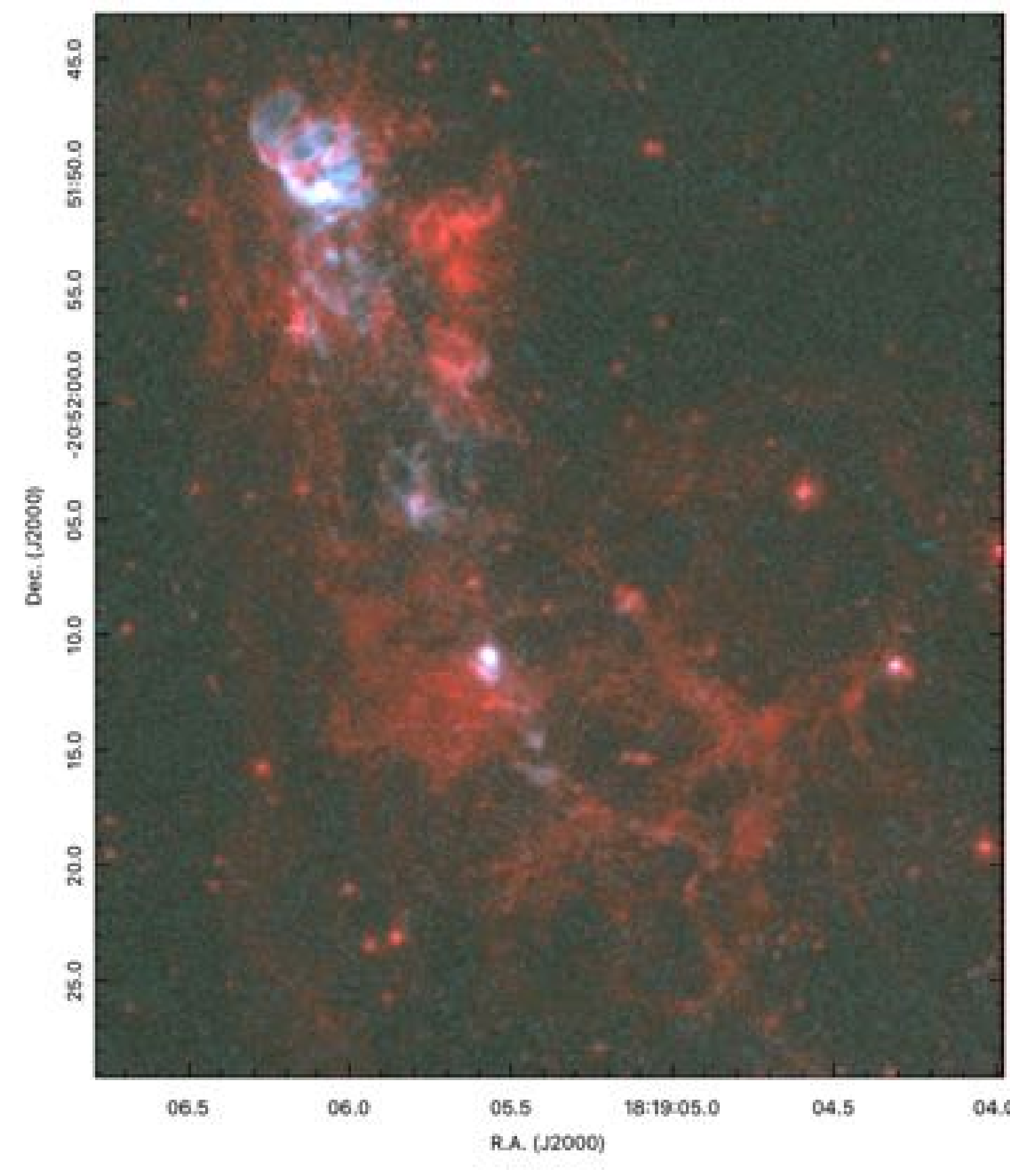}{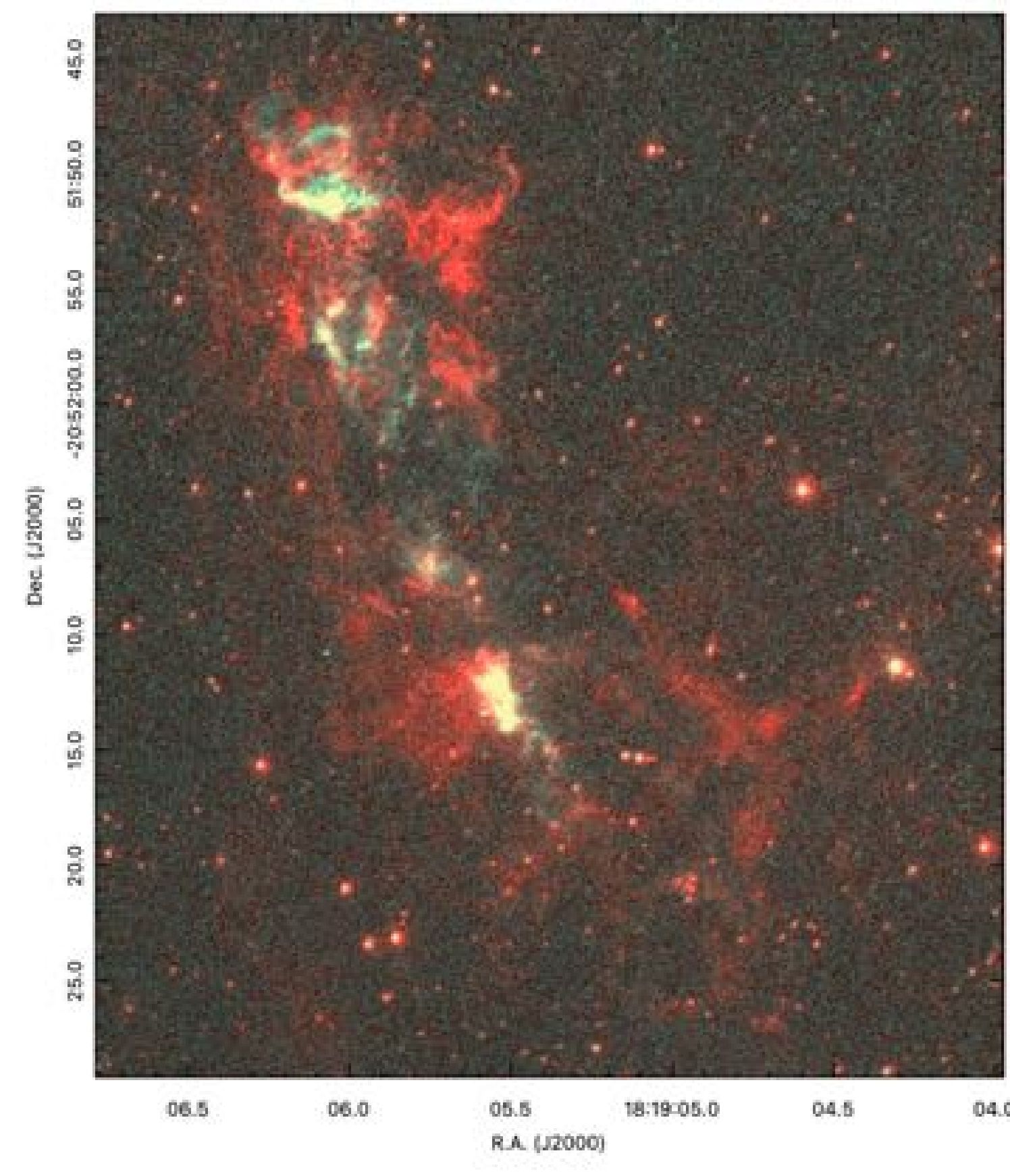}
}
\caption{HH~80 with [\Sii ] in red and [\Oiii ] in cyan in 1995 (left) and 2018 (right). An interactive version of this figure is available. Clicking on the image will switch between the 1995 and 2018 images to show the changes and motions between the 23 year span.
         }
    \label{if25}
\end{figure*}

\begin{figure*}
\center{
\plottwo{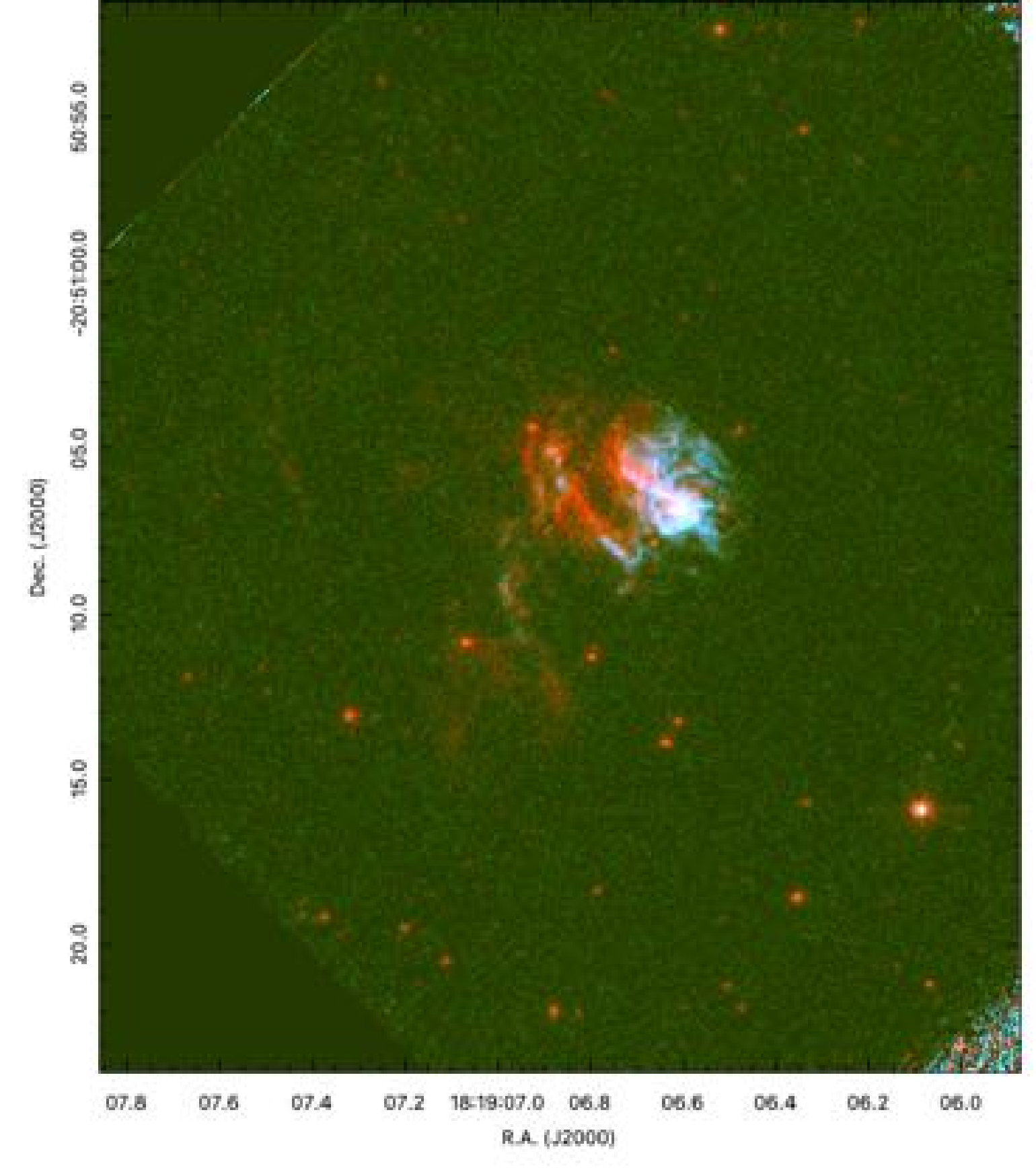}{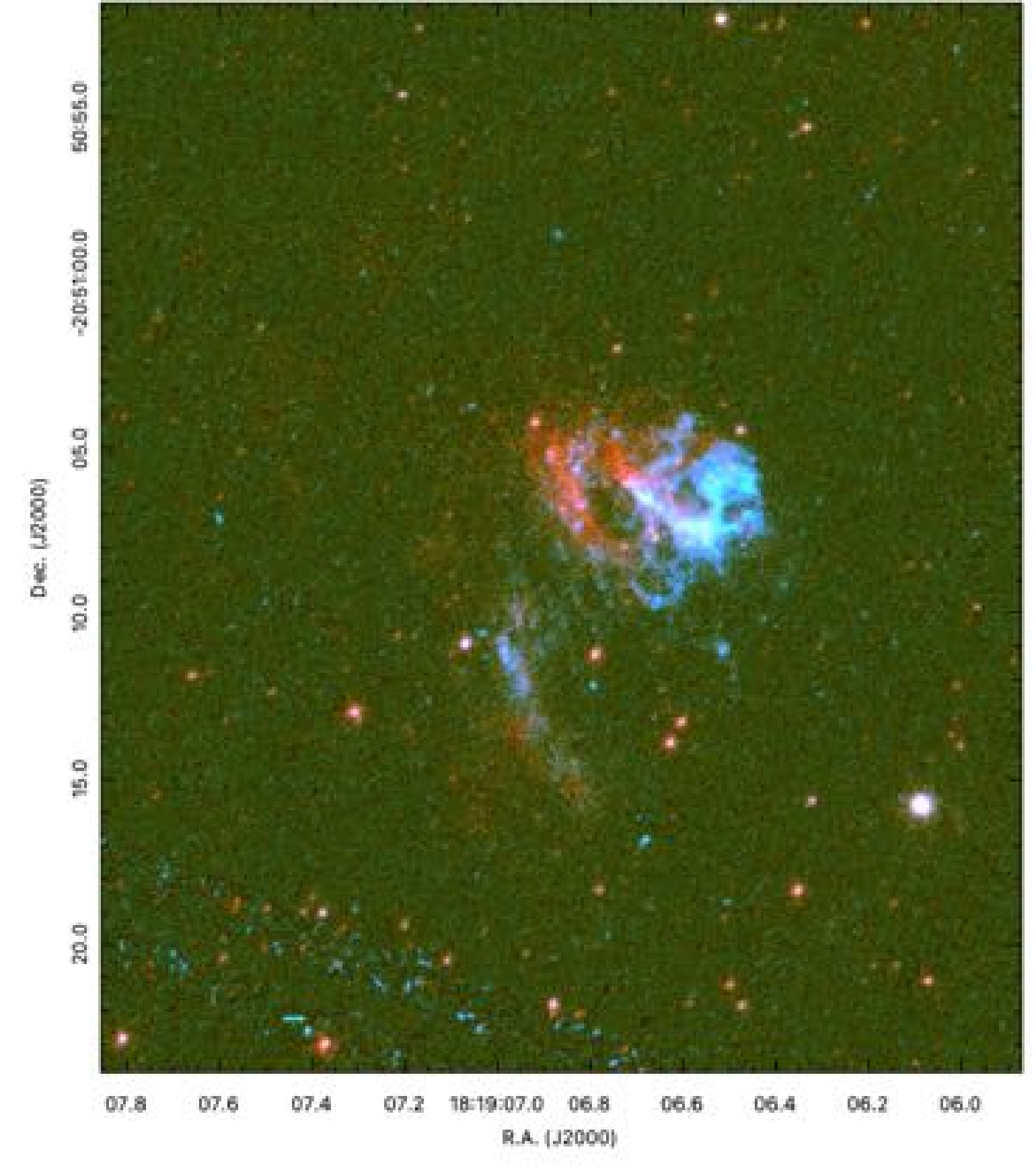}
}
\caption{HH~81 with [\Sii ] in red and [\Oiii ] in cyan in 1995 (left) and 2018 (right). An interactive version of this figure is available. Clicking on the image will switch between the 1995 and 2018 images to show the changes and motions between the 23 year span.
         }
    \label{if26}
\end{figure*}

Figures \ref{if25} and \ref{if26} are two colored images showing the changes and motions in HH~80 and HH~81 between 1995 and 2018 in which the [\Sii ] emission is shown in red and the [\Oiii ] emission is shown in cyan. 

\end{document}